\documentclass[aps,prb,reprint,superscriptaddress,longbibliography,floatfix]{revtex4-2}
\usepackage[utf8]{inputenc}
\usepackage[T1]{fontenc}
\usepackage{amsmath,amssymb,bm,braket,graphicx,xcolor}
\usepackage[hidelinks]{hyperref}
\begin{document}

\title{Topological Narrow Bands in Twisted Rhombohedral Multilayer Graphene}

\author{Alejandro Jimeno-Pozo}
\email{alejandro.jimenopozo@physics.uu.se}
\altaffiliation[Present address: ]{Department of Physics and Astronomy, Uppsala University, Box 524, 751 20 Uppsala, Sweden.}
\affiliation{Donostia International Physics Center, Paseo Manuel de Lardiz\'abal 4, San Sebasti\'an 20018, Spain}
\affiliation{Center for Theoretical Physics, Hainan University, Haikou 570228, China}
\affiliation{School of Physics and Optoelectronic Engineering, Hainan University, Haikou 570228, China}

\author{Danna Liu}
\affiliation{Center for Theoretical Physics, Hainan University, Haikou 570228, China}
\affiliation{School of Physics and Optoelectronic Engineering, Hainan University, Haikou 570228, China}

\author{Francisco Guinea}
\affiliation{Donostia International Physics Center, Paseo Manuel de Lardiz\'abal 4, San Sebasti\'an 20018, Spain}
\affiliation{IMDEA Nanoscience, C/ Faraday 9, 28049 Madrid, Spain}

\author{Pierre A. Pantale\'on}
\email{pierre.pantaleon@hainanu.edu.cn}
\affiliation{Center for Theoretical Physics, Hainan University, Haikou 570228, China}
\affiliation{School of Physics and Optoelectronic Engineering, Hainan University, Haikou 570228, China}
\affiliation{IMDEA Nanoscience, C/ Faraday 9, 28049 Madrid, Spain}

\begin{abstract}
Moiré graphene multilayers provide an atomically thin platform for engineering narrow topological bands through stacking, twist, and electrostatic control. Recent experiments on twisted rhombohedral graphene have revealed a high-Chern-number hierarchy determined by the number of rhombohedral layers. A microscopic description of this hierarchy requires connecting the ideal chiral picture to the realistic electronic structure. Here, we show that the two central moiré bands of twisted monolayer–\(N\)-layer rhombohedral graphene originate from hybridization between a monolayer Dirac cone and a rhombohedral surface state located at opposite corners of the moiré Brillouin zone. Increasing \(N\) progressively narrows these bands, whereas reversing the displacement field transfers their wave-function weight between the twisted interface and the remote surface. In spectrally isolated regimes, this redistribution switches the lower band between Chern numbers \(\mathcal{C}=N\) and \(\mathcal{C}=-1\). A projected surface theory shows how interfacial tunneling is dressed by virtual propagation through the rhombohedral stack, explaining the field-selected Chern hierarchy and the recurring threefold structure of the real-space charge textures. These results establish layer number and displacement field as complementary nanoscale controls of band flatness, spatial polarization, and topology in graphene moiré heterostructures.
\end{abstract}

\maketitle
\let\origaddcontentsline\addcontentsline
\renewcommand{\addcontentsline}[3]{}

\section{Introduction}

Moir\'e graphene multilayers provide a highly tunable setting in which band flatness, interactions, and topology can be controlled by twist angle, stacking structure, carrier density, and displacement field. This tunability has led to correlated insulating states and superconductivity in magic-angle twisted bilayer graphene~\cite{Cetal18,Cetal18b,Yankowitz2019,Lu2019_SC_TBG,Stepanov2020_TBG,Oh2021UnconvSC}, orbital magnetism and anomalous Hall responses~\cite{Sharpe2019Ferromag,serlin2020intrinsic}, and electrically controlled Chern phases in several graphene moir\'e systems~\cite{nuckolls2020strongly,wu2021chern,Das2021Chern,xie2021fractional,Youngjoon2021,choi2021correlation,saito2021hofstader,pierce2021unconventional,stepanov2021competing}. Reviews and early twistronics experiments emphasize that twist angle, stacking chirality, and electric fields provide independent knobs for organizing correlated and topological phases~\cite{RibeiroPalau2018,Chu2020Review,Zhang2021Twist,Andrei2021Marvels,Ma2023Small,Saumya2023Magic,Baek2025Moire,Ling2026Recent,Yang2026Twisting}.

Twisted monolayer--bilayer graphene is the first member of the broader monolayer--multilayer family and already shows why layer character is central. Experiments have revealed electric-field-tunable van Hove singularities, correlated metallic and insulating states, orbital magnetic order, anomalous Hall effects, and electrically switchable Chern phases~\cite{chen2020electrically,Polshyn2020,xu2021tunable,Polshyn2021,he2021competing}. In particular, the electrical reversal of orbital magnetization in twisted monolayer--bilayer graphene shows that gate fields can do more than tune bandwidths: they can change which valley-Chern sector becomes relevant once interactions polarize the valley degree of freedom~\cite{Polshyn2020}. This is directly connected to the layer-resolved structure of the low-energy bands, because the sign and magnitude of the orbital response are tied to the Chern character and to the orbital texture carried by the occupied valley. Related work has shown that spin--orbit coupling can further reorganize the correlated states of twisted monolayer--bilayer graphene~\cite{park2026tuning}. Twisted double bilayer graphene provides a complementary multilayer example, where displacement fields isolate flat bands and stabilize correlated insulating and spin-polarized phases~\cite{burg2019correlated,Rickhaus2019,shen2020correlated,Cao2020SpinPolarized,Liu2020SpinCorrelated,he2020symmetry,Kuiri2022Spontaneous}. Related mono-trilayer structures have been predicted to host isolated flat bands, especially for ABC trilayers~\cite{Goodwin2021_flatbands}, while alternating-twist and general multilayer theories show that layer number and stacking chirality can organize flat bands and valley Chern numbers~\cite{Khalaf2019_magic,Liu2019,Tarnopolsky2019_origin,Ding2023Mirror,Xie2022_alternating,Wang2026emergent}.

Recent experiments have extended this logic to monolayer--multilayer and rhombohedral graphene platforms. High-Chern-number quantum anomalous Hall states, electrically tunable orbital magnetism, and interaction-reconstructed topological phases have been reported in twisted rhombohedral and mixed multilayer systems~\cite{Wang2025_TDTG,Liu2025_TwistMonoPenta,Wang2026High,Su2025,Dong2025_TwistBiTri,Waters2024,chen2026layer, Li2026Fractional}. In particular, twisted monolayer--multilayer rhombohedral graphene has been associated with a high-Chern hierarchy controlled by the number of rhombohedral layers~\cite{Wang2026High}, while twisted monolayer--pentalayer and twisted rhombohedral trilayer--bilayer devices show that the same combination of layer number, displacement field, and topology can generate large-Chern anomalous Hall and fractional topological responses~\cite{Liu2025_TwistMonoPenta,Dong2025_TwistBiTri}. These experiments establish the phenomenology, but they also leave a microscopic theoretical question: how does the higher-order chirality of an $N$-layer rhombohedral surface state become encoded in moir\'e bands, and how does a displacement field switch the layer and Chern character of those bands? Resolving this question is essential for understanding how electrically controlled high-Chern moir\'e bands and their associated multichannel chiral edge transport emerge from the underlying moiré electronic structure.

A central lesson from these systems is that the electric field does not only tune the bandwidth; it can also change the layer and sublattice character of the low-energy states. This has been observed directly in nanoARPES measurements of twisted monolayer--bilayer graphene, where reversing the electric field selectively enhances monolayer-like or bilayer-like spectral weight~\cite{zhang2024observation}. Small-angle monolayer--multilayer systems also exhibit reconstruction, soliton networks, and strain-controlled moir\'e textures~\cite{zhang2022domino,ding2024manipulation}. The same field-controlled layer-polarization principle becomes richer when the bilayer is replaced by an $N$-layer rhombohedral stack. In the idealized strict chiral approximation, where further-neighbour interlayer hoppings and same-sublattice moir\'e tunnelling are neglected, the two low-energy surface sites are coupled by an $N$th-order amplitude proportional to $|\boldsymbol{q}|^N$. Trigonal warping and next-nearest-layer hopping add lower-order virtual propagation paths while preserving the surface origin of the low-energy sector~\cite{Min2008Chiral,Koshino2009Trigonal,Makov2024Flat}. The precise model hierarchy is introduced in the path-projected section below. Increasing $N$ therefore enhances the tendency toward narrow surface-derived bands, although the detailed dispersion is reconstructed by the additional hopping processes. Scanning tunnelling microscopy and spectroscopy (STM/STS) on untwisted rhombohedral multilayers have directly resolved this layer-dependent flattening~\cite{Zhang2024Layer}. Theoretical calculations further predict that impurity-induced interference in the surface local density of states carries information about the layer number and Bloch-state structure~\cite{Dutreix2016Friedel}.

Here we study twisted $(N+1)$-layer rhombohedral graphene, formed by placing a rotated graphene monolayer on an $N$-layer ABC-stacked slab. We find that the low-energy structure is governed by the hybridization of two inequivalent sectors located at opposite corners of the moir\'e Brillouin zone: a monolayer Dirac cone at $K'$ and a rhombohedral surface sector at $K$. We compute band structures, layer-projected spectral weights, real-space charge densities, and Chern-number phase diagrams as functions of layer number, twist angle, and displacement field. We further derive a path-projected low-energy theory in which propagation between the two rhombohedral surfaces is represented by a coherent amplitude, denoted $X_N$ and derived below, that combines ordinary Dirac propagation, trigonal warping, and next-nearest-layer hopping. Its strict chiral limit reduces to the familiar $N$th-order surface coupling and makes the competition with the unit-winding monolayer Dirac cone explicit. This framework identifies the microscopic mechanism behind the experimental high-Chern hierarchy: increasing $N$ enhances band flattening, the displacement field exchanges the twisted-interface and remote-surface character of the two central bands, and the chiral surface limit organizes the Chern-number hierarchy and the recurring angular structure of the charge textures. In regions where the two central bands touch while remaining isolated from remote bands, the topology is assigned to the complete isolated two-band subspace, as defined in the topology section below.

\section{Results}

\subsection{Twisted monolayer--rhombohedral graphene}

\begin{figure}[!t]
\centering
\includegraphics[width=\columnwidth]{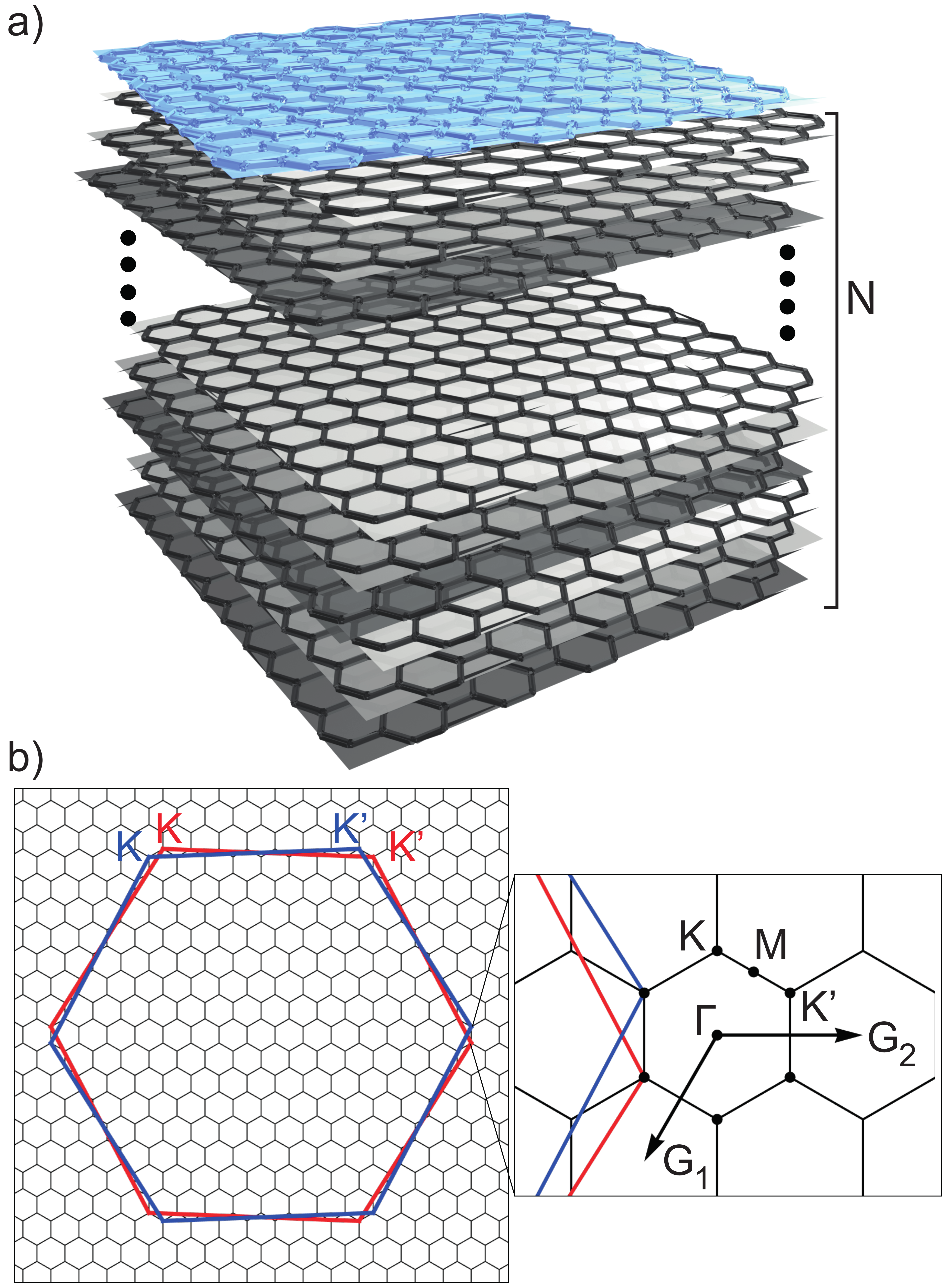}
\caption{\textbf{Twisted monolayer--rhombohedral graphene.} \textbf{a}, Schematic representation of twisted $(N+1)$-layer rhombohedral graphene, consisting of an $N$-layer ABC-stacked multilayer and a top monolayer rotated by a small angle $\theta$. \textbf{b}, Reciprocal-space construction of the moir\'e Brillouin zone. The relative rotation between the $K$ and $K'$ points of the top layer and the underlying multilayer generates the moir\'e reciprocal lattice vectors $\boldsymbol{G}_1$ and $\boldsymbol{G}_2$, and defines the high-symmetry points $\Gamma$, $K$, $K'$, and $M$.}
\label{fig:system}
\end{figure}

We consider a twisted graphene multilayer formed by placing a single graphene monolayer on top of an $N$-layer rhombohedral, or ABC-stacked, graphene slab, for a total of $N+1$ layers. We also include $N=2$, corresponding to an AB-stacked bilayer, as the simplest reference system. Rhombohedral stacking is distinguished from Bernal stacking only for $N\geq3$. The rhombohedral slab contains layers $l=1,\ldots,N$, with sublattices $A_l$ and $B_l$, arranged such that $B_l$ and $A_{l+1}$ are vertically aligned. The rotated monolayer is labeled by $l=0$. The top monolayer is rotated by $+\theta/2$, while the rhombohedral slab is rotated by $-\theta/2$, as shown in Fig.~\ref{fig:system}. The moir\'e reciprocal vectors are $\boldsymbol{G}_i=\boldsymbol{b}^{(0)}_i-\boldsymbol{b}^{(1)}_i$, with $i=1,2$. With the convention of Fig.~\ref{fig:system}, the Dirac cone of the rotated monolayer is located at the $K'$ point of the moir\'e Brillouin zone, whereas the low-energy touching of the rhombohedral stack is located at the $K$ point. The two low-energy objects hybridized by the moir\'e potential are therefore inequivalent: one is a monolayer Dirac cone and the other is an $N$th-order rhombohedral band touching. Their separation at opposite moir\'e Brillouin-zone corners is the geometric ingredient behind the band reconstruction discussed below.

The single-valley continuum Hamiltonian is written in layer-block form as
\begin{widetext}
\begin{equation}
H(\boldsymbol{q})=
\begin{pmatrix}
H_{00}(\boldsymbol{q}) & T_{01}(\boldsymbol{r}) & 0 & 0 & \cdots & 0 \\
T_{01}^{\dagger}(\boldsymbol{r}) & H_{11}(\boldsymbol{q}) & U_{12}(\boldsymbol{q}) & W_{13} & \cdots & 0 \\
0 & U_{21}(\boldsymbol{q}) & H_{22}(\boldsymbol{q}) & U_{23}(\boldsymbol{q}) & \ddots & \vdots \\
0 & W_{13}^{\dagger} & U_{32}(\boldsymbol{q}) & H_{33}(\boldsymbol{q}) & \ddots & W_{(N-2)N} \\
\vdots & \vdots & \ddots & \ddots & \ddots & U_{(N-1)N}(\boldsymbol{q}) \\
0 & 0 & \cdots & W_{(N-2)N}^{\dagger} & U_{N(N-1)}(\boldsymbol{q}) & H_{NN}(\boldsymbol{q})
\end{pmatrix}.
\label{eq:Hfull_layers_main}
\end{equation}
\end{widetext}
Here $T_{01}(\boldsymbol{r})$ is the moir\'e potential and is the only term that directly couples the rotated monolayer to the first rhombohedral layer. The blocks $U_{l,l+1}$ describe nearest-layer hopping inside the rhombohedral slab, while $W_{l,l+2}$ contains the next-nearest-layer coupling $\gamma_2$ and is present only for $N\ge3$. The effect of a perpendicular displacement field is represented by an outer-layer potential difference $V$, with onsite energies $-V/2$ on the rotated monolayer and $+V/2$ on the remote surface $L_N$. For moderate potential differences, this outer-layer profile provides a reasonable low-energy approximation, motivated by the surface character of the rhombohedral states and strong screening in ABC multilayers~\cite{Guinea2007Charge,Koshino2010Interlayer}. It retains the effective potential difference between the rotated monolayer and the remote rhombohedral surface. A self-consistent electrostatic treatment would refine the internal potential profile and the relation between the experimental displacement field and the model potential $V$. Sec.~\ref{sec:S9} shows that the Chern-number organization is preserved when the outer-layer profile is replaced by a uniform field with a linear potential drop across all layers. The explicit forms of $H_{ll}$, $U_{l,l+1}$, $W_{l,l+2}$, and $T_{01}$, together with all parameter values and the plane-wave implementation, are given in Methods and Sec.~\ref{sec:S1}.

\subsection{Moir\'e coupling transfers central-band weight to the rhombohedral stack}

The formation of the central bands can be followed directly by monitoring the layer character of the eigenstates as the moir\'e coupling is turned on. We scale the interfacial tunnelling as $T_{01}\to\alpha T_{01}$, where $\alpha=0$ gives a decoupled monolayer and rhombohedral slab and $\alpha=1$ gives the physical system~\cite{Long2024Evolution}. For each band and momentum, the layer weights are obtained from the plane-wave expansion defined in Methods by summing the eigenvector components over sublattice and moir\'e harmonics. At $\alpha=0$, the monolayer Dirac cone and the rhombohedral bands retain independent layer character. Turning on $\alpha$ couples the monolayer directly only to the first rhombohedral layer $L_1$, but the low-energy states of the ABC slab are surface states with amplitudes distributed between the interface side and the remote surface $L_N$. Consequently, the interfacial moir\'e tunnelling can redistribute wave-function weight throughout the rhombohedral slab rather than confining the hybridization to the two layers at the twisted interface.

\begin{figure*}[!t]
\centering
\includegraphics[width=\textwidth]{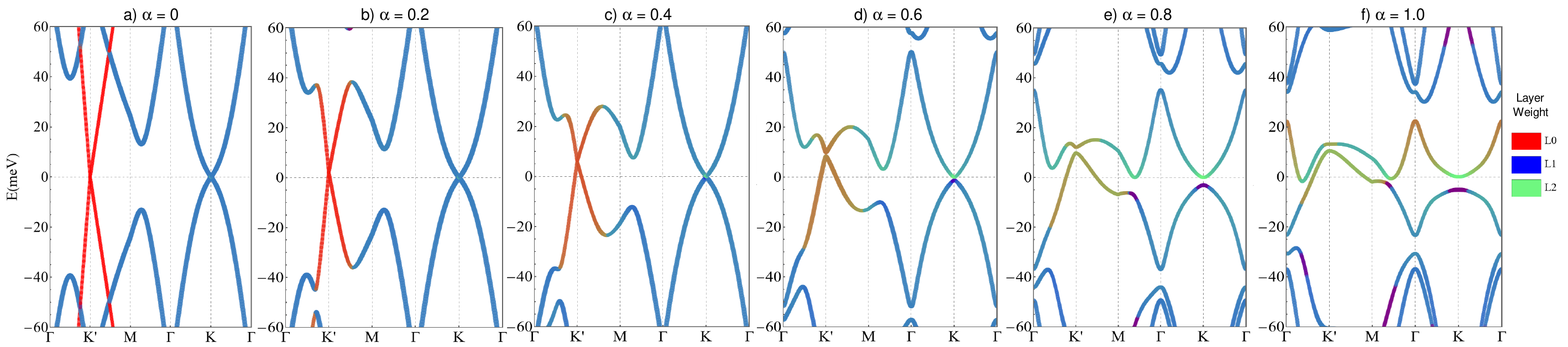}
\caption{\textbf{Layer hybridization driven by moir\'e coupling.} Evolution of the band structure of a monolayer coupled to an AB-stacked bilayer ($N=2$) at $\theta=1.05^\circ$ as the moir\'e tunnelling is scaled by $\alpha$. The value of $\alpha$ is indicated above each panel. Colour, line thickness, and opacity encode the layer composition following the conventions defined in Methods. At $\alpha=0$, the monolayer Dirac cone and AB-bilayer bands are decoupled. Increasing $\alpha$ produces avoided crossings and transfers wave-function weight from the monolayer into the bilayer and its remote surface. The extension to rhombohedral slabs with $N\geq3$ is shown in Sec.~\ref{sec:S3}.}
\label{fig:hybridization}
\end{figure*}

Figure~\ref{fig:hybridization} shows this evolution for a monolayer coupled to an AB-stacked bilayer ($N=2$) at $\theta=1.05^\circ$. This simplest reference system allows the redistribution of wave-function weight to be followed clearly as the moir\'e coupling increases. The dominant vertical hopping hybridizes $B_1$ and $A_2$ into a high-energy dimer, leaving $A_1$ and $B_2$ as the low-energy boundary sites at the interface and remote surface, respectively. At $\alpha=0$, the monolayer Dirac cone and the AB-bilayer bands cross without hybridizing and remain clearly separated by their layer character. As $\alpha$ increases, avoided crossings develop where the uncoupled branches approach one another, and the corresponding eigenstates exchange layer weight rather than undergoing a rigid energy shift. The central bands therefore evolve into mixed monolayer--bilayer states: their monolayer contribution is reduced, while an increasing fraction of the wave function is transferred first to the bilayer interface sector and subsequently to the remote surface $L_2$. At the physical coupling $\alpha=1$, neither central band is associated with a single isolated layer; instead, their layer composition reflects the coherent hybridization between the monolayer Dirac cone and the two-boundary low-energy sector of the AB bilayer. Although the detailed distribution of weight changes with layer number, see Sec.~\ref{sec:S3}, the underlying mechanism remains the same: the moir\'e tunnelling acts locally at the twisted interface, while the internal ABC coupling transmits this hybridization to the remote rhombohedral surface. The resulting interface-to-remote-surface transfer is therefore a persistent feature of the twisted monolayer--rhombohedral family~\cite{Pantaleon2026Interface}.

\subsection{Layer number and displacement field reshape the full-model bands}

\begin{figure*}[!t]
\centering
\includegraphics[width=\textwidth]{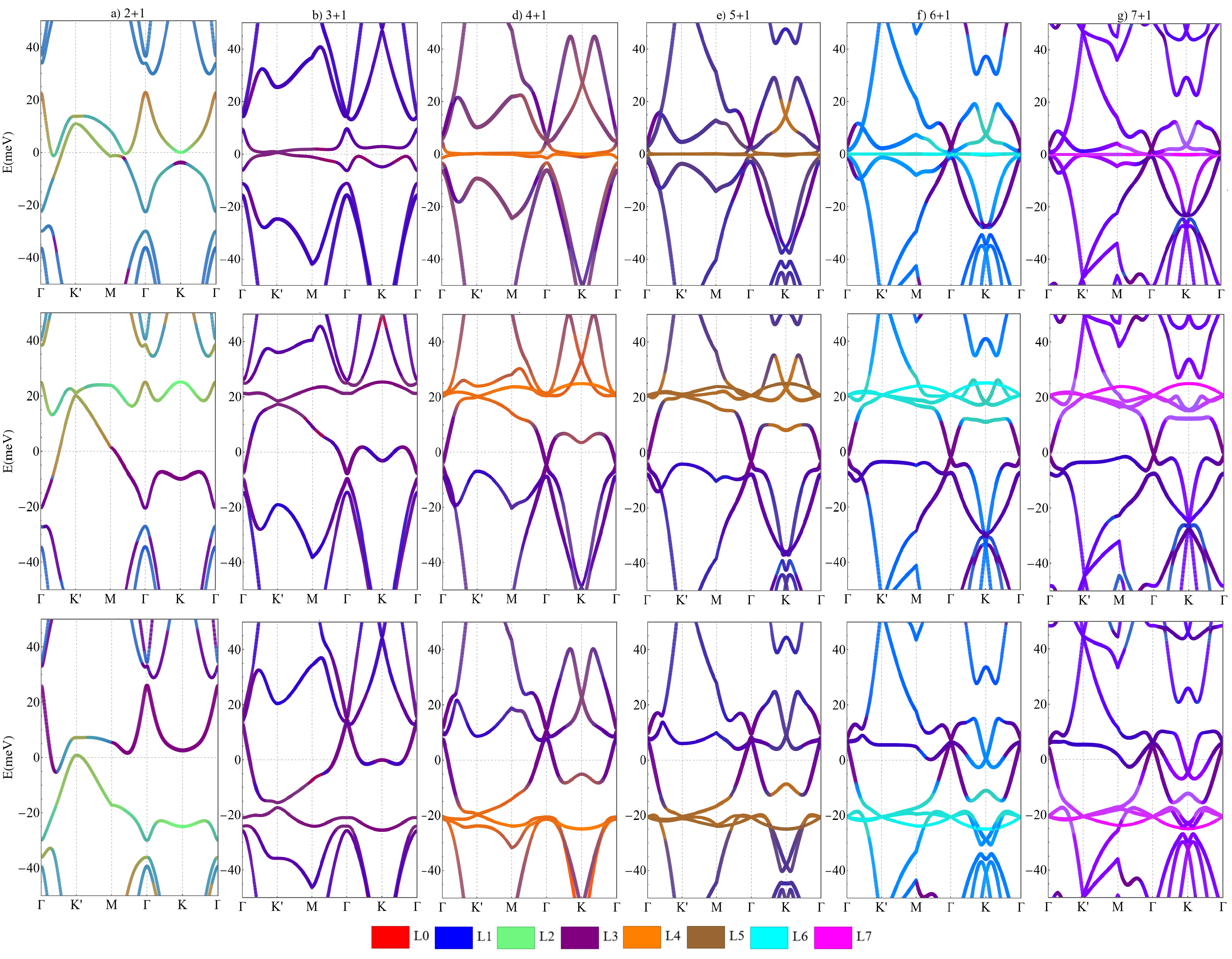}
\caption{\textbf{Layer-composition-resolved electronic structure of the full model.} Band structure of the full $(N+1)$-layer rhombohedral graphene model for $N=2$ to $N=7$ at $\theta=1.05^\circ$. The $N=2$ case is AB-bilayer, while $N\geq3$ corresponds to rhombohedral slabs. The top row corresponds to $V=0$ meV, the middle row to $V=+50$ meV, and the bottom row to $V=-50$ meV. The basis colour associated with each layer is indicated at the bottom of the figure; colour, line thickness, and opacity encode the layer composition following the conventions defined in Methods. Positive $V$ places the remote-surface character mainly in the upper central band and the twisted-interface character mainly in the lower central band, while negative $V$ reverses this layer polarization.}
\label{fig:full_bands}
\end{figure*}

We next analyse the full continuum spectra as a function of rhombohedral layer number and displacement field. Figure~\ref{fig:full_bands} shows the band structure for $N=2$ to $N=7$ at fixed twist angle $\theta=1.05^\circ$ and representative values of $V$. At zero displacement field, the two central bands become progressively narrower as the number of rhombohedral layers increases. This behaviour reflects the known low-energy surface structure of an isolated ABC stack: the interface and remote boundary sites communicate only through propagation across the intervening layers, so increasing $N$ suppresses their low-momentum coupling and makes the surface-derived dispersion progressively softer. The path-projected section below derives this statement explicitly by separating the low-energy boundary sites from the strongly hybridized interior dimers. After hybridization with the rotated monolayer, the moir\'e central bands inherit this increasingly soft rhombohedral surface dispersion. At the same time, increasing the number of layers introduces additional nearby bands and reduces their energy separation from the central manifold. Consequently, the progressive flattening of the central bands does not necessarily imply improved isolation from the remote bands.

In the screened approximation used here, the displacement field is represented by onsite potentials on the two outermost layers: the rotated monolayer $L_0$ is shifted by $-V/2$, the remote rhombohedral surface $L_N$ by $+V/2$, and the internal layers are set to zero potential. The field therefore changes the electrostatic detuning among the rotated monolayer, the interface layer $L_1$, and the remote surface $L_N$. Since the central bands are coherent mixtures of these sectors, changing $V$ does not produce a rigid displacement of the spectrum. Instead, it modifies the hybridization, avoided crossings, bandwidths, and gaps to nearby bands, producing the non-rigid reconstruction visible in Fig.~\ref{fig:full_bands}.

The layer-weighted representation makes this reconstruction directly visible. The basis colour assigned to each layer is indicated at the bottom of the figure, and the colour-blend, line-thickness, and opacity conventions are defined quantitatively in Methods. In the following, \emph{twisted-interface character} denotes a state whose leading contributions arise from the rotated monolayer $L_0$ and the first rhombohedral layer $L_1$, whereas \emph{remote-surface character} denotes a substantial contribution from the outer rhombohedral surface $L_N$.

For $V=+50$ meV, shown in the middle row of Fig.~\ref{fig:full_bands}, the upper central band is predominantly concentrated on the remote rhombohedral surface $L_N$, while the lower central band has primarily twisted-interface character arising from a mixture of $L_0$ and $L_1$. Neither band, however, is associated with a single pure layer. Their composition varies along the moir\'e Brillouin zone as different uncoupled branches approach and hybridize. In particular, near the $K'$ point, the nominally interface-dominated lower central band develops a pronounced remote-surface contribution. This momentum-dependent redistribution shows that the displacement field polarizes the central bands without suppressing the coherent hybridization between the interface and the remote rhombohedral surface.

For $V=-50$ meV, shown in the bottom row, the layer polarization is reversed. The lower central band, together with nearby lower remote bands, acquires predominantly remote-surface character, while the upper central band and nearby upper bands are concentrated mainly in the twisted-interface sector. The sign of $V$ therefore selects whether the remote-surface weight is carried mainly by the upper or lower part of the central-band manifold. This switching pattern follows from the opposite electrostatic shifts of $L_0$ and $L_N$. Nevertheless, the spectra at opposite values of $V$ are not exact particle-hole images of one another. The realistic hopping structure, the inequivalent action of the electrostatic potential on the relevant surface states, and the absence of exact particle-hole symmetry produce different bandwidths, avoided crossings, and isolation gaps for the two field orientations.

These layer-weighted spectra provide a microscopic connection to electrical switching phenomena observed in twisted monolayer--bilayer graphene and related systems, where displacement fields tune van Hove singularities, correlated phases, orbital magnetism, anomalous Hall responses, and electrically switchable topology~\cite{chen2020electrically,Polshyn2020,xu2021tunable,Polshyn2021,he2021competing,zhang2024observation}. The present calculation identifies the underlying single-particle mechanism: reversing $V$ exchanges the interface and remote-surface character carried by the upper and lower central bands. When combined with valley polarization and a field-induced change of the occupied-band Chern number, this layer exchange provides a microscopic route to electrically switching the orbital magnetization and topological response. The corresponding Chern phases are analysed below.

\subsection{A projected surface theory explains the mechanism}

To derive the surface-propagation mechanism invoked above, we now construct a low-energy theory that removes the high-energy interior states while retaining the leading realistic propagation channels. Inside the ABC slab, the dominant vertical hopping $\gamma_1$ strongly hybridizes each aligned pair $B_l$ and $A_{l+1}$. These dimerized pairs form the high-energy sector, whereas the boundary sites $A_1$ at the twisted interface and $B_N$ at the remote surface remain in the low-energy sector. We set the nearest-layer same-sublattice hopping $\gamma_4$ to zero but retain $\gamma_1$, the trigonal-warping hopping $\gamma_3$, the next-nearest-layer hopping $\gamma_2$, and both moir\'e amplitudes $u$ and $u'$. This projected construction extends the rhombohedral path expansion of Ref.~\cite{Koshino2009Trigonal} to a twisted monolayer coupled to the two surface degrees of freedom of an $N$-layer slab. For $N\ge2$, the retained basis is $\mathcal L=\{A_0,B_0,A_1,B_N\}$, and the remaining dimerized sector is eliminated to leading order in $E/\gamma_1$ using L\"owdin partitioning. The complete derivation is given in Sec.~\ref{sec:S2}.

Let $\pi=\eta q_x+iq_y$ be the Dirac operator in the slab frame and define the three elementary path amplitudes $\xi_1=v_f\pi$, $\xi_2=v_3\pi^\dagger$, and $\xi_3=\gamma_2/2$, where $v_3=(\sqrt3/2)a\gamma_3$ in units with $\hbar=1$. They advance the effective propagation from the interface surface $A_1$ toward the remote surface $B_N$ by one, two, and three layer indices, respectively. The amplitude $\xi_1$ represents ordinary Dirac propagation, $\xi_2$ the trigonal-warping path generated by $\gamma_3$, and $\xi_3$ the next-nearest-layer path generated by $\gamma_2$. Their coherent sum is represented by $X_N(\boldsymbol q)$, defined recursively as
\begin{equation}
\begin{gathered}
X_N=-\frac{1}{\gamma_1}\left(\xi_1X_{N-1}+\xi_2X_{N-2}+\xi_3X_{N-3}\right),
\\ X_0=-\gamma_1,\quad X_{N<0}=0.
\end{gathered}
\label{eq:X_recursion_main}
\end{equation}
The projected rhombohedral block in the surface basis $(A_1,B_N)$ is therefore
\begin{equation}
H_{\mathrm{ABC}}^{(N),\mathrm{path}}(\boldsymbol q)=
\begin{pmatrix}0&X_N^\dagger(\boldsymbol q)\\X_N(\boldsymbol q)&0\end{pmatrix}.
\label{eq:HABC_path_main}
\end{equation}
The ordinary Dirac paths, the trigonal-warping paths, and the next-nearest-layer paths are thus combined into a single coherent surface-to-surface coupling. The path containing only $N$ ordinary Dirac steps, which becomes the complete answer in the strict chiral limit defined below, is one contribution to this more general path-projected sum. The $\gamma_2$ and $\gamma_3$ paths reconstruct its detailed momentum dependence.

The projection also dresses the moir\'e tunnelling that initially terminates on $B_1$. By defining $D_m=-X_m/\gamma_1$, the coupling between $(A_0,B_0)$ and $(A_1,B_N)$ becomes
\begin{equation}
\widetilde T_{01}^{\mathrm{path}}(\boldsymbol r,\boldsymbol q)=
\begin{pmatrix}
T_{A_0A_1}(\boldsymbol r)&T_{A_0B_1}(\boldsymbol r)D_{N-1}^\dagger(\boldsymbol q)\\[5pt]
T_{B_0A_1}(\boldsymbol r)&T_{B_0B_1}(\boldsymbol r)D_{N-1}^\dagger(\boldsymbol q)
\end{pmatrix}.
\label{eq:T_eff_path_main}
\end{equation}
The first column describes direct tunnelling to the retained interface site $A_1$. The second describes tunnelling into $B_1$ followed by virtual propagation across the ABC slab to the remote surface $B_N$. The dressed amplitude $D_{N-1}$ therefore provides the microscopic connection between the local moir\'e coupling at the twisted interface and the remote-surface weight observed in the full-model bands.

The displacement field introduces a further correction. With $\Delta=V/2$, both sites of the remote layer carry $+\Delta$; because $A_N$ is eliminated for $N\ge2$, its potential generates an additional term within the same leading energy-independent L\"owdin projection,
\begin{equation}
\begin{gathered}
H_{\mathrm{eff},\Delta}^{\mathrm{path}}=
\begin{pmatrix}
H_{00}^{\Delta}&\widetilde T_{01}^{\mathrm{path}}\\[3pt]
(\widetilde T_{01}^{\mathrm{path}})^\dagger&H_{\mathrm{ABC},\Delta}^{(N),0}
\end{pmatrix}
+\Delta\Lambda\Lambda^\dagger,
\\
H_{\mathrm{ABC},\Delta}^{(N),0}=\begin{pmatrix}0&X_N^\dagger\\X_N&\Delta\end{pmatrix}.
\end{gathered}
\label{eq:Heff_path_main}
\end{equation}
The vector $\Lambda$ is given in Sec.~\ref{sec:S2}. The correction $\Delta\Lambda\Lambda^\dagger$ accounts for the potential acting on the eliminated remote-layer site. It does not modify $X_N$, $D_{N-1}$, or the direct onsite energy of the retained site $B_N$. 

In the strict chiral limit, $\gamma_2=\gamma_3=\gamma_4=0$ and $u=0$, while $\gamma_1$ and $u'$ remain finite. Only the path containing ordinary Dirac propagation survives, and the path amplitudes reduce to
\begin{equation}
\begin{gathered}
X_N^{\mathrm{ch}}=(-1)^{N-1}\frac{(v_f\pi)^N}{\gamma_1^{N-1}},
\\
D_{N-1}^{\dagger,\mathrm{ch}}=(-1)^{N-1}\frac{(v_f\pi^\dagger)^{N-1}}{\gamma_1^{N-1}}.
\end{gathered}
\label{eq:X_chiral_main}
\end{equation}
It is important to distinguish this strict \emph{chiral} model from a \emph{reduced} model. The latter sets $\gamma_2=\gamma_3=\gamma_4=0$ while retaining $\gamma_1$, $u$, and $u'$. Its rhombohedral surface and propagation amplitudes therefore have the pure-power form of Eq.~\eqref{eq:X_chiral_main}, but its moir\'e potential retains both tunnelling amplitudes. The strict chiral model further sets $u=0$ while retaining $\gamma_1$ and $u'$~\cite{Ledwith2022Family,Wang2022Hierarchy,SanJose2012NonAbelian}.

At $V=0$, the path-projected model reproduces the low-energy central bands of the full model with nearly identical dispersion. Differences appear mainly in the remote bands and become more visible at larger $|V|$, while the low-energy mechanism remains intact. The reduced- and strict-chiral-model bands are shown in Sec.~\ref{sec:S2}. The electronic structure can now be understood directly from the projected surface theory. In the strict chiral limit, the rhombohedral surface dispersion is controlled by $X_N^{\mathrm{ch}}\propto(v_f\pi)^N/\gamma_1^{N-1}$ and therefore becomes increasingly soft as $N$ grows. The additional $\gamma_2$ and $\gamma_3$ paths contained in $X_N$ reconstruct this pure-power dispersion while preserving its surface origin. After moir\'e hybridization, the central bands inherit this soft rhombohedral sector, explaining their progressive narrowing with increasing layer number.

The effect of the displacement field is also transparent. Near $K'$, the monolayer-dominated bands shift because the diagonal term $H_{00}^{\Delta}$ applies the same onsite potential to both monolayer sublattices. The field therefore shifts the monolayer Dirac cone without directly opening a sublattice gap, in analogy with twisted bilayer graphene~\cite{Lopes2007Graphene,Suarez2011Charge}. This shift is visible in the middle and bottom rows of Fig.~\ref{fig:full_bands}. Near $K$, the field acts differently on the two rhombohedral surface components: the retained remote site $B_N$ carries the direct potential $+\Delta$, while the eliminated site $A_N$ contributes through $\Delta\Lambda\Lambda^\dagger$. This surface detuning opens and reconstructs the gap in the rhombohedral-derived bands.

The path-projected theory therefore explains the two principal trends of Fig.~\ref{fig:full_bands}: increasing $N$ softens the rhombohedral surface dispersion, while the displacement field changes the relative alignment of the monolayer, interface, and remote-surface sectors and transfers their weight between the two central bands. The strong difference between the spectra at opposite signs of $V$ arises from the realistic hopping structure, the inequivalent action of the field on the relevant surface states, and the absence of exact particle-hole symmetry.

\subsection{Real-space density reveals moir\'e mixing and remote-surface polarization}

The layer-composition-resolved band structures diagnose the distribution of wave-function weight among the layers in momentum space. The real-space charge density provides the complementary view within the moir\'e unit cell by resolving the interference between the different plane-wave components in the expansion defined in Methods. Some terms provide a spatially uniform background, whereas interference between distinct harmonics generates modulation at moir\'e reciprocal wave vectors. Strong spatial contrast therefore signals coherent mixing between several harmonics, while a nearly uniform density indicates that the interference contributions are weak.

\begin{figure*}[!t]
\centering
\includegraphics[width=0.95\textwidth]{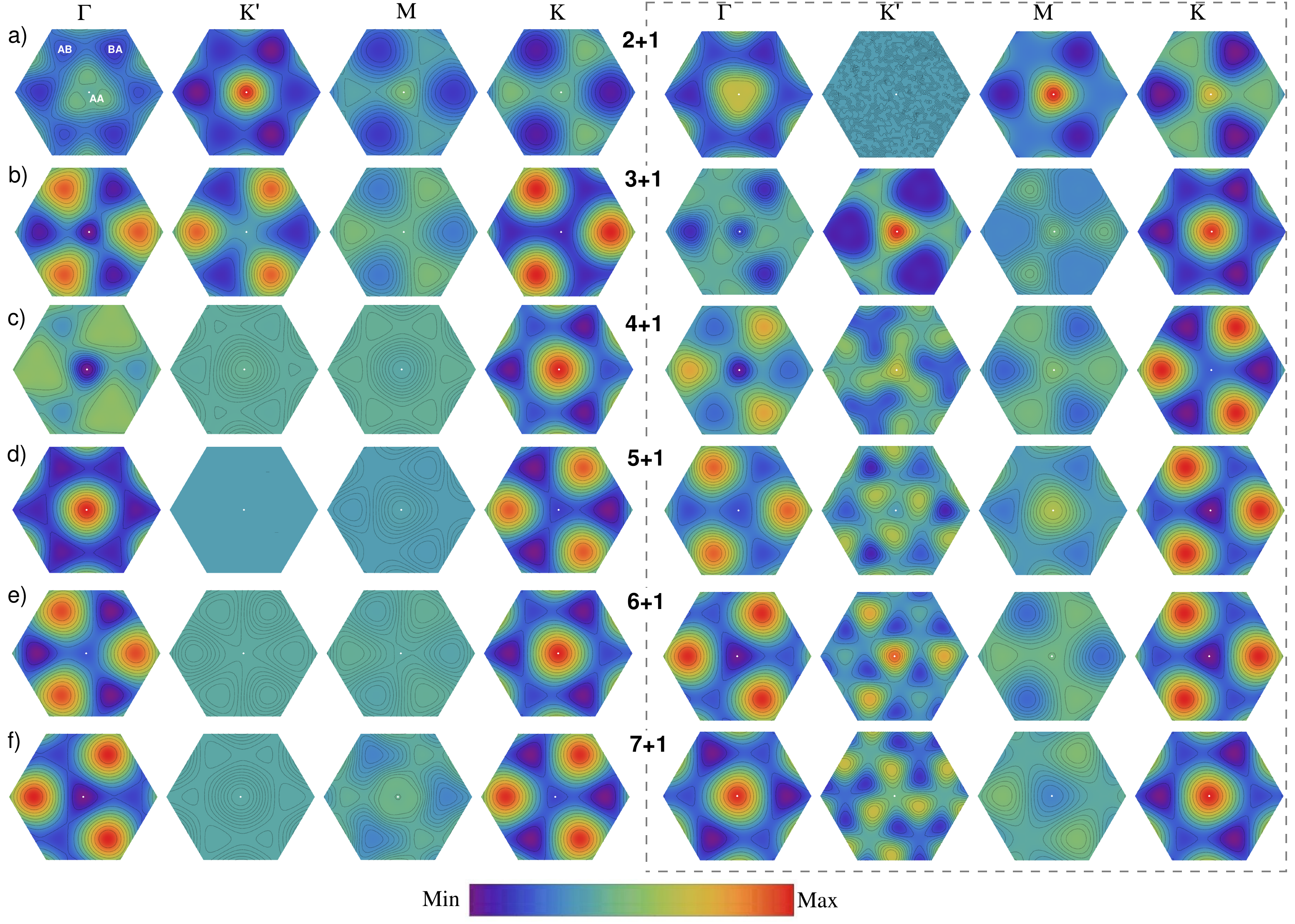}
\caption{\textbf{Real-space charge distribution in the full model.} Probability densities of the two central bands at selected high-symmetry points of the moir\'e Brillouin zone, summed over layer and sublattice. The four columns on the left correspond to the lower central band, while the four columns on the right correspond to the upper central band. Within each group, the momenta are ordered as $\Gamma$, $K'$, $M$, and $K$. Rows show results for different layer numbers, from $N=2$ to $N=7$. The colour scale is normalized using the global minimum and maximum of the total charge density over all displayed high-symmetry points. The AA, AB, and BA labels in panel \textbf{a} identify the local stacking between the rotated monolayer $L_0$ and the adjacent graphene layer $L_1$.}
\label{fig:total_density}
\end{figure*}

Figure~\ref{fig:total_density} shows the probability densities of the two central bands at selected high-symmetry points, summed over layer and sublattice describing the spatial distribution of individual states. The four columns on the left correspond to the lower central band, evaluated at $\Gamma$, $K'$, $M$, and $K$, while the four columns on the right show the corresponding density of the upper central band. The colour scale is normalized using the global minimum and maximum of the total charge density over all displayed high-symmetry points. For small $N$, particularly $N=2$ and $N=3$, both central bands exhibit pronounced triangular charge textures. These patterns reflect the $C_3$ structure imposed by the moir\'e tunnelling and demonstrate that the central-band states are coherent superpositions of several moir\'e harmonics.

As $N$ increases, the lower central band develops a marked momentum dependence. Its density becomes nearly uniform at $K'$ and $M$, while remaining strongly modulated at $\Gamma$ and $K$. Near $K$, the increasingly soft rhombohedral surface dispersion brings several plane-wave components related by moir\'e reciprocal vectors into close energetic proximity, allowing them to hybridize and generate a structured density. In the strict chiral limit, the relevant surface scale varies as $(v_f|\boldsymbol q|)^N/\gamma_1^{N-1}$, while in the path-projected theory it is controlled by $|X_N(\boldsymbol q)|$. The modulation at $\Gamma$ instead originates from the coherent interference of several symmetry-related folded momenta. The upper central band remains strongly textured over a broader set of high-symmetry points, consistent with its stronger exposure to the direct moir\'e perturbation at the twisted interface.

The total density alone does not identify which layers carry the spatial modulation. The layer-resolved charge maps in Sec.~\ref{sec:S5} provide this decomposition. For $N=2,3,4$, the strongest moir\'e texture generally appears on the two layers forming the twisted interface, $L_0$ and $L_1$, because they are directly connected by $T_{01}$. The remaining layers respond indirectly as part of the rhombohedral stack under the moir\'e perturbation generated at the interface. For $N=2$, this makes clear that the system is neither a conventional twisted bilayer nor an isolated AB bilayer, but a hybrid structure in which the directly coupled $L_0$--$L_1$ interface coexists with the low-energy boundary sector of the AB bilayer.

For $N=3$ and $N=4$, the layer-resolved maps reveal a progressive separation between spatial texture and integrated layer weight. The monolayer $L_0$ and the interface layer $L_1$ carry the strongest spatial modulation, whereas the lower central band develops substantial weight on the remote rhombohedral surface $L_N$. In the projected description, this remote-surface component is carried mainly by $B_N$. Unlike $A_1$, which is coupled directly to the rotated monolayer, $B_N$ is reached through the dressed propagation amplitude $D_{N-1}$. For the parameters considered here, a state dominated by $B_N$ can therefore carry substantial integrated charge on the remote surface while remaining only weakly modulated within the moir\'e unit cell~\cite{Pantaleon2026Interface}. This explains why some total-density maps of the lower central band become smooth at selected momenta even though the corresponding states remain strongly rhombohedral. Related surface-dependent contrast has been observed by STM/STS in twisted monolayer--bilayer graphene, where the flat-band LDOS is concentrated around selected stacking regions on the twisted side but comparatively uniform on the remote bilayer-side surface~\cite{Tong2022Spectroscopic}.

The displacement-field dependence gives a direct real-space view of the same layer-switching mechanism. The representative $N=3$ maps in Sec.~\ref{sec:S6} show that for $V=-50$ meV the lower central band is concentrated predominantly on the remote surface $L_N$, with a smaller contribution from the adjacent layer $L_{N-1}$. The upper central band is distributed over several layers and exhibits a more structured density profile, particularly at $\Gamma$ and $K'$. For $V=+50$ meV, the layer assignment is reversed: the upper central band acquires the dominant remote-surface contribution, while the lower central band becomes concentrated mainly in the twisted-interface region formed by $L_0$ and $L_1$. Residual weight remains on the opposite side of the stack at selected momenta, reflecting the coherent hybridization of the central bands rather than a complete separation into pure interface and remote-surface states. This real-space redistribution is fully consistent with the momentum-space layer composition shown in Fig.~\ref{fig:full_bands}.

The additional displacement-field maps in Sec.~\ref{sec:S6} show the same exchange for $N=2$ and $N=4$, while Sec.~\ref{sec:S4} shows the corresponding field-driven redistribution in the layer-composition-resolved band structure for $N=4$. Thus, reversing $V$ selects which central band carries the dominant remote-surface $L_N$ character and which remains more strongly hybridized with the moir\'e interface. The real-space densities and momentum-space layer weights therefore provide two complementary representations of the same mechanism: the displacement field transfers central-band character between the twisted interface and the remote rhombohedral surface.

These density maps also provide an intuitive guide to possible STM/STS textures. Since all states contributing within the probed energy window enter the LDOS, a window dominated by states near $\Gamma$ can display a spatial pattern resembling their density distribution. When states near both $K$ and $K'$ contribute, their patterns combine with their respective spectral weights. In the weak-tunnelling regime, the differential conductance approximately follows the LDOS at the tip, so the relevant comparison is with the layer-resolved density on the exposed surface. The displacement-field-driven redistribution between the interface and remote surface can therefore change the surface spectral weight and spatial contrast. Gate-dependent conductance maps in twisted monolayer--bilayer graphene have resolved a transition between localized and more uniform flat-band spectral weight as carrier density and displacement field are varied together~\cite{Li2023Gate}. In rhombohedral trilayer graphene aligned with hBN, STM/STS measurements and surface-projected LDOS calculations similarly relate gate-dependent peak intensities to layer polarization~\cite{Seewald2025Mapping}. Our maps therefore identify qualitative spectroscopic tendencies; a quantitative conductance prediction requires the full energy-resolved momentum sum and the experimental tunnelling conditions.

In addition, and as mentioned before, the real-space switching with the field provides a microscopic connection to electrically controlled orbital responses in related twisted monolayer--bilayer systems. The displacement field changes the relative energy of the monolayer and remote-surface sectors and thereby selects which central band carries the dominant $B_N$ character. Although interactions are not included in the present calculation, an interaction-driven spin or valley polarization combined with occupation of a field-selected Chern band can produce an orbital magnetic response whose sign is electrically switchable~\cite{Polshyn2020,zhang2024observation}. In the present system, this mechanism is enriched by the layer-dependent rhombohedral surface structure, whose strict-chiral contribution carries an $N$-fold phase winding.

\subsection{The $N\bmod 3$ organization of charge textures}

\begin{figure*}[!t]
\centering
\includegraphics[width=0.95\textwidth]{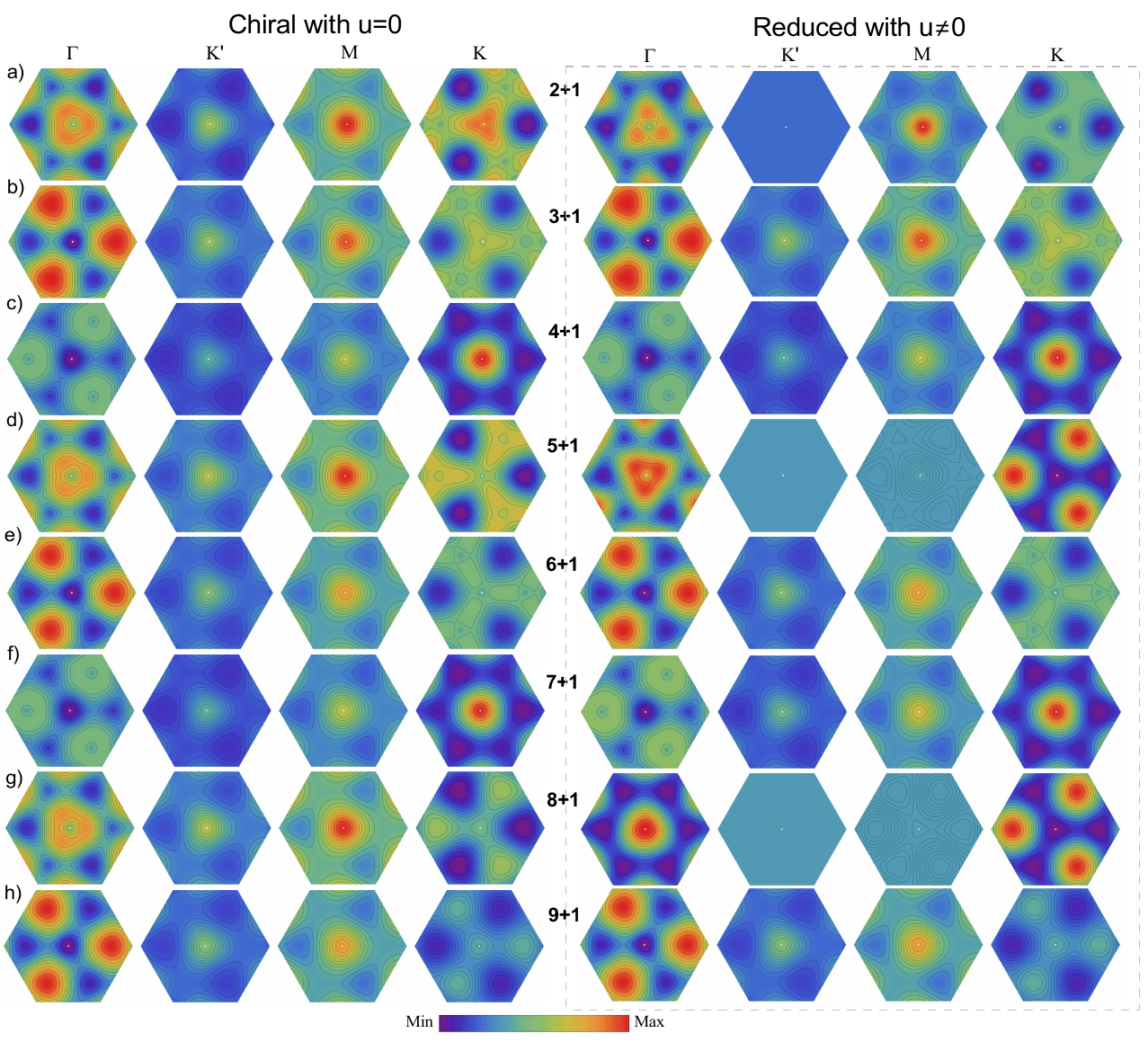}
\caption{\textbf{Real-space density in the strict chiral and reduced models.} Probability densities of the lower and upper central bands evaluated at $\Gamma$, $K'$, $M$, and $K$ for $N=2$ to $N=9$. At momenta where the displayed central state belongs to a degenerate subspace, the density is averaged over the corresponding orthonormal states; the $M$-point density is additionally averaged over the three $C_3$-related momenta. The recurring angular textures organize into the three residue classes of $N\bmod3$.}
\label{fig:mod3_density}
\end{figure*}

The total-density maps and Fig.~\ref{fig:mod3_density} reveal a recurrence that is not obvious from a single layer number. Similar angular patterns reappear when the number of rhombohedral layers is changed by three, and the maps separate naturally into the families $N=2,5,8,\ldots$, $N=3,6,9,\ldots$, and $N=4,7,10,\ldots$. This grouping is not imposed by the numerical procedure; it emerges from the phase structure of the projected wave functions.

Figure~\ref{fig:mod3_density} compares the reduced continuum model and the strict chiral model. Both set $\gamma_2=\gamma_3=\gamma_4=0$; their difference is the same-sublattice moir\'e amplitude $u$, which is retained in the reduced model and removed in the strict chiral model. The differences are most visible at $\Gamma$, where $u$ produces a ring-like structure, and at $K'$, where it enhances the central density, in analogy with twisted bilayer graphene~\cite{Rademaker2018Charge,Guinea2018Electrostatic}. This sensitivity is especially clear for $N=2,5,8$. Apart from these features, the persistence of the same angular organization in both descriptions shows that the main recurrence is already encoded in the projected low-energy structure rather than being generated by $\gamma_2$, $\gamma_3$, or $\gamma_4$.

The angular recurrence follows from the strict chiral limit of the path theory. Writing $\pi=|\pi|e^{i\varphi_{\boldsymbol q}}$, the upper-right rhombohedral matrix element and the remote-surface tunnelling dressing contain
\begin{equation}
\left(X_N^{\mathrm{ch}}\right)^\dagger\propto |\pi|^N e^{-iN\varphi_{\boldsymbol q}},
\qquad
D_{N-1}^{\dagger,\mathrm{ch}}\propto |\pi|^{N-1}e^{-i(N-1)\varphi_{\boldsymbol q}}.
\label{eq:phase_winding_path_main}
\end{equation}
Under a $C_3$ rotation, these factors acquire the phases $e^{-i2\pi N/3}$ and $e^{-i2\pi(N-1)/3}$. Both are unchanged under $N\to N+3$, giving the three residue classes
\begin{equation}
N=3m,\qquad N=3m+1,\qquad N=3m+2.
\end{equation}
The classification concerns the dominant angular phase structure. The radial amplitudes still scale with different powers of $|\pi|/\gamma_1$, so members of the same residue class need not have identical bandwidth, layer polarization, or density contrast.

The triangular charge textures are related to patterns generated by other moir\'e perturbations of multilayer graphene, including rhombohedral graphene on hBN and twisted graphene--graphite interfaces~\cite{Pantaleon2021Narrow,Zhang2025_R8GhBN,Wu2014,Waters2023,Waters2024}. In the present system, however, the periodic perturbation is not produced by an external hBN potential but by the moir\'e hybridization between the rotated monolayer and the rhombohedral stack. The resulting textures therefore provide a real-space signature of the same interface-to-remote-surface coupling that controls band flattening and displacement-field-driven layer switching.

\subsection{The Chern hierarchy is organized by the monolayer and rhombohedral surface sectors}

The band structures and charge densities show that the two central bands arise from the hybridization of a monolayer Dirac cone and the rhombohedral surface sector described by $X_N$ in the preceding section. When one central band is isolated from both the other central band and the remote bands, its topology is characterized by an Abelian Chern number $\mathcal C$. When the two central bands touch each other while remaining separated from the remote bands, their individual Chern numbers are no longer defined. The invariant must then be assigned to the complete isolated two-band subspace and calculated using the non-Abelian Berry connection~\cite{Palumbo2021NonAbelian}; here, non-Abelian refers only to the matrix-valued Berry connection within that subspace. The numerical construction is given in Methods. This situation is especially transparent in the strict chiral model at $V=0$, where the two central bands touch while the pair remains isolated from the remote spectrum.

\begin{figure*}[!t]
\centering
\includegraphics[width=0.98\textwidth]{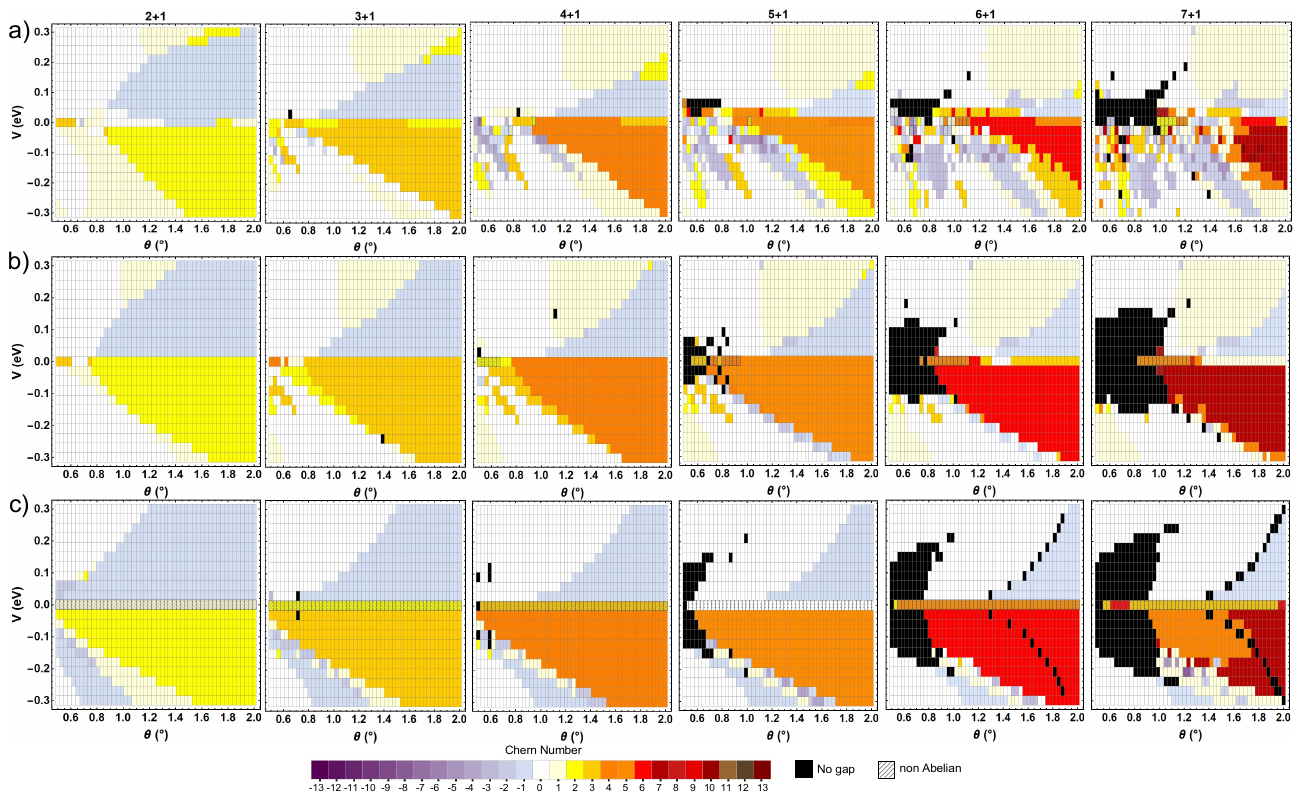}
\caption{\textbf{Chern-number phase diagram.} Comparison between the full, reduced, and strict chiral models for $(N+1)$-layer rhombohedral graphene from $N=2$ to $N=7$ for the lower central band. The Chern number is encoded in the colour map. Coloured regions indicate $(\theta,V)$ points where the lower central band is isolated and assigned an Abelian Chern number. Hatched regions indicate that the two central bands touch but remain isolated from the remote bands, so that the Chern number is assigned to the isolated two-band subspace using the non-Abelian Berry connection. Black regions correspond to parameter regimes where the central-band sector is not isolated from the remote bands.}
\label{fig:chern_lower}
\end{figure*}

The Chern-number phase diagrams of the lower central band are shown in Fig.~\ref{fig:chern_lower} as a function of twist angle and displacement field for $N=2$ to $N=7$. The corresponding upper-band phase diagrams are shown in Sec.~\ref{sec:S8}. Coloured regions indicate parameter points where the selected central band is isolated and is assigned an Abelian Chern number. Hatched regions indicate that the two central bands touch but remain isolated as a two-band subspace, while black regions indicate that the central-band sector is not isolated from the remote bands.

Figure~\ref{fig:chern_lower} compares the full, reduced, and strict chiral continuum models defined above. The path-projected theory provides the analytical connection between these descriptions: it retains the $\gamma_2$ and $\gamma_3$ propagation channels in the surface amplitude $X_N$ while omitting $\gamma_4$ and the high-energy dimer bands. It therefore explains how the additional rhombohedral paths deform the strict chiral surface sector before the complete remote spectrum is restored in the full model.

The cleanest interpretation applies when the central bands remain isolated from the remote spectrum and retain predominantly monolayer-like or rhombohedral-like character. In the strict chiral limit, Eq.~\eqref{eq:X_chiral_main} combines a unit-winding monolayer Dirac cone with an $N$th-order rhombohedral surface sector. The displacement field changes their relative energy and selects which sector lies below the central gap. Moir\'e hybridization then converts these two momentum-space sectors into the isolated central minibands. In the single valley and field convention used here, the lower central band has
\begin{equation}
\mathcal C=N
\end{equation}
for negative displacement field and
\begin{equation}
\mathcal C=-1
\end{equation}
for positive displacement field. The upper central band shows the complementary exchange. These integer sequences are the ideal chiral reference established for twisted chiral multilayers~\cite{Ledwith2022Family,Wang2022Hierarchy}. The new role of the full calculation is to identify how the same sectors are selected, layer polarized, and deformed by realistic hopping and the complete remote spectrum. Changing between the two isolated-band sequences requires a closing of the gap between the central bands. In the strict chiral model, this closing occurs at $V=0$ while the two-band manifold remains isolated from the remote spectrum; the invariant at the closing is therefore the Chern number of the isolated two-band subspace calculated using the non-Abelian Berry connection.

The path amplitude $X_N$ also makes clear why the realistic rhombohedral sector cannot generally be identified with a single $N$th-order monomial. The $\gamma_2$ and $\gamma_3$ propagation paths split and reshape the low-energy nodes, while $\gamma_4$ and the complete remote-band spectrum introduce particle-hole asymmetry and additional avoided crossings. Consequently, the phase with $\mathcal C=N$ in the realistic calculations should not be inferred solely from a local winding of $X_N$. It is more accurately identified as the isolated-band phase adiabatically connected to the rhombohedral branch of the strict chiral model. In the same way, the $\mathcal C=-1$ sequence is adiabatically connected to the unit-winding monolayer branch. This distinction preserves the physical interpretation provided by the strict chiral limit while allowing the realistic hopping processes to deform the nodes and phase boundaries.

The opposite signs and unequal magnitudes of the two field-selected sequences cannot be understood by treating the central pair as a closed two-band Hamiltonian. For $N\ge2$, the central bands originate from a larger multiband system and can exchange Berry curvature with the remote bands. Their combined Chern number is therefore not required to vanish. Time-reversal symmetry relates the two graphene valleys and reverses all Chern numbers in the opposite valley.

At smaller twist angles or larger $N$, the central bands approach the remote spectrum and the phase diagrams become increasingly fragmented. Additional band inversions transfer Berry curvature between the central and remote bands, producing narrow regions with larger positive or negative Chern numbers. Their detailed distribution is sensitive to the further-neighbour hopping parameters and to microscopic ingredients not included in the present continuum model. The robust result is the field-driven exchange between the $\mathcal C=N$ and $\mathcal C=-1$ isolated-band sequences. The strict chiral surface sectors organize this hierarchy, while the reduced, path-projected, and full models show how it is continuously deformed by realistic moir\'e and rhombohedral hopping processes.

\section{Conclusions}

Our results establish a unified low-energy picture of twisted monolayer--rhombohedral graphene in which band flattening, layer polarization, real-space charge texture, and topology originate from the hybridization of two inequivalent sectors: a rotated monolayer Dirac cone and the surface states of an $N$-layer ABC stack. Increasing $N$ softens the rhombohedral surface dispersion and progressively narrows the central bands. This enhanced flattening also brings the central manifold closer to the remote spectrum, making its isolation and layer composition increasingly sensitive to twist angle, displacement field, and interlayer hopping processes.

The displacement field acts primarily by changing the relative energy of the rotated monolayer $L_0$ and the remote rhombohedral surface $L_N$. Because the central states are coherent mixtures of the monolayer, interface, and remote-surface sectors, the field does not produce a rigid spectral shift. Instead, it reconstructs the avoided crossings and transfers wave-function weight between the twisted $L_0$--$L_1$ interface and $L_N$. The interface layer $L_1$ is coupled directly to the monolayer through $T_{01}$, whereas the remote surface is reached through propagation across the ABC stack, encoded in the projected theory by $X_N$ and the dressed moir\'e amplitude $D_{N-1}$. This asymmetric coupling explains why one central band can carry predominantly remote-surface character while the other remains more strongly hybridized with the twisted interface, and why reversing $V$ exchanges these roles.

The real-space charge densities reveal a complementary aspect of the same mechanism. The strongest moir\'e modulation generally remains at the directly coupled interface, while a large fraction of the integrated wave-function weight can reside on the remote surface. This apparent separation follows because integrated layer weights contain only the diagonal probabilities of the moir\'e harmonics, whereas the real-space density also contains their interference. A $B_N$-dominated state can therefore be strongly rhombohedral and weakly modulated at the same time. The recurring $N\bmod3$ angular textures provide an additional wave-function fingerprint of the pure-power phase factors inherited from the projected surface and tunnelling amplitudes. 

The topology follows from the same two-sector structure. In the strict chiral limit, the monolayer Dirac cone carries unit winding, whereas the rhombohedral surface coupling is of $N$th-order. These ideal high-Chern sectors are known from chiral twisted-multilayer theory~\cite{Ledwith2022Family,Wang2022Hierarchy}. The contribution of the present model hierarchy is to show how displacement field selects their layer-resolved descendants and how $\gamma_2$, $\gamma_3$, $\gamma_4$, same-sublattice moir\'e tunnelling, particle-hole asymmetry, and the complete remote spectrum deform them. When an individual central band remains isolated, the full calculations retain the field-selected hierarchy adiabatically connected to the strict chiral sectors. For the lower central band in the valley and field convention used here, negative $V$ selects the sequence $\mathcal C=N$, while positive $V$ selects $\mathcal C=-1$; the upper central band displays the complementary exchange. When the central bands touch but remain isolated from remote bands, their individual invariants are undefined and the Chern number must instead be assigned to the isolated two-band subspace using the non-Abelian Berry connection.

When the central manifold approaches the remote spectrum, particularly at smaller twist angles and larger $N$, additional avoided crossings and band inversions transfer Berry curvature between central and remote bands. The phase diagrams then become fragmented and contain narrow regions with larger positive or negative Chern numbers. Their detailed distribution is sensitive to the graphite hopping parametrization, lattice relaxation, strain, electrostatic screening, and other microscopic effects not included in the present single-particle continuum model. These ingredients can shift quantitative phase boundaries and modify narrow high-$|\mathcal C|$ regions, but they do not alter the organizing mechanism: the central bands originate from monolayer--surface hybridization, their layer character is exchanged by the displacement field, and the robust isolated-band sequences remain adiabatically connected to the unit-winding and $N$th-order strict chiral sectors.

This framework provides a single-particle basis for interpreting recent experiments on twisted monolayer--multilayer and rhombohedral graphene, where high-Chern quantum anomalous Hall states, electrically tunable orbital magnetism, and interaction-driven topological phases have been reported~\cite{Wang2025_TDTG,Liu2025_TwistMonoPenta,Wang2026High,Su2025,Dong2025_TwistBiTri,chen2026layer,Li2026Fractional}. Our calculations do not describe the interacting ordered states themselves. They identify the narrow, layer-polarized Chern bands on which such states can develop. In particular, the reported layer-number hierarchy of high-Chern orbital magnets is naturally associated with the field-selected $N$th-order rhombohedral sector~\cite{Wang2026High}. Once interactions select a spin- or valley-polarized state, occupation of this sector can generate a strong orbital response whose sign depends on the occupied valley and on the displacement-field-selected layer character. STM/STS measurements in twisted monolayer--bilayer graphene further show that interactions can redistribute electronic spectral weight within the moir\'e unit cell~\cite{Li2022Imaging}, emphasizing the importance of the spatial structure of the states.

Establishing the non-interacting dispersion, topology, and wave-function structure provides the reference needed to identify how interactions reconstruct the central bands. Bandwidths and isolation gaps set the kinetic-energy scales, while the layer- and sublattice-resolved wave functions determine the Coulomb form factors. The occupied-state charge distribution generates the Hartree potential, while the electrostatic shift of an individual state depends on its spatial overlap with that potential. For a given Hartree potential, states with similar layer-resolved densities experience similar shifts, whereas different distributions can produce momentum-dependent shifts and modify the dispersion. As demonstrated in our previous work~\cite{Pantaleon2021Narrow}, a narrow bandwidth alone is therefore insufficient to anticipate a strong Hartree reconstruction; the spatial structure of the states is also essential. Consequently, the displacement-field-driven redistribution of wave-function weight changes both the single-particle spectrum and the interaction matrix elements. These results provide a microscopic starting point for an interacting theory; determining the ordered ground state or the stability of correlated topological phases requires a dedicated many-body calculation.

The same mechanism provides a microscopic interpretation of the opposite-bias spectral response observed by nanoARPES in twisted monolayer--bilayer graphene~\cite{zhang2024observation}. In the present language, that system is the $N=2$ member of the family: reversing the displacement field changes the relative alignment of the monolayer Dirac cone and the AB-bilayer low-energy sector and therefore exchanges their layer and momentum character within the central bands. For larger $N$, the same wave-function redistribution produces remote-surface polarization and organizes the field-selected Chern hierarchy.

The model hierarchy separates this robust mechanism from quantitative corrections. The strict chiral model isolates the pure-power surface path, while the reduced model restores the same-sublattice moir\'e amplitude $u$. The path-projected theory retains the $\gamma_2$ and $\gamma_3$ propagation channels through $X_N$ and $D_{N-1}$, and the full model restores $\gamma_4$, the high-energy dimer bands, particle-hole asymmetry, and the complete remote spectrum. These additions reshape the detailed dispersions and topological phase boundaries without changing their common origin. Twisted monolayer--rhombohedral graphene therefore provides a platform in which layer number controls the available surface propagation paths, displacement field selects the active layer and topological sector, and twist angle controls the degree of moir\'e hybridization and band isolation.

\section{Methods}

\subsection{Continuum-model implementation}

The numerical calculations use the single-valley continuum Hamiltonian introduced in Eq.~\eqref{eq:Hfull_layers_main}. The basis states are labeled by layer $l$, sublattice $\alpha=A,B$, and plane-wave momentum $\boldsymbol{k}+\boldsymbol{G}$, where $\boldsymbol{G}$ belongs to the truncated moir\'e reciprocal lattice. The intralayer Dirac blocks, nearest- and next-nearest-layer hopping matrices, and interfacial moir\'e tunnelling matrices are given explicitly in Sec.~\ref{sec:S1}. The displacement field is represented by the screened outer-layer potential difference $V$, with onsite energies $-V/2$ on $L_0$ and $+V/2$ on $L_N$, while the internal layers are assigned zero onsite potential.

Unless otherwise stated, the calculations use $\theta=1.05^\circ$ and {the interlayer hopping parameters are adopted from the DFT-based ABC-trilayer fit of Ref.~\cite{Zhang2010Band} and used consistently across layer numbers}. We use  $\gamma_0=3.16$~eV, $\gamma_1=0.502$~eV, $\gamma_2=-0.0171$~eV, $\gamma_3=-0.377$~eV, and $\gamma_4=-0.099$~eV, together with the moir\'e tunnelling amplitudes $u=0.0797$~eV and $u'=0.0975$~eV~\cite{Zhang2010Band,Bistritzer2011Moire,Koshino2015Interlayer,Koshino2018Maximally,Pantaleon2021Narrow}. The interlayer terms follow the corrugated twisted-bilayer parametrization of Ref.~\cite{Koshino2018Maximally}.

\subsection{Continuum-model hierarchy and surface projection}

We use four related descriptions. The full model retains $\gamma_1$, $\gamma_2$, $\gamma_3$, $\gamma_4$, $u$, and $u'$, together with the complete layer and plane-wave basis. The reduced model sets $\gamma_2=\gamma_3=\gamma_4=0$ while retaining $\gamma_1$, $u$, and $u'$. The strict chiral model further sets $u=0$, retaining only $\gamma_1$ and the opposite-sublattice moir\'e tunnelling $u'$ in addition to the intralayer Dirac terms. These limits separate the surface-hybridization mechanism from the corrections introduced by additional hopping processes. Their comparison with the full model identifies the persistence of central-band narrowing, field-driven layer switching, and the characteristic isolated-band Chern sequences.  To test the sensitivity to the microscopic input, Sec.~\ref{sec:S9} repeats the full-model calculations for an alternative hopping parametrization of rhombohedral graphene~\cite{ParraMartinez2025,Zhou2021} and for a linear displacement-field profile.

The analytical path-projected model is distinct from both the reduced and strict chiral models. It sets $\gamma_4=0$ but retains $\gamma_1$, $\gamma_2$, $\gamma_3$, $u$, and $u'$ in the low-energy basis $(A_0,B_0,A_1,B_N)$. The high-energy dimerized sector is eliminated to leading order in $E/\gamma_1$ using L\"owdin partitioning. The resulting surface-to-surface propagation amplitude $X_N$, the dressed moir\'e amplitude $D_{N-1}$, and the finite-field correction $\Delta\Lambda\Lambda^\dagger$ are derived in Sec.~\ref{sec:S2}.

\subsection{Layer and sublattice weights}

For a normalized eigenstate of band $n$ at momentum $\boldsymbol{k}$,
\begin{equation}
|\psi_{n\boldsymbol{k}}\rangle=
\sum_{l,\alpha,\boldsymbol{G}}
C_{l\alpha,\boldsymbol{G}}^{(n)}(\boldsymbol{k})
|l,\alpha,\boldsymbol{k}+\boldsymbol{G}\rangle,
\end{equation}
the layer- and sublattice-resolved weight is
\begin{equation}
W_{l\alpha}^{(n)}(\boldsymbol{k})=
\sum_{\boldsymbol G}\left|C_{l\alpha,\boldsymbol G}^{(n)}(\boldsymbol k)\right|^2,
\qquad
\sum_{l,\alpha}W_{l\alpha}^{(n)}(\boldsymbol k)=1.
\label{eq:layer_sublattice_weight}
\end{equation}
The total weight on layer $L_l$ is
\begin{equation}
\begin{gathered}
W_l^{(n)}(\boldsymbol{k})=
\sum_{\alpha=A,B}W_{l\alpha}^{(n)}(\boldsymbol{k})
=
\sum_{\alpha=A,B}\sum_{\boldsymbol{G}}
\left|C_{l\alpha,\boldsymbol{G}}^{(n)}(\boldsymbol{k})\right|^2,
\\
\sum_lW_l^{(n)}(\boldsymbol{k})=1.
\end{gathered}
\label{eq:layer_weight_methods}
\end{equation}
For the layer-composition-resolved band structures, let $a_1\ge a_2$ be the two largest layer weights, carried by layers $\ell_1$ and $\ell_2$. The displayed colour is obtained from a continuous blend of their basis colours using the fraction $a_1/(a_1+a_2)$. When $a_2/(a_1+a_2)<0.10$, the pure colour of the dominant layer $\ell_1$ is used. The line thickness increases with $a_1$. The opacity is determined by the confidence
\begin{equation}
\begin{gathered}
Q=\sqrt{PS},\\
P=\frac{N_L\sum_l\left(W_l^{(n)}\right)^2-1}{N_L-1},
\\
S=\frac{a_1-a_2}{a_1+a_2},
\end{gathered}
\end{equation}
where $N_L=N+1$ is the total number of layers. Here $P$ measures the total layer purity and $S$ measures the separation between the two leading layer weights. A thin grey skeleton indicates the underlying dispersion independently of the layer assignment.

\subsection{Real-space charge density}

The layer- and sublattice-resolved real-space amplitude is reconstructed from the plane-wave coefficients as
\begin{equation}
\psi_{l\alpha}^{(n)}(\boldsymbol{r};\boldsymbol{k})=
\sum_{\boldsymbol{G}}
C_{l\alpha,\boldsymbol{G}}^{(n)}(\boldsymbol{k})
e^{i(\boldsymbol{k}+\boldsymbol{G})\cdot\boldsymbol{r}}.
\end{equation}
The corresponding density components are
\begin{equation}
\begin{gathered}
\rho_{l\alpha}^{(n)}(\boldsymbol{r};\boldsymbol{k})=
\left|\psi_{l\alpha}^{(n)}(\boldsymbol{r};\boldsymbol{k})\right|^2,
\\
\rho_l^{(n)}=\sum_{\alpha=A,B}\rho_{l\alpha}^{(n)},
\\
\rho_{\mathrm{tot}}^{(n)}=\sum_l\rho_l^{(n)}.
\end{gathered}
\label{eq:realspace_density_methods}
\end{equation}
Unlike the integrated weights in Eq.~\eqref{eq:layer_weight_methods}, the real-space density contains interference terms between different moir\'e harmonics. The total-density maps are obtained by summing over all layers and sublattices, while the layer-resolved maps retain the layer index before the sublattice sum. Where a displayed map represents a degenerate central subspace, the density is averaged over the corresponding orthonormal eigenstates. Additional symmetry averaging is specified in the relevant figure captions. The maps shown together use a common normalization fixed by the global minimum and maximum of the total density over the displayed high-symmetry points.

\subsection{Chern-number calculation}

For a non-degenerate band isolated from all other bands, the Berry curvature is calculated as
\begin{equation}
\Omega_n(\boldsymbol{k})=
-2\,\mathrm{Im}\sum_{m\neq n}
\frac{
\langle u_{n\boldsymbol{k}}|\partial_{k_x}H|u_{m\boldsymbol{k}}\rangle
\langle u_{m\boldsymbol{k}}|\partial_{k_y}H|u_{n\boldsymbol{k}}\rangle
}{
\left(E_m-E_n\right)^2
},
\label{eq:berry_curvature_methods}
\end{equation}
and the Abelian Chern number is
\begin{equation}
\mathcal C_n=
\frac{1}{2\pi}
\int_{\mathrm{mBZ}}
\Omega_n(\boldsymbol{k})\,d^2\boldsymbol{k}.
\label{eq:chern_methods}
\end{equation}

When the two central bands touch but remain isolated from the remote spectrum, the individual-band Chern numbers are not defined. We then retain the two eigenvectors as a common subspace, collected in the matrix $\Psi(\boldsymbol{k})$. The overlap matrix between neighbouring momentum points is
\begin{equation}
M_\mu(\boldsymbol{k})=
\Psi^\dagger(\boldsymbol{k})
\Psi(\boldsymbol{k}+\Delta\boldsymbol{k}_\mu).
\end{equation}
For the singular-value decomposition $M_\mu=U_\mu\Sigma_\mu V_\mu^\dagger$, the corresponding unitary link matrix is
\begin{equation}
\mathcal U_\mu(\boldsymbol{k})=U_\mu V_\mu^\dagger.
\label{eq:nonabelian_link_methods}
\end{equation}
The Wilson loop around a triangular plaquette is the ordered product of its three link matrices,
\begin{equation}
\mathcal W_\triangle=
\mathcal U_{12}\mathcal U_{23}\mathcal U_{31},
\qquad
\Phi_\triangle=\arg\det\mathcal W_\triangle,
\end{equation}
and the Chern number of the isolated two-band subspace is
\begin{equation}
\mathcal C=
\frac{1}{2\pi}
\sum_\triangle\Phi_\triangle.
\label{eq:nonabelian_chern_methods}
\end{equation}
This is the non-Abelian extension of the gauge-invariant lattice formulation of Ref.~\cite{Fukui2005}. In the phase diagrams, coloured regions denote isolated individual bands with Abelian Chern number, hatched regions denote an isolated two-band subspace requiring the non-Abelian Berry connection, and black regions denote parameter points where the central-band sector is not isolated from remote bands.

\section*{Data availability}
The numerical data underlying the main and Supplementary figures will be available from the corresponding authors upon reasonable request.

\begin{acknowledgments}
We thank Gerardo Naumis and Federico Escudero for useful discussions. This research is supported by the National Natural Science Foundation of China (NSFC) Grant. No. 125471103. A.J.-P. thanks the Center for Theoretical Physics and the School of Physics and Optoelectronic Engineering at Hainan University for their hospitality, where part of this work was carried out during his visit. A.J.-P. and F.G. acknowledge funding from the EU NextGenerationEU/PRTR-C17.I1 and the IKUR Strategy under the collaboration agreement between Ikerbasque Foundation and DIPC on behalf of the Department of Education of the Basque Government. D.L. acknowledges support from the China Scholarship Council program, project ID 202406960092. P.A.P. and F.G acknowledges support from the ``Severo Ochoa'' Programme for Centres of Excellence in R\&D (CEX2020-001039-S/AEI/10.13039/501100011033), NOVMOMAT Grant PID2022-142162NB-I00 funded by MCIN/AEI/10.13039/501100011033 and by ``ERDF A way of making Europe''.
\end{acknowledgments}

\section*{Author contributions}
A.J.-P. and P.A.P. conceived the project. A.J.-P. performed the numerical calculations. A.J.-P., D.L., F.G. and P.A.P. analysed the results and contributed to their interpretation. A.J.-P. and P.A.P. wrote the manuscript with input from all authors.

\section*{Competing interests}
The authors declare no competing interests.

%


\begin{thebibliography}{92}%
\makeatletter
\providecommand \@ifxundefined [1]{%
 \@ifx{#1\undefined}
}%
\providecommand \@ifnum [1]{%
 \ifnum #1\expandafter \@firstoftwo
 \else \expandafter \@secondoftwo
 \fi
}%
\providecommand \@ifx [1]{%
 \ifx #1\expandafter \@firstoftwo
 \else \expandafter \@secondoftwo
 \fi
}%
\providecommand \natexlab [1]{#1}%
\providecommand \enquote  [1]{``#1''}%
\providecommand \bibnamefont  [1]{#1}%
\providecommand \bibfnamefont [1]{#1}%
\providecommand \citenamefont [1]{#1}%
\providecommand \href@noop [0]{\@secondoftwo}%
\providecommand \href [0]{\begingroup \@sanitize@url \@href}%
\providecommand \@href[1]{\@@startlink{#1}\@@href}%
\providecommand \@@href[1]{\endgroup#1\@@endlink}%
\providecommand \@sanitize@url [0]{\catcode `\\12\catcode `\$12\catcode `\&12\catcode `\#12\catcode `\^12\catcode `\_12\catcode `\%12\relax}%
\providecommand \@@startlink[1]{}%
\providecommand \@@endlink[0]{}%
\providecommand \url  [0]{\begingroup\@sanitize@url \@url }%
\providecommand \@url [1]{\endgroup\@href {#1}{\urlprefix }}%
\providecommand \urlprefix  [0]{URL }%
\providecommand \Eprint [0]{\href }%
\providecommand \doibase [0]{https://doi.org/}%
\providecommand \selectlanguage [0]{\@gobble}%
\providecommand \bibinfo  [0]{\@secondoftwo}%
\providecommand \bibfield  [0]{\@secondoftwo}%
\providecommand \translation [1]{[#1]}%
\providecommand \BibitemOpen [0]{}%
\providecommand \bibitemStop [0]{}%
\providecommand \bibitemNoStop [0]{.\EOS\space}%
\providecommand \EOS [0]{\spacefactor3000\relax}%
\providecommand \BibitemShut  [1]{\csname bibitem#1\endcsname}%
\let\auto@bib@innerbib\@empty
\bibitem [{\citenamefont {Cao}\ \emph {et~al.}(2018{\natexlab{a}})\citenamefont {Cao}, \citenamefont {Fatemi}, \citenamefont {Demir}, \citenamefont {Fang}, \citenamefont {Tomarken}, \citenamefont {Luo}, \citenamefont {Sanchez-Yamagishi}, \citenamefont {Watanabe}, \citenamefont {Taniguchi}, \citenamefont {Kaxiras}, \citenamefont {Ashoori},\ and\ \citenamefont {Jarillo-Herrero}}]{Cetal18}%
  \BibitemOpen
  \bibfield  {author} {\bibinfo {author} {\bibfnamefont {Y.}~\bibnamefont {Cao}}, \bibinfo {author} {\bibfnamefont {V.}~\bibnamefont {Fatemi}}, \bibinfo {author} {\bibfnamefont {A.}~\bibnamefont {Demir}}, \bibinfo {author} {\bibfnamefont {S.}~\bibnamefont {Fang}}, \bibinfo {author} {\bibfnamefont {S.~L.}\ \bibnamefont {Tomarken}}, \bibinfo {author} {\bibfnamefont {J.~Y.}\ \bibnamefont {Luo}}, \bibinfo {author} {\bibfnamefont {J.~D.}\ \bibnamefont {Sanchez-Yamagishi}}, \bibinfo {author} {\bibfnamefont {K.}~\bibnamefont {Watanabe}}, \bibinfo {author} {\bibfnamefont {T.}~\bibnamefont {Taniguchi}}, \bibinfo {author} {\bibfnamefont {E.}~\bibnamefont {Kaxiras}}, \bibinfo {author} {\bibfnamefont {R.~C.}\ \bibnamefont {Ashoori}},\ and\ \bibinfo {author} {\bibfnamefont {P.}~\bibnamefont {Jarillo-Herrero}},\ }\bibfield  {title} {\bibinfo {title} {{Correlated insulator behaviour at half-filling in magic-angle graphene superlattices}},\ }\href {https://doi.org/10.1038/nature26154} {\bibfield  {journal} {\bibinfo
  {journal} {Nature}\ }\textbf {\bibinfo {volume} {556}},\ \bibinfo {pages} {80} (\bibinfo {year} {2018}{\natexlab{a}})}\BibitemShut {NoStop}%
\bibitem [{\citenamefont {Cao}\ \emph {et~al.}(2018{\natexlab{b}})\citenamefont {Cao}, \citenamefont {Fatemi}, \citenamefont {Fang}, \citenamefont {Watanabe}, \citenamefont {Taniguchi}, \citenamefont {Kaxiras},\ and\ \citenamefont {Jarillo-Herrero}}]{Cetal18b}%
  \BibitemOpen
  \bibfield  {author} {\bibinfo {author} {\bibfnamefont {Y.}~\bibnamefont {Cao}}, \bibinfo {author} {\bibfnamefont {V.}~\bibnamefont {Fatemi}}, \bibinfo {author} {\bibfnamefont {S.}~\bibnamefont {Fang}}, \bibinfo {author} {\bibfnamefont {K.}~\bibnamefont {Watanabe}}, \bibinfo {author} {\bibfnamefont {T.}~\bibnamefont {Taniguchi}}, \bibinfo {author} {\bibfnamefont {E.}~\bibnamefont {Kaxiras}},\ and\ \bibinfo {author} {\bibfnamefont {P.}~\bibnamefont {Jarillo-Herrero}},\ }\bibfield  {title} {\bibinfo {title} {Unconventional superconductivity in magic-angle graphene superlattices},\ }\href {https://doi.org/10.1038/nature26160} {\bibfield  {journal} {\bibinfo  {journal} {Nature}\ }\textbf {\bibinfo {volume} {556}},\ \bibinfo {pages} {43} (\bibinfo {year} {2018}{\natexlab{b}})}\BibitemShut {NoStop}%
\bibitem [{\citenamefont {Yankowitz}\ \emph {et~al.}(2019)\citenamefont {Yankowitz}, \citenamefont {Chen}, \citenamefont {Polshyn}, \citenamefont {Zhang}, \citenamefont {Watanabe}, \citenamefont {Taniguchi}, \citenamefont {Graf}, \citenamefont {Young},\ and\ \citenamefont {Dean}}]{Yankowitz2019}%
  \BibitemOpen
  \bibfield  {author} {\bibinfo {author} {\bibfnamefont {M.}~\bibnamefont {Yankowitz}}, \bibinfo {author} {\bibfnamefont {S.}~\bibnamefont {Chen}}, \bibinfo {author} {\bibfnamefont {H.}~\bibnamefont {Polshyn}}, \bibinfo {author} {\bibfnamefont {Y.}~\bibnamefont {Zhang}}, \bibinfo {author} {\bibfnamefont {K.}~\bibnamefont {Watanabe}}, \bibinfo {author} {\bibfnamefont {T.}~\bibnamefont {Taniguchi}}, \bibinfo {author} {\bibfnamefont {D.}~\bibnamefont {Graf}}, \bibinfo {author} {\bibfnamefont {A.~F.}\ \bibnamefont {Young}},\ and\ \bibinfo {author} {\bibfnamefont {C.~R.}\ \bibnamefont {Dean}},\ }\bibfield  {title} {\bibinfo {title} {Tuning superconductivity in twisted bilayer graphene},\ }\href {https://doi.org/10.1126/science.aav1910} {\bibfield  {journal} {\bibinfo  {journal} {Science}\ }\textbf {\bibinfo {volume} {363}},\ \bibinfo {pages} {1059} (\bibinfo {year} {2019})}\BibitemShut {NoStop}%
\bibitem [{\citenamefont {Lu}\ \emph {et~al.}(2019)\citenamefont {Lu}, \citenamefont {Stepanov}, \citenamefont {Yang}, \citenamefont {Xie}, \citenamefont {Aamir}, \citenamefont {Das}, \citenamefont {Urgell}, \citenamefont {Watanabe}, \citenamefont {Taniguchi}, \citenamefont {Zhang}, \citenamefont {Bachtold}, \citenamefont {MacDonald},\ and\ \citenamefont {Efetov}}]{Lu2019_SC_TBG}%
  \BibitemOpen
  \bibfield  {author} {\bibinfo {author} {\bibfnamefont {X.}~\bibnamefont {Lu}}, \bibinfo {author} {\bibfnamefont {P.}~\bibnamefont {Stepanov}}, \bibinfo {author} {\bibfnamefont {W.}~\bibnamefont {Yang}}, \bibinfo {author} {\bibfnamefont {M.}~\bibnamefont {Xie}}, \bibinfo {author} {\bibfnamefont {M.~A.}\ \bibnamefont {Aamir}}, \bibinfo {author} {\bibfnamefont {I.}~\bibnamefont {Das}}, \bibinfo {author} {\bibfnamefont {C.}~\bibnamefont {Urgell}}, \bibinfo {author} {\bibfnamefont {K.}~\bibnamefont {Watanabe}}, \bibinfo {author} {\bibfnamefont {T.}~\bibnamefont {Taniguchi}}, \bibinfo {author} {\bibfnamefont {G.}~\bibnamefont {Zhang}}, \bibinfo {author} {\bibfnamefont {A.}~\bibnamefont {Bachtold}}, \bibinfo {author} {\bibfnamefont {A.~H.}\ \bibnamefont {MacDonald}},\ and\ \bibinfo {author} {\bibfnamefont {D.~K.}\ \bibnamefont {Efetov}},\ }\bibfield  {title} {\bibinfo {title} {Superconductors, orbital magnets and correlated states in magic-angle bilayer graphene},\ }\href
  {https://doi.org/10.1038/s41586-019-1695-0} {\bibfield  {journal} {\bibinfo  {journal} {Nature}\ }\textbf {\bibinfo {volume} {574}},\ \bibinfo {pages} {653} (\bibinfo {year} {2019})}\BibitemShut {NoStop}%
\bibitem [{\citenamefont {Stepanov}\ \emph {et~al.}(2020)\citenamefont {Stepanov}, \citenamefont {Das}, \citenamefont {Lu}, \citenamefont {Fahimniya}, \citenamefont {Watanabe}, \citenamefont {Taniguchi}, \citenamefont {Koppens}, \citenamefont {Lischner}, \citenamefont {Levitov},\ and\ \citenamefont {Efetov}}]{Stepanov2020_TBG}%
  \BibitemOpen
  \bibfield  {author} {\bibinfo {author} {\bibfnamefont {P.}~\bibnamefont {Stepanov}}, \bibinfo {author} {\bibfnamefont {I.}~\bibnamefont {Das}}, \bibinfo {author} {\bibfnamefont {X.}~\bibnamefont {Lu}}, \bibinfo {author} {\bibfnamefont {A.}~\bibnamefont {Fahimniya}}, \bibinfo {author} {\bibfnamefont {K.}~\bibnamefont {Watanabe}}, \bibinfo {author} {\bibfnamefont {T.}~\bibnamefont {Taniguchi}}, \bibinfo {author} {\bibfnamefont {F.~H.~L.}\ \bibnamefont {Koppens}}, \bibinfo {author} {\bibfnamefont {J.}~\bibnamefont {Lischner}}, \bibinfo {author} {\bibfnamefont {L.}~\bibnamefont {Levitov}},\ and\ \bibinfo {author} {\bibfnamefont {D.~K.}\ \bibnamefont {Efetov}},\ }\bibfield  {title} {\bibinfo {title} {Untying the insulating and superconducting orders in magic-angle graphene},\ }\href {https://doi.org/10.1038/s41586-020-2459-6} {\bibfield  {journal} {\bibinfo  {journal} {Nature}\ }\textbf {\bibinfo {volume} {583}},\ \bibinfo {pages} {375} (\bibinfo {year} {2020})}\BibitemShut {NoStop}%
\bibitem [{\citenamefont {Oh}\ \emph {et~al.}(2021)\citenamefont {Oh}, \citenamefont {Nuckolls}, \citenamefont {Wong}, \citenamefont {Lee}, \citenamefont {Liu}, \citenamefont {Watanabe}, \citenamefont {Taniguchi},\ and\ \citenamefont {Yazdani}}]{Oh2021UnconvSC}%
  \BibitemOpen
  \bibfield  {author} {\bibinfo {author} {\bibfnamefont {M.}~\bibnamefont {Oh}}, \bibinfo {author} {\bibfnamefont {K.~P.}\ \bibnamefont {Nuckolls}}, \bibinfo {author} {\bibfnamefont {D.}~\bibnamefont {Wong}}, \bibinfo {author} {\bibfnamefont {R.~L.}\ \bibnamefont {Lee}}, \bibinfo {author} {\bibfnamefont {X.}~\bibnamefont {Liu}}, \bibinfo {author} {\bibfnamefont {K.}~\bibnamefont {Watanabe}}, \bibinfo {author} {\bibfnamefont {T.}~\bibnamefont {Taniguchi}},\ and\ \bibinfo {author} {\bibfnamefont {A.}~\bibnamefont {Yazdani}},\ }\bibfield  {title} {\bibinfo {title} {Evidence for unconventional superconductivity in twisted bilayer graphene},\ }\href {https://doi.org/10.1038/s41586-021-04121-x} {\bibfield  {journal} {\bibinfo  {journal} {Nature}\ }\textbf {\bibinfo {volume} {600}},\ \bibinfo {pages} {240–245} (\bibinfo {year} {2021})}\BibitemShut {NoStop}%
\bibitem [{\citenamefont {Sharpe}\ \emph {et~al.}(2019)\citenamefont {Sharpe}, \citenamefont {Fox}, \citenamefont {Barnard}, \citenamefont {Finney}, \citenamefont {Watanabe}, \citenamefont {Taniguchi}, \citenamefont {Kastner},\ and\ \citenamefont {Goldhaber-Gordon}}]{Sharpe2019Ferromag}%
  \BibitemOpen
  \bibfield  {author} {\bibinfo {author} {\bibfnamefont {A.~L.}\ \bibnamefont {Sharpe}}, \bibinfo {author} {\bibfnamefont {E.~J.}\ \bibnamefont {Fox}}, \bibinfo {author} {\bibfnamefont {A.~W.}\ \bibnamefont {Barnard}}, \bibinfo {author} {\bibfnamefont {J.}~\bibnamefont {Finney}}, \bibinfo {author} {\bibfnamefont {K.}~\bibnamefont {Watanabe}}, \bibinfo {author} {\bibfnamefont {T.}~\bibnamefont {Taniguchi}}, \bibinfo {author} {\bibfnamefont {M.~A.}\ \bibnamefont {Kastner}},\ and\ \bibinfo {author} {\bibfnamefont {D.}~\bibnamefont {Goldhaber-Gordon}},\ }\bibfield  {title} {\bibinfo {title} {Emergent ferromagnetism near three-quarters filling in twisted bilayer graphene},\ }\href {https://doi.org/10.1126/science.aaw3780} {\bibfield  {journal} {\bibinfo  {journal} {Science}\ }\textbf {\bibinfo {volume} {365}},\ \bibinfo {pages} {605–608} (\bibinfo {year} {2019})}\BibitemShut {NoStop}%
\bibitem [{\citenamefont {Serlin}\ \emph {et~al.}(2020)\citenamefont {Serlin}, \citenamefont {Tschirhart}, \citenamefont {Polshyn}, \citenamefont {Zhang}, \citenamefont {Zhu}, \citenamefont {Watanabe}, \citenamefont {Taniguchi}, \citenamefont {Balents},\ and\ \citenamefont {Young}}]{serlin2020intrinsic}%
  \BibitemOpen
  \bibfield  {author} {\bibinfo {author} {\bibfnamefont {M.}~\bibnamefont {Serlin}}, \bibinfo {author} {\bibfnamefont {C.~L.}\ \bibnamefont {Tschirhart}}, \bibinfo {author} {\bibfnamefont {H.}~\bibnamefont {Polshyn}}, \bibinfo {author} {\bibfnamefont {Y.}~\bibnamefont {Zhang}}, \bibinfo {author} {\bibfnamefont {J.}~\bibnamefont {Zhu}}, \bibinfo {author} {\bibfnamefont {K.}~\bibnamefont {Watanabe}}, \bibinfo {author} {\bibfnamefont {T.}~\bibnamefont {Taniguchi}}, \bibinfo {author} {\bibfnamefont {L.}~\bibnamefont {Balents}},\ and\ \bibinfo {author} {\bibfnamefont {A.~F.}\ \bibnamefont {Young}},\ }\bibfield  {title} {\bibinfo {title} {Intrinsic quantized anomalous hall effect in a moir{\'{e}} heterostructure},\ }\href {https://doi.org/10.1126/science.aay5533} {\bibfield  {journal} {\bibinfo  {journal} {Science}\ }\textbf {\bibinfo {volume} {367}},\ \bibinfo {pages} {900} (\bibinfo {year} {2020})}\BibitemShut {NoStop}%
\bibitem [{\citenamefont {Nuckolls}\ \emph {et~al.}(2020)\citenamefont {Nuckolls}, \citenamefont {Oh}, \citenamefont {Wong}, \citenamefont {Lian}, \citenamefont {Watanabe}, \citenamefont {Taniguchi}, \citenamefont {Bernevig},\ and\ \citenamefont {Yazdani}}]{nuckolls2020strongly}%
  \BibitemOpen
  \bibfield  {author} {\bibinfo {author} {\bibfnamefont {K.~P.}\ \bibnamefont {Nuckolls}}, \bibinfo {author} {\bibfnamefont {M.}~\bibnamefont {Oh}}, \bibinfo {author} {\bibfnamefont {D.}~\bibnamefont {Wong}}, \bibinfo {author} {\bibfnamefont {B.}~\bibnamefont {Lian}}, \bibinfo {author} {\bibfnamefont {K.}~\bibnamefont {Watanabe}}, \bibinfo {author} {\bibfnamefont {T.}~\bibnamefont {Taniguchi}}, \bibinfo {author} {\bibfnamefont {B.~A.}\ \bibnamefont {Bernevig}},\ and\ \bibinfo {author} {\bibfnamefont {A.}~\bibnamefont {Yazdani}},\ }\bibfield  {title} {\bibinfo {title} {Strongly correlated chern insulators in magic-angle twisted bilayer graphene},\ }\href {https://doi.org/10.1038/s41586-020-3028-8} {\bibfield  {journal} {\bibinfo  {journal} {Nature}\ }\textbf {\bibinfo {volume} {588}},\ \bibinfo {pages} {610–615} (\bibinfo {year} {2020})}\BibitemShut {NoStop}%
\bibitem [{\citenamefont {Wu}\ \emph {et~al.}(2021)\citenamefont {Wu}, \citenamefont {Zhang}, \citenamefont {Watanabe}, \citenamefont {Taniguchi},\ and\ \citenamefont {Andrei}}]{wu2021chern}%
  \BibitemOpen
  \bibfield  {author} {\bibinfo {author} {\bibfnamefont {S.}~\bibnamefont {Wu}}, \bibinfo {author} {\bibfnamefont {Z.}~\bibnamefont {Zhang}}, \bibinfo {author} {\bibfnamefont {K.}~\bibnamefont {Watanabe}}, \bibinfo {author} {\bibfnamefont {T.}~\bibnamefont {Taniguchi}},\ and\ \bibinfo {author} {\bibfnamefont {E.~Y.}\ \bibnamefont {Andrei}},\ }\bibfield  {title} {\bibinfo {title} {Chern insulators, van hove singularities and topological flat bands in magic-angle twisted bilayer graphene},\ }\href {https://doi.org/10.1038/s41563-020-00911-2} {\bibfield  {journal} {\bibinfo  {journal} {Nature Materials}\ }\textbf {\bibinfo {volume} {20}},\ \bibinfo {pages} {488} (\bibinfo {year} {2021})}\BibitemShut {NoStop}%
\bibitem [{\citenamefont {Das}\ \emph {et~al.}(2021)\citenamefont {Das}, \citenamefont {Lu}, \citenamefont {Herzog-Arbeitman}, \citenamefont {Song}, \citenamefont {Watanabe}, \citenamefont {Taniguchi}, \citenamefont {Bernevig},\ and\ \citenamefont {Efetov}}]{Das2021Chern}%
  \BibitemOpen
  \bibfield  {author} {\bibinfo {author} {\bibfnamefont {I.}~\bibnamefont {Das}}, \bibinfo {author} {\bibfnamefont {X.}~\bibnamefont {Lu}}, \bibinfo {author} {\bibfnamefont {J.}~\bibnamefont {Herzog-Arbeitman}}, \bibinfo {author} {\bibfnamefont {Z.-D.}\ \bibnamefont {Song}}, \bibinfo {author} {\bibfnamefont {K.}~\bibnamefont {Watanabe}}, \bibinfo {author} {\bibfnamefont {T.}~\bibnamefont {Taniguchi}}, \bibinfo {author} {\bibfnamefont {B.~A.}\ \bibnamefont {Bernevig}},\ and\ \bibinfo {author} {\bibfnamefont {D.~K.}\ \bibnamefont {Efetov}},\ }\bibfield  {title} {\bibinfo {title} {Symmetry-broken chern insulators and rashba-like landau-level crossings in magic-angle bilayer graphene},\ }\href {https://doi.org/10.1038/s41567-021-01186-3} {\bibfield  {journal} {\bibinfo  {journal} {Nature Physics}\ }\textbf {\bibinfo {volume} {17}},\ \bibinfo {pages} {710–714} (\bibinfo {year} {2021})}\BibitemShut {NoStop}%
\bibitem [{\citenamefont {Xie}\ \emph {et~al.}(2021)\citenamefont {Xie}, \citenamefont {Pierce}, \citenamefont {Park}, \citenamefont {Parker}, \citenamefont {Khalaf}, \citenamefont {Ledwith}, \citenamefont {Cao}, \citenamefont {Lee}, \citenamefont {Chen}, \citenamefont {Forrester}, \citenamefont {Watanabe}, \citenamefont {Taniguchi}, \citenamefont {Vishwanath}, \citenamefont {Jarillo-Herrero},\ and\ \citenamefont {Yacoby}}]{xie2021fractional}%
  \BibitemOpen
  \bibfield  {author} {\bibinfo {author} {\bibfnamefont {Y.}~\bibnamefont {Xie}}, \bibinfo {author} {\bibfnamefont {A.~T.}\ \bibnamefont {Pierce}}, \bibinfo {author} {\bibfnamefont {J.~M.}\ \bibnamefont {Park}}, \bibinfo {author} {\bibfnamefont {D.~E.}\ \bibnamefont {Parker}}, \bibinfo {author} {\bibfnamefont {E.}~\bibnamefont {Khalaf}}, \bibinfo {author} {\bibfnamefont {P.}~\bibnamefont {Ledwith}}, \bibinfo {author} {\bibfnamefont {Y.}~\bibnamefont {Cao}}, \bibinfo {author} {\bibfnamefont {S.~H.}\ \bibnamefont {Lee}}, \bibinfo {author} {\bibfnamefont {S.}~\bibnamefont {Chen}}, \bibinfo {author} {\bibfnamefont {P.~R.}\ \bibnamefont {Forrester}}, \bibinfo {author} {\bibfnamefont {K.}~\bibnamefont {Watanabe}}, \bibinfo {author} {\bibfnamefont {T.}~\bibnamefont {Taniguchi}}, \bibinfo {author} {\bibfnamefont {A.}~\bibnamefont {Vishwanath}}, \bibinfo {author} {\bibfnamefont {P.}~\bibnamefont {Jarillo-Herrero}},\ and\ \bibinfo {author} {\bibfnamefont {A.}~\bibnamefont {Yacoby}},\ }\bibfield  {title} {\bibinfo
  {title} {Fractional chern insulators in magic-angle twisted bilayer graphene},\ }\href {https://doi.org/10.1038/s41586-021-04002-3} {\bibfield  {journal} {\bibinfo  {journal} {Nature}\ }\textbf {\bibinfo {volume} {600}},\ \bibinfo {pages} {439} (\bibinfo {year} {2021})}\BibitemShut {NoStop}%
\bibitem [{\citenamefont {Choi}\ \emph {et~al.}(2021{\natexlab{a}})\citenamefont {Choi}, \citenamefont {Kim}, \citenamefont {Lewandowski}, \citenamefont {Peng}, \citenamefont {Thomson}, \citenamefont {Polski}, \citenamefont {Zhang}, \citenamefont {Watanabe}, \citenamefont {Taniguchi}, \citenamefont {Alicea},\ and\ \citenamefont {Nadj-Perge}}]{Youngjoon2021}%
  \BibitemOpen
  \bibfield  {author} {\bibinfo {author} {\bibfnamefont {Y.}~\bibnamefont {Choi}}, \bibinfo {author} {\bibfnamefont {H.}~\bibnamefont {Kim}}, \bibinfo {author} {\bibfnamefont {C.}~\bibnamefont {Lewandowski}}, \bibinfo {author} {\bibfnamefont {Y.}~\bibnamefont {Peng}}, \bibinfo {author} {\bibfnamefont {A.}~\bibnamefont {Thomson}}, \bibinfo {author} {\bibfnamefont {R.}~\bibnamefont {Polski}}, \bibinfo {author} {\bibfnamefont {Y.}~\bibnamefont {Zhang}}, \bibinfo {author} {\bibfnamefont {K.}~\bibnamefont {Watanabe}}, \bibinfo {author} {\bibfnamefont {T.}~\bibnamefont {Taniguchi}}, \bibinfo {author} {\bibfnamefont {J.}~\bibnamefont {Alicea}},\ and\ \bibinfo {author} {\bibfnamefont {S.}~\bibnamefont {Nadj-Perge}},\ }\bibfield  {title} {\bibinfo {title} {Interaction-driven band flattening and correlated phases in twisted bilayer graphene},\ }\href {https://doi.org/10.1038/s41567-021-01359-0} {\bibfield  {journal} {\bibinfo  {journal} {Nature Physics}\ }\textbf {\bibinfo {volume} {17}},\ \bibinfo {pages} {1375}
  (\bibinfo {year} {2021}{\natexlab{a}})}\BibitemShut {NoStop}%
\bibitem [{\citenamefont {Choi}\ \emph {et~al.}(2021{\natexlab{b}})\citenamefont {Choi}, \citenamefont {Kim}, \citenamefont {Peng}, \citenamefont {Thomson}, \citenamefont {Lewandowski}, \citenamefont {Polski}, \citenamefont {Zhang}, \citenamefont {Arora}, \citenamefont {Watanabe}, \citenamefont {Taniguchi}, \citenamefont {Alicea},\ and\ \citenamefont {Nadj-Perge}}]{choi2021correlation}%
  \BibitemOpen
  \bibfield  {author} {\bibinfo {author} {\bibfnamefont {Y.}~\bibnamefont {Choi}}, \bibinfo {author} {\bibfnamefont {H.}~\bibnamefont {Kim}}, \bibinfo {author} {\bibfnamefont {Y.}~\bibnamefont {Peng}}, \bibinfo {author} {\bibfnamefont {A.}~\bibnamefont {Thomson}}, \bibinfo {author} {\bibfnamefont {C.}~\bibnamefont {Lewandowski}}, \bibinfo {author} {\bibfnamefont {R.}~\bibnamefont {Polski}}, \bibinfo {author} {\bibfnamefont {Y.}~\bibnamefont {Zhang}}, \bibinfo {author} {\bibfnamefont {H.~S.}\ \bibnamefont {Arora}}, \bibinfo {author} {\bibfnamefont {K.}~\bibnamefont {Watanabe}}, \bibinfo {author} {\bibfnamefont {T.}~\bibnamefont {Taniguchi}}, \bibinfo {author} {\bibfnamefont {J.}~\bibnamefont {Alicea}},\ and\ \bibinfo {author} {\bibfnamefont {S.}~\bibnamefont {Nadj-Perge}},\ }\bibfield  {title} {\bibinfo {title} {Correlation-driven topological phases in magic-angle twisted bilayer graphene},\ }\href {https://doi.org/10.1038/s41586-020-03159-7} {\bibfield  {journal} {\bibinfo  {journal} {Nature}\ }\textbf
  {\bibinfo {volume} {589}},\ \bibinfo {pages} {536} (\bibinfo {year} {2021}{\natexlab{b}})}\BibitemShut {NoStop}%
\bibitem [{\citenamefont {Saito}\ \emph {et~al.}(2021)\citenamefont {Saito}, \citenamefont {Ge}, \citenamefont {Rademaker}, \citenamefont {Watanabe}, \citenamefont {Taniguchi}, \citenamefont {Abanin},\ and\ \citenamefont {Young}}]{saito2021hofstader}%
  \BibitemOpen
  \bibfield  {author} {\bibinfo {author} {\bibfnamefont {Y.}~\bibnamefont {Saito}}, \bibinfo {author} {\bibfnamefont {J.}~\bibnamefont {Ge}}, \bibinfo {author} {\bibfnamefont {L.}~\bibnamefont {Rademaker}}, \bibinfo {author} {\bibfnamefont {K.}~\bibnamefont {Watanabe}}, \bibinfo {author} {\bibfnamefont {T.}~\bibnamefont {Taniguchi}}, \bibinfo {author} {\bibfnamefont {D.~A.}\ \bibnamefont {Abanin}},\ and\ \bibinfo {author} {\bibfnamefont {A.~F.}\ \bibnamefont {Young}},\ }\bibfield  {title} {\bibinfo {title} {Hofstadter subband ferromagnetism and symmetry-broken chern insulators in twisted bilayer graphene},\ }\href {https://doi.org/10.1038/s41567-020-01129-4} {\bibfield  {journal} {\bibinfo  {journal} {Nature Physics}\ }\textbf {\bibinfo {volume} {17}},\ \bibinfo {pages} {478} (\bibinfo {year} {2021})}\BibitemShut {NoStop}%
\bibitem [{\citenamefont {Pierce}\ \emph {et~al.}(2021)\citenamefont {Pierce}, \citenamefont {Xie}, \citenamefont {Park}, \citenamefont {Khalaf}, \citenamefont {Lee}, \citenamefont {Cao}, \citenamefont {Parker}, \citenamefont {Forrester}, \citenamefont {Chen}, \citenamefont {Watanabe}, \citenamefont {Taniguchi}, \citenamefont {Vishwanath}, \citenamefont {Jarillo-Herrero},\ and\ \citenamefont {Yacoby}}]{pierce2021unconventional}%
  \BibitemOpen
  \bibfield  {author} {\bibinfo {author} {\bibfnamefont {A.~T.}\ \bibnamefont {Pierce}}, \bibinfo {author} {\bibfnamefont {Y.}~\bibnamefont {Xie}}, \bibinfo {author} {\bibfnamefont {J.~M.}\ \bibnamefont {Park}}, \bibinfo {author} {\bibfnamefont {E.}~\bibnamefont {Khalaf}}, \bibinfo {author} {\bibfnamefont {S.~H.}\ \bibnamefont {Lee}}, \bibinfo {author} {\bibfnamefont {Y.}~\bibnamefont {Cao}}, \bibinfo {author} {\bibfnamefont {D.~E.}\ \bibnamefont {Parker}}, \bibinfo {author} {\bibfnamefont {P.~R.}\ \bibnamefont {Forrester}}, \bibinfo {author} {\bibfnamefont {S.}~\bibnamefont {Chen}}, \bibinfo {author} {\bibfnamefont {K.}~\bibnamefont {Watanabe}}, \bibinfo {author} {\bibfnamefont {T.}~\bibnamefont {Taniguchi}}, \bibinfo {author} {\bibfnamefont {A.}~\bibnamefont {Vishwanath}}, \bibinfo {author} {\bibfnamefont {P.}~\bibnamefont {Jarillo-Herrero}},\ and\ \bibinfo {author} {\bibfnamefont {A.}~\bibnamefont {Yacoby}},\ }\bibfield  {title} {\bibinfo {title} {Unconventional sequence of correlated chern insulators in
  magic-angle twisted bilayer graphene},\ }\href {https://doi.org/10.1038/s41567-021-01347-4} {\bibfield  {journal} {\bibinfo  {journal} {Nature Physics}\ }\textbf {\bibinfo {volume} {17}},\ \bibinfo {pages} {1210} (\bibinfo {year} {2021})}\BibitemShut {NoStop}%
\bibitem [{\citenamefont {Stepanov}\ \emph {et~al.}(2021)\citenamefont {Stepanov}, \citenamefont {Xie}, \citenamefont {Taniguchi}, \citenamefont {Watanabe}, \citenamefont {Lu}, \citenamefont {MacDonald}, \citenamefont {Bernevig},\ and\ \citenamefont {Efetov}}]{stepanov2021competing}%
  \BibitemOpen
  \bibfield  {author} {\bibinfo {author} {\bibfnamefont {P.}~\bibnamefont {Stepanov}}, \bibinfo {author} {\bibfnamefont {M.}~\bibnamefont {Xie}}, \bibinfo {author} {\bibfnamefont {T.}~\bibnamefont {Taniguchi}}, \bibinfo {author} {\bibfnamefont {K.}~\bibnamefont {Watanabe}}, \bibinfo {author} {\bibfnamefont {X.}~\bibnamefont {Lu}}, \bibinfo {author} {\bibfnamefont {A.~H.}\ \bibnamefont {MacDonald}}, \bibinfo {author} {\bibfnamefont {B.~A.}\ \bibnamefont {Bernevig}},\ and\ \bibinfo {author} {\bibfnamefont {D.~K.}\ \bibnamefont {Efetov}},\ }\bibfield  {title} {\bibinfo {title} {Competing zero-field chern insulators in superconducting twisted bilayer graphene},\ }\href {https://doi.org/10.1103/PhysRevLett.127.197701} {\bibfield  {journal} {\bibinfo  {journal} {Physical Review Letters}\ }\textbf {\bibinfo {volume} {127}},\ \bibinfo {pages} {197701} (\bibinfo {year} {2021})}\BibitemShut {NoStop}%
\bibitem [{\citenamefont {Ribeiro-Palau}\ \emph {et~al.}(2018)\citenamefont {Ribeiro-Palau}, \citenamefont {Zhang}, \citenamefont {Watanabe}, \citenamefont {Taniguchi}, \citenamefont {Hone},\ and\ \citenamefont {Dean}}]{RibeiroPalau2018}%
  \BibitemOpen
  \bibfield  {author} {\bibinfo {author} {\bibfnamefont {R.}~\bibnamefont {Ribeiro-Palau}}, \bibinfo {author} {\bibfnamefont {C.}~\bibnamefont {Zhang}}, \bibinfo {author} {\bibfnamefont {K.}~\bibnamefont {Watanabe}}, \bibinfo {author} {\bibfnamefont {T.}~\bibnamefont {Taniguchi}}, \bibinfo {author} {\bibfnamefont {J.}~\bibnamefont {Hone}},\ and\ \bibinfo {author} {\bibfnamefont {C.~R.}\ \bibnamefont {Dean}},\ }\bibfield  {title} {\bibinfo {title} {Twistable electronics with dynamically rotatable heterostructures},\ }\href {https://doi.org/10.1126/science.aat6981} {\bibfield  {journal} {\bibinfo  {journal} {Science}\ }\textbf {\bibinfo {volume} {361}},\ \bibinfo {pages} {690–693} (\bibinfo {year} {2018})}\BibitemShut {NoStop}%
\bibitem [{\citenamefont {Chu}\ \emph {et~al.}(2020)\citenamefont {Chu}, \citenamefont {Liu}, \citenamefont {Yuan}, \citenamefont {Shen}, \citenamefont {Yang}, \citenamefont {Shi}, \citenamefont {Yang},\ and\ \citenamefont {Zhang}}]{Chu2020Review}%
  \BibitemOpen
  \bibfield  {author} {\bibinfo {author} {\bibfnamefont {Y.}~\bibnamefont {Chu}}, \bibinfo {author} {\bibfnamefont {L.}~\bibnamefont {Liu}}, \bibinfo {author} {\bibfnamefont {Y.}~\bibnamefont {Yuan}}, \bibinfo {author} {\bibfnamefont {C.}~\bibnamefont {Shen}}, \bibinfo {author} {\bibfnamefont {R.}~\bibnamefont {Yang}}, \bibinfo {author} {\bibfnamefont {D.}~\bibnamefont {Shi}}, \bibinfo {author} {\bibfnamefont {W.}~\bibnamefont {Yang}},\ and\ \bibinfo {author} {\bibfnamefont {G.}~\bibnamefont {Zhang}},\ }\bibfield  {title} {\bibinfo {title} {A review of experimental advances in twisted graphene moiré superlattice*},\ }\href {https://doi.org/10.1088/1674-1056/abb221} {\bibfield  {journal} {\bibinfo  {journal} {Chinese Physics B}\ }\textbf {\bibinfo {volume} {29}},\ \bibinfo {pages} {128104} (\bibinfo {year} {2020})}\BibitemShut {NoStop}%
\bibitem [{\citenamefont {Zhang}\ \emph {et~al.}(2021)\citenamefont {Zhang}, \citenamefont {Lin}, \citenamefont {Lei}, \citenamefont {Qi}, \citenamefont {Ban}, \citenamefont {Vinu}, \citenamefont {Yi},\ and\ \citenamefont {Liu}}]{Zhang2021Twist}%
  \BibitemOpen
  \bibfield  {author} {\bibinfo {author} {\bibfnamefont {E.}~\bibnamefont {Zhang}}, \bibinfo {author} {\bibfnamefont {F.}~\bibnamefont {Lin}}, \bibinfo {author} {\bibfnamefont {Z.}~\bibnamefont {Lei}}, \bibinfo {author} {\bibfnamefont {S.}~\bibnamefont {Qi}}, \bibinfo {author} {\bibfnamefont {S.}~\bibnamefont {Ban}}, \bibinfo {author} {\bibfnamefont {A.}~\bibnamefont {Vinu}}, \bibinfo {author} {\bibfnamefont {J.}~\bibnamefont {Yi}},\ and\ \bibinfo {author} {\bibfnamefont {Y.}~\bibnamefont {Liu}},\ }\bibfield  {title} {\bibinfo {title} {Twist the doorknob to open the electronic properties of graphene-based van der waals structure},\ }\href {https://doi.org/10.1016/j.matt.2021.08.020} {\bibfield  {journal} {\bibinfo  {journal} {Matter}\ }\textbf {\bibinfo {volume} {4}},\ \bibinfo {pages} {3444–3482} (\bibinfo {year} {2021})}\BibitemShut {NoStop}%
\bibitem [{\citenamefont {Andrei}\ \emph {et~al.}(2021)\citenamefont {Andrei}, \citenamefont {Efetov}, \citenamefont {Jarillo-Herrero}, \citenamefont {MacDonald}, \citenamefont {Mak}, \citenamefont {Senthil}, \citenamefont {Tutuc}, \citenamefont {Yazdani},\ and\ \citenamefont {Young}}]{Andrei2021Marvels}%
  \BibitemOpen
  \bibfield  {author} {\bibinfo {author} {\bibfnamefont {E.~Y.}\ \bibnamefont {Andrei}}, \bibinfo {author} {\bibfnamefont {D.~K.}\ \bibnamefont {Efetov}}, \bibinfo {author} {\bibfnamefont {P.}~\bibnamefont {Jarillo-Herrero}}, \bibinfo {author} {\bibfnamefont {A.~H.}\ \bibnamefont {MacDonald}}, \bibinfo {author} {\bibfnamefont {K.~F.}\ \bibnamefont {Mak}}, \bibinfo {author} {\bibfnamefont {T.}~\bibnamefont {Senthil}}, \bibinfo {author} {\bibfnamefont {E.}~\bibnamefont {Tutuc}}, \bibinfo {author} {\bibfnamefont {A.}~\bibnamefont {Yazdani}},\ and\ \bibinfo {author} {\bibfnamefont {A.~F.}\ \bibnamefont {Young}},\ }\bibfield  {title} {\bibinfo {title} {The marvels of moiré materials},\ }\href {https://doi.org/10.1038/s41578-021-00284-1} {\bibfield  {journal} {\bibinfo  {journal} {Nature Reviews Materials}\ }\textbf {\bibinfo {volume} {6}},\ \bibinfo {pages} {201–206} (\bibinfo {year} {2021})}\BibitemShut {NoStop}%
\bibitem [{\citenamefont {Ma}\ \emph {et~al.}(2023)\citenamefont {Ma}, \citenamefont {Zhang}, \citenamefont {Li}, \citenamefont {Geng},\ and\ \citenamefont {Hu}}]{Ma2023Small}%
  \BibitemOpen
  \bibfield  {author} {\bibinfo {author} {\bibfnamefont {W.}~\bibnamefont {Ma}}, \bibinfo {author} {\bibfnamefont {Q.}~\bibnamefont {Zhang}}, \bibinfo {author} {\bibfnamefont {L.}~\bibnamefont {Li}}, \bibinfo {author} {\bibfnamefont {D.}~\bibnamefont {Geng}},\ and\ \bibinfo {author} {\bibfnamefont {W.}~\bibnamefont {Hu}},\ }\bibfield  {title} {\bibinfo {title} {Small twist, big miracle—recent progress in the fabrication of twisted 2d materials},\ }\href {https://doi.org/10.1039/d3tc02660d} {\bibfield  {journal} {\bibinfo  {journal} {Journal of Materials Chemistry C}\ }\textbf {\bibinfo {volume} {11}},\ \bibinfo {pages} {15793–15816} (\bibinfo {year} {2023})}\BibitemShut {NoStop}%
\bibitem [{\citenamefont {Saumya}\ \emph {et~al.}(2023)\citenamefont {Saumya}, \citenamefont {Naskar},\ and\ \citenamefont {Mukhopadhyay}}]{Saumya2023Magic}%
  \BibitemOpen
  \bibfield  {author} {\bibinfo {author} {\bibfnamefont {K.}~\bibnamefont {Saumya}}, \bibinfo {author} {\bibfnamefont {S.}~\bibnamefont {Naskar}},\ and\ \bibinfo {author} {\bibfnamefont {T.}~\bibnamefont {Mukhopadhyay}},\ }\bibfield  {title} {\bibinfo {title} {‘magic’ of twisted multi-layered graphene and 2d nano-heterostructures},\ }\href {https://doi.org/10.1088/2399-1984/acf0a9} {\bibfield  {journal} {\bibinfo  {journal} {Nano Futures}\ }\textbf {\bibinfo {volume} {7}},\ \bibinfo {pages} {032005} (\bibinfo {year} {2023})}\BibitemShut {NoStop}%
\bibitem [{\citenamefont {Baek}\ \emph {et~al.}(2025)\citenamefont {Baek}, \citenamefont {Choi},\ and\ \citenamefont {Lee}}]{Baek2025Moire}%
  \BibitemOpen
  \bibfield  {author} {\bibinfo {author} {\bibfnamefont {J.-H.}\ \bibnamefont {Baek}}, \bibinfo {author} {\bibfnamefont {W.}~\bibnamefont {Choi}},\ and\ \bibinfo {author} {\bibfnamefont {G.-H.}\ \bibnamefont {Lee}},\ }\bibfield  {title} {\bibinfo {title} {Moiré-driven band renormalization and quantum transport in twisted 2d materials},\ }\bibfield  {journal} {\bibinfo  {journal} {npj 2D Materials and Applications}\ }\textbf {\bibinfo {volume} {10}},\ \href {https://doi.org/10.1038/s41699-025-00642-5} {10.1038/s41699-025-00642-5} (\bibinfo {year} {2025})\BibitemShut {NoStop}%
\bibitem [{\citenamefont {Tong}\ \emph {et~al.}(2026)\citenamefont {Tong}, \citenamefont {Xu},\ and\ \citenamefont {Yin}}]{Ling2026Recent}%
  \BibitemOpen
  \bibfield  {author} {\bibinfo {author} {\bibfnamefont {L.-H.}\ \bibnamefont {Tong}}, \bibinfo {author} {\bibfnamefont {C.-C.}\ \bibnamefont {Xu}},\ and\ \bibinfo {author} {\bibfnamefont {L.-J.}\ \bibnamefont {Yin}},\ }\bibfield  {title} {\bibinfo {title} {Recent progress in scanning tunneling microscopy/spectroscopy characterization for magic-angle twisted multilayer graphene},\ }\href {https://doi.org/10.15302/frontphys.2026.065301} {\bibfield  {journal} {\bibinfo  {journal} {Frontiers of Physics}\ }\textbf {\bibinfo {volume} {21}},\ \bibinfo {pages} {065301} (\bibinfo {year} {2026})}\BibitemShut {NoStop}%
\bibitem [{\citenamefont {Yang}\ and\ \citenamefont {Shen}(2026)}]{Yang2026Twisting}%
  \BibitemOpen
  \bibfield  {author} {\bibinfo {author} {\bibfnamefont {S.-Y.}\ \bibnamefont {Yang}}\ and\ \bibinfo {author} {\bibfnamefont {C.}~\bibnamefont {Shen}},\ }\bibfield  {title} {\bibinfo {title} {Twisting graphene into correlation and topology},\ }\bibfield  {journal} {\bibinfo  {journal} {National Science Review}\ }\href {https://doi.org/10.1093/nsr/nwag363} {10.1093/nsr/nwag363} (\bibinfo {year} {2026})\BibitemShut {NoStop}%
\bibitem [{\citenamefont {Chen}\ \emph {et~al.}(2020)\citenamefont {Chen}, \citenamefont {He}, \citenamefont {Zhang}, \citenamefont {Hsieh}, \citenamefont {Fei}, \citenamefont {Watanabe}, \citenamefont {Taniguchi}, \citenamefont {Cobden}, \citenamefont {Xu}, \citenamefont {Dean},\ and\ \citenamefont {Yankowitz}}]{chen2020electrically}%
  \BibitemOpen
  \bibfield  {author} {\bibinfo {author} {\bibfnamefont {S.}~\bibnamefont {Chen}}, \bibinfo {author} {\bibfnamefont {M.}~\bibnamefont {He}}, \bibinfo {author} {\bibfnamefont {Y.-H.}\ \bibnamefont {Zhang}}, \bibinfo {author} {\bibfnamefont {V.}~\bibnamefont {Hsieh}}, \bibinfo {author} {\bibfnamefont {Z.}~\bibnamefont {Fei}}, \bibinfo {author} {\bibfnamefont {K.}~\bibnamefont {Watanabe}}, \bibinfo {author} {\bibfnamefont {T.}~\bibnamefont {Taniguchi}}, \bibinfo {author} {\bibfnamefont {D.~H.}\ \bibnamefont {Cobden}}, \bibinfo {author} {\bibfnamefont {X.}~\bibnamefont {Xu}}, \bibinfo {author} {\bibfnamefont {C.~R.}\ \bibnamefont {Dean}},\ and\ \bibinfo {author} {\bibfnamefont {M.}~\bibnamefont {Yankowitz}},\ }\bibfield  {title} {\bibinfo {title} {Electrically tunable correlated and topological states in twisted monolayer–bilayer graphene},\ }\href {https://doi.org/10.1038/s41567-020-01062-6} {\bibfield  {journal} {\bibinfo  {journal} {Nat. Phys.}\ }\textbf {\bibinfo {volume} {17}},\ \bibinfo {pages} {374}
  (\bibinfo {year} {2020})}\BibitemShut {NoStop}%
\bibitem [{\citenamefont {Polshyn}\ \emph {et~al.}(2020)\citenamefont {Polshyn}, \citenamefont {Zhu}, \citenamefont {Kumar}, \citenamefont {Zhang}, \citenamefont {Yang}, \citenamefont {Tschirhart}, \citenamefont {Serlin}, \citenamefont {Watanabe}, \citenamefont {Taniguchi}, \citenamefont {MacDonald},\ and\ \citenamefont {Young}}]{Polshyn2020}%
  \BibitemOpen
  \bibfield  {author} {\bibinfo {author} {\bibfnamefont {H.}~\bibnamefont {Polshyn}}, \bibinfo {author} {\bibfnamefont {J.}~\bibnamefont {Zhu}}, \bibinfo {author} {\bibfnamefont {M.~A.}\ \bibnamefont {Kumar}}, \bibinfo {author} {\bibfnamefont {Y.}~\bibnamefont {Zhang}}, \bibinfo {author} {\bibfnamefont {F.}~\bibnamefont {Yang}}, \bibinfo {author} {\bibfnamefont {C.~L.}\ \bibnamefont {Tschirhart}}, \bibinfo {author} {\bibfnamefont {M.}~\bibnamefont {Serlin}}, \bibinfo {author} {\bibfnamefont {K.}~\bibnamefont {Watanabe}}, \bibinfo {author} {\bibfnamefont {T.}~\bibnamefont {Taniguchi}}, \bibinfo {author} {\bibfnamefont {A.~H.}\ \bibnamefont {MacDonald}},\ and\ \bibinfo {author} {\bibfnamefont {A.~F.}\ \bibnamefont {Young}},\ }\bibfield  {title} {\bibinfo {title} {Electrical switching of magnetic order in an orbital chern insulator},\ }\href {https://doi.org/10.1038/s41586-020-2963-8} {\bibfield  {journal} {\bibinfo  {journal} {Nature}\ }\textbf {\bibinfo {volume} {588}},\ \bibinfo {pages} {66–70} (\bibinfo
  {year} {2020})}\BibitemShut {NoStop}%
\bibitem [{\citenamefont {Xu}\ \emph {et~al.}(2021)\citenamefont {Xu}, \citenamefont {Al~Ezzi}, \citenamefont {Balakrishnan}, \citenamefont {Garcia-Ruiz}, \citenamefont {Tsim}, \citenamefont {Mullan}, \citenamefont {Barrier}, \citenamefont {Xin}, \citenamefont {Piot}, \citenamefont {Taniguchi}, \citenamefont {Watanabe}, \citenamefont {Carvalho}, \citenamefont {Mishchenko}, \citenamefont {Geim}, \citenamefont {Fal’ko}, \citenamefont {Adam}, \citenamefont {Neto}, \citenamefont {Novoselov},\ and\ \citenamefont {Shi}}]{xu2021tunable}%
  \BibitemOpen
  \bibfield  {author} {\bibinfo {author} {\bibfnamefont {S.}~\bibnamefont {Xu}}, \bibinfo {author} {\bibfnamefont {M.~M.}\ \bibnamefont {Al~Ezzi}}, \bibinfo {author} {\bibfnamefont {N.}~\bibnamefont {Balakrishnan}}, \bibinfo {author} {\bibfnamefont {A.}~\bibnamefont {Garcia-Ruiz}}, \bibinfo {author} {\bibfnamefont {B.}~\bibnamefont {Tsim}}, \bibinfo {author} {\bibfnamefont {C.}~\bibnamefont {Mullan}}, \bibinfo {author} {\bibfnamefont {J.}~\bibnamefont {Barrier}}, \bibinfo {author} {\bibfnamefont {N.}~\bibnamefont {Xin}}, \bibinfo {author} {\bibfnamefont {B.~A.}\ \bibnamefont {Piot}}, \bibinfo {author} {\bibfnamefont {T.}~\bibnamefont {Taniguchi}}, \bibinfo {author} {\bibfnamefont {K.}~\bibnamefont {Watanabe}}, \bibinfo {author} {\bibfnamefont {A.}~\bibnamefont {Carvalho}}, \bibinfo {author} {\bibfnamefont {A.}~\bibnamefont {Mishchenko}}, \bibinfo {author} {\bibfnamefont {A.~K.}\ \bibnamefont {Geim}}, \bibinfo {author} {\bibfnamefont {V.~I.}\ \bibnamefont {Fal’ko}}, \bibinfo {author} {\bibfnamefont
  {S.}~\bibnamefont {Adam}}, \bibinfo {author} {\bibfnamefont {A.~H.~C.}\ \bibnamefont {Neto}}, \bibinfo {author} {\bibfnamefont {K.~S.}\ \bibnamefont {Novoselov}},\ and\ \bibinfo {author} {\bibfnamefont {Y.}~\bibnamefont {Shi}},\ }\bibfield  {title} {\bibinfo {title} {Tunable van hove singularities and correlated states in twisted monolayer–bilayer graphene},\ }\href {https://doi.org/10.1038/s41567-021-01172-9} {\bibfield  {journal} {\bibinfo  {journal} {Nat. Phys.}\ }\textbf {\bibinfo {volume} {17}},\ \bibinfo {pages} {619} (\bibinfo {year} {2021})}\BibitemShut {NoStop}%
\bibitem [{\citenamefont {Polshyn}\ \emph {et~al.}(2021)\citenamefont {Polshyn}, \citenamefont {Zhang}, \citenamefont {Kumar}, \citenamefont {Soejima}, \citenamefont {Ledwith}, \citenamefont {Watanabe}, \citenamefont {Taniguchi}, \citenamefont {Vishwanath}, \citenamefont {Zaletel},\ and\ \citenamefont {Young}}]{Polshyn2021}%
  \BibitemOpen
  \bibfield  {author} {\bibinfo {author} {\bibfnamefont {H.}~\bibnamefont {Polshyn}}, \bibinfo {author} {\bibfnamefont {Y.}~\bibnamefont {Zhang}}, \bibinfo {author} {\bibfnamefont {M.~A.}\ \bibnamefont {Kumar}}, \bibinfo {author} {\bibfnamefont {T.}~\bibnamefont {Soejima}}, \bibinfo {author} {\bibfnamefont {P.}~\bibnamefont {Ledwith}}, \bibinfo {author} {\bibfnamefont {K.}~\bibnamefont {Watanabe}}, \bibinfo {author} {\bibfnamefont {T.}~\bibnamefont {Taniguchi}}, \bibinfo {author} {\bibfnamefont {A.}~\bibnamefont {Vishwanath}}, \bibinfo {author} {\bibfnamefont {M.~P.}\ \bibnamefont {Zaletel}},\ and\ \bibinfo {author} {\bibfnamefont {A.~F.}\ \bibnamefont {Young}},\ }\bibfield  {title} {\bibinfo {title} {Topological charge density waves at half-integer filling of a moiré superlattice},\ }\href {https://doi.org/10.1038/s41567-021-01418-6} {\bibfield  {journal} {\bibinfo  {journal} {Nature Physics}\ }\textbf {\bibinfo {volume} {18}},\ \bibinfo {pages} {42–47} (\bibinfo {year} {2021})}\BibitemShut {NoStop}%
\bibitem [{\citenamefont {He}\ \emph {et~al.}(2021)\citenamefont {He}, \citenamefont {Zhang}, \citenamefont {Li}, \citenamefont {Fei}, \citenamefont {Watanabe}, \citenamefont {Taniguchi}, \citenamefont {Xu},\ and\ \citenamefont {Yankowitz}}]{he2021competing}%
  \BibitemOpen
  \bibfield  {author} {\bibinfo {author} {\bibfnamefont {M.}~\bibnamefont {He}}, \bibinfo {author} {\bibfnamefont {Y.-H.}\ \bibnamefont {Zhang}}, \bibinfo {author} {\bibfnamefont {Y.}~\bibnamefont {Li}}, \bibinfo {author} {\bibfnamefont {Z.}~\bibnamefont {Fei}}, \bibinfo {author} {\bibfnamefont {K.}~\bibnamefont {Watanabe}}, \bibinfo {author} {\bibfnamefont {T.}~\bibnamefont {Taniguchi}}, \bibinfo {author} {\bibfnamefont {X.}~\bibnamefont {Xu}},\ and\ \bibinfo {author} {\bibfnamefont {M.}~\bibnamefont {Yankowitz}},\ }\bibfield  {title} {\bibinfo {title} {Competing correlated states and abundant orbital magnetism in twisted monolayer-bilayer graphene},\ }\href {https://doi.org/10.1038/s41467-021-25044-1} {\bibfield  {journal} {\bibinfo  {journal} {Nat. Commun.}\ }\textbf {\bibinfo {volume} {12}},\ \bibinfo {pages} {4727} (\bibinfo {year} {2021})}\BibitemShut {NoStop}%
\bibitem [{\citenamefont {Park}\ \emph {et~al.}(2026)\citenamefont {Park}, \citenamefont {Luo}, \citenamefont {Rademaker}, \citenamefont {Smet}, \citenamefont {Scheurer},\ and\ \citenamefont {Classen}}]{park2026tuning}%
  \BibitemOpen
  \bibfield  {author} {\bibinfo {author} {\bibfnamefont {J.}~\bibnamefont {Park}}, \bibinfo {author} {\bibfnamefont {M.}~\bibnamefont {Luo}}, \bibinfo {author} {\bibfnamefont {L.}~\bibnamefont {Rademaker}}, \bibinfo {author} {\bibfnamefont {J.}~\bibnamefont {Smet}}, \bibinfo {author} {\bibfnamefont {M.~S.}\ \bibnamefont {Scheurer}},\ and\ \bibinfo {author} {\bibfnamefont {L.}~\bibnamefont {Classen}},\ }\bibfield  {title} {\bibinfo {title} {Tuning correlated states of twisted mono-bilayer graphene with proximity-induced spin-orbit coupling},\ }\href {https://doi.org/10.1103/xr1t-173x} {\bibfield  {journal} {\bibinfo  {journal} {Physical Review Research}\ }\textbf {\bibinfo {volume} {8}},\ \bibinfo {pages} {033225} (\bibinfo {year} {2026})}\BibitemShut {NoStop}%
\bibitem [{\citenamefont {Burg}\ \emph {et~al.}(2019)\citenamefont {Burg}, \citenamefont {Zhu}, \citenamefont {Taniguchi}, \citenamefont {Watanabe}, \citenamefont {MacDonald},\ and\ \citenamefont {Tutuc}}]{burg2019correlated}%
  \BibitemOpen
  \bibfield  {author} {\bibinfo {author} {\bibfnamefont {G.~W.}\ \bibnamefont {Burg}}, \bibinfo {author} {\bibfnamefont {J.}~\bibnamefont {Zhu}}, \bibinfo {author} {\bibfnamefont {T.}~\bibnamefont {Taniguchi}}, \bibinfo {author} {\bibfnamefont {K.}~\bibnamefont {Watanabe}}, \bibinfo {author} {\bibfnamefont {A.~H.}\ \bibnamefont {MacDonald}},\ and\ \bibinfo {author} {\bibfnamefont {E.}~\bibnamefont {Tutuc}},\ }\bibfield  {title} {\bibinfo {title} {Correlated insulating states in twisted double bilayer graphene},\ }\bibfield  {journal} {\bibinfo  {journal} {Phys. Rev. Lett.}\ }\textbf {\bibinfo {volume} {123}},\ \href {https://doi.org/10.1103/physrevlett.123.197702} {10.1103/physrevlett.123.197702} (\bibinfo {year} {2019})\BibitemShut {NoStop}%
\bibitem [{\citenamefont {Rickhaus}\ \emph {et~al.}(2019)\citenamefont {Rickhaus}, \citenamefont {Zheng}, \citenamefont {Lado}, \citenamefont {Lee}, \citenamefont {Kurzmann}, \citenamefont {Eich}, \citenamefont {Pisoni}, \citenamefont {Tong}, \citenamefont {Garreis}, \citenamefont {Gold}, \citenamefont {Masseroni}, \citenamefont {Taniguchi}, \citenamefont {Wantanabe}, \citenamefont {Ihn},\ and\ \citenamefont {Ensslin}}]{Rickhaus2019}%
  \BibitemOpen
  \bibfield  {author} {\bibinfo {author} {\bibfnamefont {P.}~\bibnamefont {Rickhaus}}, \bibinfo {author} {\bibfnamefont {G.}~\bibnamefont {Zheng}}, \bibinfo {author} {\bibfnamefont {J.~L.}\ \bibnamefont {Lado}}, \bibinfo {author} {\bibfnamefont {Y.}~\bibnamefont {Lee}}, \bibinfo {author} {\bibfnamefont {A.}~\bibnamefont {Kurzmann}}, \bibinfo {author} {\bibfnamefont {M.}~\bibnamefont {Eich}}, \bibinfo {author} {\bibfnamefont {R.}~\bibnamefont {Pisoni}}, \bibinfo {author} {\bibfnamefont {C.}~\bibnamefont {Tong}}, \bibinfo {author} {\bibfnamefont {R.}~\bibnamefont {Garreis}}, \bibinfo {author} {\bibfnamefont {C.}~\bibnamefont {Gold}}, \bibinfo {author} {\bibfnamefont {M.}~\bibnamefont {Masseroni}}, \bibinfo {author} {\bibfnamefont {T.}~\bibnamefont {Taniguchi}}, \bibinfo {author} {\bibfnamefont {K.}~\bibnamefont {Wantanabe}}, \bibinfo {author} {\bibfnamefont {T.}~\bibnamefont {Ihn}},\ and\ \bibinfo {author} {\bibfnamefont {K.}~\bibnamefont {Ensslin}},\ }\bibfield  {title} {\bibinfo {title} {Gap opening in
  twisted double bilayer graphene by crystal fields},\ }\href {https://doi.org/10.1021/acs.nanolett.9b03660} {\bibfield  {journal} {\bibinfo  {journal} {Nano Letters}\ }\textbf {\bibinfo {volume} {19}},\ \bibinfo {pages} {8821–8828} (\bibinfo {year} {2019})}\BibitemShut {NoStop}%
\bibitem [{\citenamefont {Shen}\ \emph {et~al.}(2020)\citenamefont {Shen}, \citenamefont {Chu}, \citenamefont {Wu}, \citenamefont {Li}, \citenamefont {Wang}, \citenamefont {Zhao}, \citenamefont {Tang}, \citenamefont {Liu}, \citenamefont {Tian}, \citenamefont {Watanabe}, \citenamefont {Taniguchi}, \citenamefont {Yang}, \citenamefont {Meng}, \citenamefont {Shi}, \citenamefont {Yazyev},\ and\ \citenamefont {Zhang}}]{shen2020correlated}%
  \BibitemOpen
  \bibfield  {author} {\bibinfo {author} {\bibfnamefont {C.}~\bibnamefont {Shen}}, \bibinfo {author} {\bibfnamefont {Y.}~\bibnamefont {Chu}}, \bibinfo {author} {\bibfnamefont {Q.}~\bibnamefont {Wu}}, \bibinfo {author} {\bibfnamefont {N.}~\bibnamefont {Li}}, \bibinfo {author} {\bibfnamefont {S.}~\bibnamefont {Wang}}, \bibinfo {author} {\bibfnamefont {Y.}~\bibnamefont {Zhao}}, \bibinfo {author} {\bibfnamefont {J.}~\bibnamefont {Tang}}, \bibinfo {author} {\bibfnamefont {J.}~\bibnamefont {Liu}}, \bibinfo {author} {\bibfnamefont {J.}~\bibnamefont {Tian}}, \bibinfo {author} {\bibfnamefont {K.}~\bibnamefont {Watanabe}}, \bibinfo {author} {\bibfnamefont {T.}~\bibnamefont {Taniguchi}}, \bibinfo {author} {\bibfnamefont {R.}~\bibnamefont {Yang}}, \bibinfo {author} {\bibfnamefont {Z.~Y.}\ \bibnamefont {Meng}}, \bibinfo {author} {\bibfnamefont {D.}~\bibnamefont {Shi}}, \bibinfo {author} {\bibfnamefont {O.~V.}\ \bibnamefont {Yazyev}},\ and\ \bibinfo {author} {\bibfnamefont {G.}~\bibnamefont {Zhang}},\ }\bibfield  {title}
  {\bibinfo {title} {Correlated states in twisted double bilayer graphene},\ }\href {https://doi.org/10.1038/s41567-020-0825-9} {\bibfield  {journal} {\bibinfo  {journal} {Nat. Phys.}\ }\textbf {\bibinfo {volume} {16}},\ \bibinfo {pages} {520} (\bibinfo {year} {2020})}\BibitemShut {NoStop}%
\bibitem [{\citenamefont {Cao}\ \emph {et~al.}(2020)\citenamefont {Cao}, \citenamefont {Rodan-Legrain}, \citenamefont {Rubies-Bigorda}, \citenamefont {Park}, \citenamefont {Watanabe}, \citenamefont {Taniguchi},\ and\ \citenamefont {Jarillo-Herrero}}]{Cao2020SpinPolarized}%
  \BibitemOpen
  \bibfield  {author} {\bibinfo {author} {\bibfnamefont {Y.}~\bibnamefont {Cao}}, \bibinfo {author} {\bibfnamefont {D.}~\bibnamefont {Rodan-Legrain}}, \bibinfo {author} {\bibfnamefont {O.}~\bibnamefont {Rubies-Bigorda}}, \bibinfo {author} {\bibfnamefont {J.~M.}\ \bibnamefont {Park}}, \bibinfo {author} {\bibfnamefont {K.}~\bibnamefont {Watanabe}}, \bibinfo {author} {\bibfnamefont {T.}~\bibnamefont {Taniguchi}},\ and\ \bibinfo {author} {\bibfnamefont {P.}~\bibnamefont {Jarillo-Herrero}},\ }\bibfield  {title} {\bibinfo {title} {Tunable correlated states and spin-polarized phases in twisted bilayer-bilayer graphene},\ }\href {https://doi.org/10.1038/s41586-020-2260-6} {\bibfield  {journal} {\bibinfo  {journal} {Nature}\ }\textbf {\bibinfo {volume} {583}},\ \bibinfo {pages} {215} (\bibinfo {year} {2020})}\BibitemShut {NoStop}%
\bibitem [{\citenamefont {Liu}\ \emph {et~al.}(2020)\citenamefont {Liu}, \citenamefont {Hao}, \citenamefont {Khalaf}, \citenamefont {Lee}, \citenamefont {Ronen}, \citenamefont {Yoo}, \citenamefont {Najafabadi}, \citenamefont {Watanabe}, \citenamefont {Taniguchi}, \citenamefont {Vishwanath},\ and\ \citenamefont {Kim}}]{Liu2020SpinCorrelated}%
  \BibitemOpen
  \bibfield  {author} {\bibinfo {author} {\bibfnamefont {X.}~\bibnamefont {Liu}}, \bibinfo {author} {\bibfnamefont {Z.}~\bibnamefont {Hao}}, \bibinfo {author} {\bibfnamefont {E.}~\bibnamefont {Khalaf}}, \bibinfo {author} {\bibfnamefont {J.~Y.}\ \bibnamefont {Lee}}, \bibinfo {author} {\bibfnamefont {Y.}~\bibnamefont {Ronen}}, \bibinfo {author} {\bibfnamefont {H.}~\bibnamefont {Yoo}}, \bibinfo {author} {\bibfnamefont {D.~H.}\ \bibnamefont {Najafabadi}}, \bibinfo {author} {\bibfnamefont {K.}~\bibnamefont {Watanabe}}, \bibinfo {author} {\bibfnamefont {T.}~\bibnamefont {Taniguchi}}, \bibinfo {author} {\bibfnamefont {A.}~\bibnamefont {Vishwanath}},\ and\ \bibinfo {author} {\bibfnamefont {P.}~\bibnamefont {Kim}},\ }\bibfield  {title} {\bibinfo {title} {Tunable spin-polarized correlated states in twisted double bilayer graphene},\ }\href {https://doi.org/10.1038/s41586-020-2458-7} {\bibfield  {journal} {\bibinfo  {journal} {Nature}\ }\textbf {\bibinfo {volume} {583}},\ \bibinfo {pages} {221} (\bibinfo {year}
  {2020})}\BibitemShut {NoStop}%
\bibitem [{\citenamefont {He}\ \emph {et~al.}(2020)\citenamefont {He}, \citenamefont {Li}, \citenamefont {Cai}, \citenamefont {Liu}, \citenamefont {Watanabe}, \citenamefont {Taniguchi}, \citenamefont {Xu},\ and\ \citenamefont {Yankowitz}}]{he2020symmetry}%
  \BibitemOpen
  \bibfield  {author} {\bibinfo {author} {\bibfnamefont {M.}~\bibnamefont {He}}, \bibinfo {author} {\bibfnamefont {Y.}~\bibnamefont {Li}}, \bibinfo {author} {\bibfnamefont {J.}~\bibnamefont {Cai}}, \bibinfo {author} {\bibfnamefont {Y.}~\bibnamefont {Liu}}, \bibinfo {author} {\bibfnamefont {K.}~\bibnamefont {Watanabe}}, \bibinfo {author} {\bibfnamefont {T.}~\bibnamefont {Taniguchi}}, \bibinfo {author} {\bibfnamefont {X.}~\bibnamefont {Xu}},\ and\ \bibinfo {author} {\bibfnamefont {M.}~\bibnamefont {Yankowitz}},\ }\bibfield  {title} {\bibinfo {title} {Symmetry breaking in twisted double bilayer graphene},\ }\href {https://doi.org/10.1038/s41567-020-1030-6} {\bibfield  {journal} {\bibinfo  {journal} {Nat. Phys.}\ }\textbf {\bibinfo {volume} {17}},\ \bibinfo {pages} {26} (\bibinfo {year} {2020})}\BibitemShut {NoStop}%
\bibitem [{\citenamefont {Kuiri}\ \emph {et~al.}(2022)\citenamefont {Kuiri}, \citenamefont {Coleman}, \citenamefont {Gao}, \citenamefont {Vishnuradhan}, \citenamefont {Watanabe}, \citenamefont {Taniguchi}, \citenamefont {Zhu}, \citenamefont {MacDonald},\ and\ \citenamefont {Folk}}]{Kuiri2022Spontaneous}%
  \BibitemOpen
  \bibfield  {author} {\bibinfo {author} {\bibfnamefont {M.}~\bibnamefont {Kuiri}}, \bibinfo {author} {\bibfnamefont {C.}~\bibnamefont {Coleman}}, \bibinfo {author} {\bibfnamefont {Z.}~\bibnamefont {Gao}}, \bibinfo {author} {\bibfnamefont {A.}~\bibnamefont {Vishnuradhan}}, \bibinfo {author} {\bibfnamefont {K.}~\bibnamefont {Watanabe}}, \bibinfo {author} {\bibfnamefont {T.}~\bibnamefont {Taniguchi}}, \bibinfo {author} {\bibfnamefont {J.}~\bibnamefont {Zhu}}, \bibinfo {author} {\bibfnamefont {A.~H.}\ \bibnamefont {MacDonald}},\ and\ \bibinfo {author} {\bibfnamefont {J.}~\bibnamefont {Folk}},\ }\bibfield  {title} {\bibinfo {title} {Spontaneous time-reversal symmetry breaking in twisted double bilayer graphene},\ }\href {https://doi.org/10.1038/s41467-022-34192-x} {\bibfield  {journal} {\bibinfo  {journal} {Nature Communications}\ }\textbf {\bibinfo {volume} {13}},\ \bibinfo {pages} {6468} (\bibinfo {year} {2022})}\BibitemShut {NoStop}%
\bibitem [{\citenamefont {Goodwin}\ \emph {et~al.}(2021)\citenamefont {Goodwin}, \citenamefont {Klebl}, \citenamefont {Vitale}, \citenamefont {Liang}, \citenamefont {Gogtay}, \citenamefont {van Gorp}, \citenamefont {Kennes}, \citenamefont {Mostofi},\ and\ \citenamefont {Lischner}}]{Goodwin2021_flatbands}%
  \BibitemOpen
  \bibfield  {author} {\bibinfo {author} {\bibfnamefont {Z.~A.~H.}\ \bibnamefont {Goodwin}}, \bibinfo {author} {\bibfnamefont {L.}~\bibnamefont {Klebl}}, \bibinfo {author} {\bibfnamefont {V.}~\bibnamefont {Vitale}}, \bibinfo {author} {\bibfnamefont {X.}~\bibnamefont {Liang}}, \bibinfo {author} {\bibfnamefont {V.}~\bibnamefont {Gogtay}}, \bibinfo {author} {\bibfnamefont {X.}~\bibnamefont {van Gorp}}, \bibinfo {author} {\bibfnamefont {D.~M.}\ \bibnamefont {Kennes}}, \bibinfo {author} {\bibfnamefont {A.~A.}\ \bibnamefont {Mostofi}},\ and\ \bibinfo {author} {\bibfnamefont {J.}~\bibnamefont {Lischner}},\ }\bibfield  {title} {\bibinfo {title} {Flat bands, electron interactions, and magnetic order in magic-angle mono-trilayer graphene},\ }\href {https://doi.org/10.1103/PhysRevMaterials.5.084008} {\bibfield  {journal} {\bibinfo  {journal} {Physical Review Materials}\ }\textbf {\bibinfo {volume} {5}},\ \bibinfo {pages} {084008} (\bibinfo {year} {2021})}\BibitemShut {NoStop}%
\bibitem [{\citenamefont {Khalaf}\ \emph {et~al.}(2019)\citenamefont {Khalaf}, \citenamefont {Kruchkov}, \citenamefont {Tarnopolsky},\ and\ \citenamefont {Vishwanath}}]{Khalaf2019_magic}%
  \BibitemOpen
  \bibfield  {author} {\bibinfo {author} {\bibfnamefont {E.}~\bibnamefont {Khalaf}}, \bibinfo {author} {\bibfnamefont {A.~J.}\ \bibnamefont {Kruchkov}}, \bibinfo {author} {\bibfnamefont {G.}~\bibnamefont {Tarnopolsky}},\ and\ \bibinfo {author} {\bibfnamefont {A.}~\bibnamefont {Vishwanath}},\ }\bibfield  {title} {\bibinfo {title} {Magic angle hierarchy in twisted graphene multilayers},\ }\href {https://doi.org/10.1103/PhysRevB.100.085109} {\bibfield  {journal} {\bibinfo  {journal} {Physical Review B}\ }\textbf {\bibinfo {volume} {100}},\ \bibinfo {pages} {085109} (\bibinfo {year} {2019})}\BibitemShut {NoStop}%
\bibitem [{\citenamefont {Liu}\ \emph {et~al.}(2019)\citenamefont {Liu}, \citenamefont {Ma}, \citenamefont {Gao},\ and\ \citenamefont {Dai}}]{Liu2019}%
  \BibitemOpen
  \bibfield  {author} {\bibinfo {author} {\bibfnamefont {J.}~\bibnamefont {Liu}}, \bibinfo {author} {\bibfnamefont {Z.}~\bibnamefont {Ma}}, \bibinfo {author} {\bibfnamefont {J.}~\bibnamefont {Gao}},\ and\ \bibinfo {author} {\bibfnamefont {X.}~\bibnamefont {Dai}},\ }\bibfield  {title} {\bibinfo {title} {Quantum valley hall effect, orbital magnetism, and anomalous hall effect in twisted multilayer graphene systems},\ }\bibfield  {journal} {\bibinfo  {journal} {Physical Review X}\ }\textbf {\bibinfo {volume} {9}},\ \href {https://doi.org/10.1103/physrevx.9.031021} {10.1103/physrevx.9.031021} (\bibinfo {year} {2019})\BibitemShut {NoStop}%
\bibitem [{\citenamefont {Tarnopolsky}\ \emph {et~al.}(2019)\citenamefont {Tarnopolsky}, \citenamefont {Kruchkov},\ and\ \citenamefont {Vishwanath}}]{Tarnopolsky2019_origin}%
  \BibitemOpen
  \bibfield  {author} {\bibinfo {author} {\bibfnamefont {G.}~\bibnamefont {Tarnopolsky}}, \bibinfo {author} {\bibfnamefont {A.~J.}\ \bibnamefont {Kruchkov}},\ and\ \bibinfo {author} {\bibfnamefont {A.}~\bibnamefont {Vishwanath}},\ }\bibfield  {title} {\bibinfo {title} {Origin of magic angles in twisted bilayer graphene},\ }\bibfield  {journal} {\bibinfo  {journal} {Physical Review Letters}\ }\textbf {\bibinfo {volume} {122}},\ \href {https://doi.org/10.1103/physrevlett.122.106405} {10.1103/physrevlett.122.106405} (\bibinfo {year} {2019})\BibitemShut {NoStop}%
\bibitem [{\citenamefont {Ding}\ \emph {et~al.}(2023)\citenamefont {Ding}, \citenamefont {Liang}, \citenamefont {Ma}, \citenamefont {L\"u},\ and\ \citenamefont {Gao}}]{Ding2023Mirror}%
  \BibitemOpen
  \bibfield  {author} {\bibinfo {author} {\bibfnamefont {S.-P.}\ \bibnamefont {Ding}}, \bibinfo {author} {\bibfnamefont {M.}~\bibnamefont {Liang}}, \bibinfo {author} {\bibfnamefont {Z.}~\bibnamefont {Ma}}, \bibinfo {author} {\bibfnamefont {J.-T.}\ \bibnamefont {L\"u}},\ and\ \bibinfo {author} {\bibfnamefont {J.-H.}\ \bibnamefont {Gao}},\ }\bibfield  {title} {\bibinfo {title} {Mirror symmetry decomposition in double-twisted multilayer graphene systems},\ }\href {https://doi.org/10.1103/PhysRevB.108.195119} {\bibfield  {journal} {\bibinfo  {journal} {Phys. Rev. B}\ }\textbf {\bibinfo {volume} {108}},\ \bibinfo {pages} {195119} (\bibinfo {year} {2023})}\BibitemShut {NoStop}%
\bibitem [{\citenamefont {Xie}\ \emph {et~al.}(2022)\citenamefont {Xie}, \citenamefont {Peng}, \citenamefont {Zhang},\ and\ \citenamefont {Liu}}]{Xie2022_alternating}%
  \BibitemOpen
  \bibfield  {author} {\bibinfo {author} {\bibfnamefont {B.}~\bibnamefont {Xie}}, \bibinfo {author} {\bibfnamefont {R.}~\bibnamefont {Peng}}, \bibinfo {author} {\bibfnamefont {S.}~\bibnamefont {Zhang}},\ and\ \bibinfo {author} {\bibfnamefont {J.}~\bibnamefont {Liu}},\ }\bibfield  {title} {\bibinfo {title} {Alternating twisted multilayer graphene: generic partition rules, double flat bands, and orbital magnetoelectric effect},\ }\href {https://doi.org/10.1038/s41524-022-00789-5} {\bibfield  {journal} {\bibinfo  {journal} {npj Computational Materials}\ }\textbf {\bibinfo {volume} {8}},\ \bibinfo {pages} {110} (\bibinfo {year} {2022})}\BibitemShut {NoStop}%
\bibitem [{\citenamefont {Wang}\ \emph {et~al.}(2026{\natexlab{a}})\citenamefont {Wang}, \citenamefont {Xie}, \citenamefont {Wang}, \citenamefont {Zhang}, \citenamefont {Gao}, \citenamefont {Zhang}, \citenamefont {Yi}, \citenamefont {Xie}, \citenamefont {Shi}, \citenamefont {Zhao}, \citenamefont {Xiong}, \citenamefont {Watanabe}, \citenamefont {Taniguchi}, \citenamefont {Song}, \citenamefont {Lu},\ and\ \citenamefont {Chen}}]{Wang2026emergent}%
  \BibitemOpen
  \bibfield  {author} {\bibinfo {author} {\bibfnamefont {Y.}~\bibnamefont {Wang}}, \bibinfo {author} {\bibfnamefont {J.}~\bibnamefont {Xie}}, \bibinfo {author} {\bibfnamefont {Y.-J.}\ \bibnamefont {Wang}}, \bibinfo {author} {\bibfnamefont {J.}~\bibnamefont {Zhang}}, \bibinfo {author} {\bibfnamefont {Y.}~\bibnamefont {Gao}}, \bibinfo {author} {\bibfnamefont {Z.}~\bibnamefont {Zhang}}, \bibinfo {author} {\bibfnamefont {D.}~\bibnamefont {Yi}}, \bibinfo {author} {\bibfnamefont {Y.}~\bibnamefont {Xie}}, \bibinfo {author} {\bibfnamefont {J.}~\bibnamefont {Shi}}, \bibinfo {author} {\bibfnamefont {G.}~\bibnamefont {Zhao}}, \bibinfo {author} {\bibfnamefont {C.}~\bibnamefont {Xiong}}, \bibinfo {author} {\bibfnamefont {K.}~\bibnamefont {Watanabe}}, \bibinfo {author} {\bibfnamefont {T.}~\bibnamefont {Taniguchi}}, \bibinfo {author} {\bibfnamefont {Z.-D.}\ \bibnamefont {Song}}, \bibinfo {author} {\bibfnamefont {X.}~\bibnamefont {Lu}},\ and\ \bibinfo {author} {\bibfnamefont {Y.}~\bibnamefont {Chen}},\ }\href
  {https://doi.org/10.48550/ARXIV.2608.12478} {\bibinfo {title} {Emergent trans-moiré orbitals and topology in rhombohedral graphene}},\ \bibinfo {howpublished} {arXiv} (\bibinfo {year} {2026}{\natexlab{a}}),\ \bibinfo {note} {(accessed November 24, 2026)},\ \Eprint {https://arxiv.org/abs/2608.12478} {arXiv:2608.12478} \BibitemShut {NoStop}%
\bibitem [{\citenamefont {Wang}\ \emph {et~al.}(2026{\natexlab{b}})\citenamefont {Wang}, \citenamefont {Wang}, \citenamefont {Zhang}, \citenamefont {Huo}, \citenamefont {Zhou}, \citenamefont {Zhang}, \citenamefont {Watanabe}, \citenamefont {Taniguchi}, \citenamefont {Yang}, \citenamefont {Dai}, \citenamefont {Xie}, \citenamefont {Liu}, \citenamefont {Song},\ and\ \citenamefont {Lu}}]{Wang2025_TDTG}%
  \BibitemOpen
  \bibfield  {author} {\bibinfo {author} {\bibfnamefont {W.}~\bibnamefont {Wang}}, \bibinfo {author} {\bibfnamefont {Y.}~\bibnamefont {Wang}}, \bibinfo {author} {\bibfnamefont {Z.}~\bibnamefont {Zhang}}, \bibinfo {author} {\bibfnamefont {Z.}~\bibnamefont {Huo}}, \bibinfo {author} {\bibfnamefont {G.}~\bibnamefont {Zhou}}, \bibinfo {author} {\bibfnamefont {S.}~\bibnamefont {Zhang}}, \bibinfo {author} {\bibfnamefont {K.}~\bibnamefont {Watanabe}}, \bibinfo {author} {\bibfnamefont {T.}~\bibnamefont {Taniguchi}}, \bibinfo {author} {\bibfnamefont {X.}~\bibnamefont {Yang}}, \bibinfo {author} {\bibfnamefont {Q.}~\bibnamefont {Dai}}, \bibinfo {author} {\bibfnamefont {X.~C.}\ \bibnamefont {Xie}}, \bibinfo {author} {\bibfnamefont {K.}~\bibnamefont {Liu}}, \bibinfo {author} {\bibfnamefont {Z.}~\bibnamefont {Song}},\ and\ \bibinfo {author} {\bibfnamefont {X.}~\bibnamefont {Lu}},\ }\bibfield  {title} {\bibinfo {title} {Programmable quantum anomalous hall insulator in twisted crystalline flatbands},\ }\href
  {https://doi.org/10.1103/7m1b-hxy8} {\bibfield  {journal} {\bibinfo  {journal} {Physical Review X}\ }\textbf {\bibinfo {volume} {16}},\ \bibinfo {pages} {011015} (\bibinfo {year} {2026}{\natexlab{b}})}\BibitemShut {NoStop}%
\bibitem [{\citenamefont {Liu}\ \emph {et~al.}(2025)\citenamefont {Liu}, \citenamefont {Chen}, \citenamefont {Ding}, \citenamefont {Zhou}, \citenamefont {Xiang}, \citenamefont {Fang}, \citenamefont {Wu}, \citenamefont {Zhan}, \citenamefont {Zhang}, \citenamefont {Chen}, \citenamefont {Watanabe}, \citenamefont {Taniguchi}, \citenamefont {Xin},\ and\ \citenamefont {Xu}}]{Liu2025_TwistMonoPenta}%
  \BibitemOpen
  \bibfield  {author} {\bibinfo {author} {\bibfnamefont {N.}~\bibnamefont {Liu}}, \bibinfo {author} {\bibfnamefont {Z.}~\bibnamefont {Chen}}, \bibinfo {author} {\bibfnamefont {J.}~\bibnamefont {Ding}}, \bibinfo {author} {\bibfnamefont {W.}~\bibnamefont {Zhou}}, \bibinfo {author} {\bibfnamefont {H.}~\bibnamefont {Xiang}}, \bibinfo {author} {\bibfnamefont {X.}~\bibnamefont {Fang}}, \bibinfo {author} {\bibfnamefont {L.}~\bibnamefont {Wu}}, \bibinfo {author} {\bibfnamefont {X.}~\bibnamefont {Zhan}}, \bibinfo {author} {\bibfnamefont {L.}~\bibnamefont {Zhang}}, \bibinfo {author} {\bibfnamefont {Q.}~\bibnamefont {Chen}}, \bibinfo {author} {\bibfnamefont {K.}~\bibnamefont {Watanabe}}, \bibinfo {author} {\bibfnamefont {T.}~\bibnamefont {Taniguchi}}, \bibinfo {author} {\bibfnamefont {N.}~\bibnamefont {Xin}},\ and\ \bibinfo {author} {\bibfnamefont {S.}~\bibnamefont {Xu}},\ }\href {https://doi.org/10.48550/arxiv.2507.11347} {\bibinfo {title} {Diverse high-chern-number quantum anomalous hall insulators in twisted
  rhombohedral graphene}},\ \bibinfo {howpublished} {arXiv} (\bibinfo {year} {2025}),\ \bibinfo {note} {(accessed November 24, 2026)},\ \Eprint {https://arxiv.org/abs/2507.11347} {arXiv:2507.11347} \BibitemShut {NoStop}%
\bibitem [{\citenamefont {Wang}\ \emph {et~al.}(2026{\natexlab{c}})\citenamefont {Wang}, \citenamefont {Benitez}, \citenamefont {Rao}, \citenamefont {Phong}, \citenamefont {Chu}, \citenamefont {Watanabe}, \citenamefont {Taniguchi}, \citenamefont {Lewandowski},\ and\ \citenamefont {Jarillo-Herrero}}]{Wang2026High}%
  \BibitemOpen
  \bibfield  {author} {\bibinfo {author} {\bibfnamefont {X.}~\bibnamefont {Wang}}, \bibinfo {author} {\bibfnamefont {L.~A.}\ \bibnamefont {Benitez}}, \bibinfo {author} {\bibfnamefont {S.}~\bibnamefont {Rao}}, \bibinfo {author} {\bibfnamefont {V.~T.}\ \bibnamefont {Phong}}, \bibinfo {author} {\bibfnamefont {W.~I.}\ \bibnamefont {Chu}}, \bibinfo {author} {\bibfnamefont {K.}~\bibnamefont {Watanabe}}, \bibinfo {author} {\bibfnamefont {T.}~\bibnamefont {Taniguchi}}, \bibinfo {author} {\bibfnamefont {C.}~\bibnamefont {Lewandowski}},\ and\ \bibinfo {author} {\bibfnamefont {P.}~\bibnamefont {Jarillo-Herrero}},\ }\bibfield  {title} {\bibinfo {title} {High-chern-number orbital magnetism in twisted rhombohedral graphene},\ }\bibfield  {journal} {\bibinfo  {journal} {Nature Materials}\ }\href {https://doi.org/10.1038/s41563-026-02659-7} {10.1038/s41563-026-02659-7} (\bibinfo {year} {2026}{\natexlab{c}})\BibitemShut {NoStop}%
\bibitem [{\citenamefont {Su}\ \emph {et~al.}(2025)\citenamefont {Su}, \citenamefont {Waters}, \citenamefont {Zhou}, \citenamefont {Watanabe}, \citenamefont {Taniguchi}, \citenamefont {Zhang}, \citenamefont {Yankowitz},\ and\ \citenamefont {Folk}}]{Su2025}%
  \BibitemOpen
  \bibfield  {author} {\bibinfo {author} {\bibfnamefont {R.}~\bibnamefont {Su}}, \bibinfo {author} {\bibfnamefont {D.}~\bibnamefont {Waters}}, \bibinfo {author} {\bibfnamefont {B.}~\bibnamefont {Zhou}}, \bibinfo {author} {\bibfnamefont {K.}~\bibnamefont {Watanabe}}, \bibinfo {author} {\bibfnamefont {T.}~\bibnamefont {Taniguchi}}, \bibinfo {author} {\bibfnamefont {Y.-H.}\ \bibnamefont {Zhang}}, \bibinfo {author} {\bibfnamefont {M.}~\bibnamefont {Yankowitz}},\ and\ \bibinfo {author} {\bibfnamefont {J.}~\bibnamefont {Folk}},\ }\bibfield  {title} {\bibinfo {title} {Moiré-driven topological electronic crystals in twisted graphene},\ }\href {https://doi.org/10.1038/s41586-024-08239-6} {\bibfield  {journal} {\bibinfo  {journal} {Nature}\ }\textbf {\bibinfo {volume} {637}},\ \bibinfo {pages} {1084–1089} (\bibinfo {year} {2025})}\BibitemShut {NoStop}%
\bibitem [{\citenamefont {Dong}\ \emph {et~al.}(2025)\citenamefont {Dong}, \citenamefont {Liu}, \citenamefont {Zhu}, \citenamefont {Pan}, \citenamefont {Hong}, \citenamefont {Wang}, \citenamefont {Ren}, \citenamefont {Jia}, \citenamefont {Watanabe}, \citenamefont {Taniguchi}, \citenamefont {Du}, \citenamefont {Shi}, \citenamefont {Yang},\ and\ \citenamefont {Zhang}}]{Dong2025_TwistBiTri}%
  \BibitemOpen
  \bibfield  {author} {\bibinfo {author} {\bibfnamefont {J.}~\bibnamefont {Dong}}, \bibinfo {author} {\bibfnamefont {L.}~\bibnamefont {Liu}}, \bibinfo {author} {\bibfnamefont {J.}~\bibnamefont {Zhu}}, \bibinfo {author} {\bibfnamefont {Z.}~\bibnamefont {Pan}}, \bibinfo {author} {\bibfnamefont {Y.}~\bibnamefont {Hong}}, \bibinfo {author} {\bibfnamefont {F.}~\bibnamefont {Wang}}, \bibinfo {author} {\bibfnamefont {Z.}~\bibnamefont {Ren}}, \bibinfo {author} {\bibfnamefont {Z.}~\bibnamefont {Jia}}, \bibinfo {author} {\bibfnamefont {K.}~\bibnamefont {Watanabe}}, \bibinfo {author} {\bibfnamefont {T.}~\bibnamefont {Taniguchi}}, \bibinfo {author} {\bibfnamefont {L.}~\bibnamefont {Du}}, \bibinfo {author} {\bibfnamefont {D.}~\bibnamefont {Shi}}, \bibinfo {author} {\bibfnamefont {W.}~\bibnamefont {Yang}},\ and\ \bibinfo {author} {\bibfnamefont {G.}~\bibnamefont {Zhang}},\ }\href {https://doi.org/10.48550/arxiv.2507.09908} {\bibinfo {title} {Observation of integer and fractional chern insulators in high chern number
  flatbands}},\ \bibinfo {howpublished} {arXiv} (\bibinfo {year} {2025}),\ \bibinfo {note} {(accessed November 24, 2026)},\ \Eprint {https://arxiv.org/abs/2507.09908} {arXiv:2507.09908} \BibitemShut {NoStop}%
\bibitem [{\citenamefont {Waters}\ \emph {et~al.}(2024)\citenamefont {Waters}, \citenamefont {Su}, \citenamefont {Thompson}, \citenamefont {Okounkova}, \citenamefont {Arreguin-Martinez}, \citenamefont {He}, \citenamefont {Hinds}, \citenamefont {Watanabe}, \citenamefont {Taniguchi}, \citenamefont {Xu}, \citenamefont {Zhang}, \citenamefont {Folk},\ and\ \citenamefont {Yankowitz}}]{Waters2024}%
  \BibitemOpen
  \bibfield  {author} {\bibinfo {author} {\bibfnamefont {D.}~\bibnamefont {Waters}}, \bibinfo {author} {\bibfnamefont {R.}~\bibnamefont {Su}}, \bibinfo {author} {\bibfnamefont {E.}~\bibnamefont {Thompson}}, \bibinfo {author} {\bibfnamefont {A.}~\bibnamefont {Okounkova}}, \bibinfo {author} {\bibfnamefont {E.}~\bibnamefont {Arreguin-Martinez}}, \bibinfo {author} {\bibfnamefont {M.}~\bibnamefont {He}}, \bibinfo {author} {\bibfnamefont {K.}~\bibnamefont {Hinds}}, \bibinfo {author} {\bibfnamefont {K.}~\bibnamefont {Watanabe}}, \bibinfo {author} {\bibfnamefont {T.}~\bibnamefont {Taniguchi}}, \bibinfo {author} {\bibfnamefont {X.}~\bibnamefont {Xu}}, \bibinfo {author} {\bibfnamefont {Y.-H.}\ \bibnamefont {Zhang}}, \bibinfo {author} {\bibfnamefont {J.}~\bibnamefont {Folk}},\ and\ \bibinfo {author} {\bibfnamefont {M.}~\bibnamefont {Yankowitz}},\ }\bibfield  {title} {\bibinfo {title} {Topological flat bands in a family of multilayer graphene moiré lattices},\ }\bibfield  {journal} {\bibinfo  {journal} {Nature
  Communications}\ }\textbf {\bibinfo {volume} {15}},\ \href {https://doi.org/10.1038/s41467-024-55001-7} {10.1038/s41467-024-55001-7} (\bibinfo {year} {2024})\BibitemShut {NoStop}%
\bibitem [{\citenamefont {Chen}\ \emph {et~al.}(2026)\citenamefont {Chen}, \citenamefont {Liu}, \citenamefont {Hua}, \citenamefont {Xiang}, \citenamefont {Zhou}, \citenamefont {Ding}, \citenamefont {Fang}, \citenamefont {Wu}, \citenamefont {Zhang}, \citenamefont {Chen}, \citenamefont {Chen}, \citenamefont {Watanabe}, \citenamefont {Taniguchi}, \citenamefont {Xin}, \citenamefont {Zhu},\ and\ \citenamefont {Xu}}]{chen2026layer}%
  \BibitemOpen
  \bibfield  {author} {\bibinfo {author} {\bibfnamefont {Z.}~\bibnamefont {Chen}}, \bibinfo {author} {\bibfnamefont {N.}~\bibnamefont {Liu}}, \bibinfo {author} {\bibfnamefont {J.}~\bibnamefont {Hua}}, \bibinfo {author} {\bibfnamefont {H.}~\bibnamefont {Xiang}}, \bibinfo {author} {\bibfnamefont {W.}~\bibnamefont {Zhou}}, \bibinfo {author} {\bibfnamefont {J.}~\bibnamefont {Ding}}, \bibinfo {author} {\bibfnamefont {X.}~\bibnamefont {Fang}}, \bibinfo {author} {\bibfnamefont {L.}~\bibnamefont {Wu}}, \bibinfo {author} {\bibfnamefont {L.}~\bibnamefont {Zhang}}, \bibinfo {author} {\bibfnamefont {Q.}~\bibnamefont {Chen}}, \bibinfo {author} {\bibfnamefont {X.}~\bibnamefont {Chen}}, \bibinfo {author} {\bibfnamefont {K.}~\bibnamefont {Watanabe}}, \bibinfo {author} {\bibfnamefont {T.}~\bibnamefont {Taniguchi}}, \bibinfo {author} {\bibfnamefont {N.}~\bibnamefont {Xin}}, \bibinfo {author} {\bibfnamefont {W.}~\bibnamefont {Zhu}},\ and\ \bibinfo {author} {\bibfnamefont {S.}~\bibnamefont {Xu}},\ }\bibfield  {title} {\bibinfo
  {title} {Layer-engineered quantum anomalous hall effect in twisted rhombohedral graphene},\ }\bibfield  {journal} {\bibinfo  {journal} {Nature Materials}\ }\href {https://doi.org/10.1038/s41563-026-02711-6} {10.1038/s41563-026-02711-6} (\bibinfo {year} {2026})\BibitemShut {NoStop}%
\bibitem [{\citenamefont {Li}\ \emph {et~al.}(2026)\citenamefont {Li}, \citenamefont {Wang}, \citenamefont {Wang}, \citenamefont {Zhang}, \citenamefont {Yang}, \citenamefont {Watanabe}, \citenamefont {Taniguchi}, \citenamefont {Xie}, \citenamefont {Wang}, \citenamefont {Liu}, \citenamefont {Song},\ and\ \citenamefont {Lu}}]{Li2026Fractional}%
  \BibitemOpen
  \bibfield  {author} {\bibinfo {author} {\bibfnamefont {Z.}~\bibnamefont {Li}}, \bibinfo {author} {\bibfnamefont {W.}~\bibnamefont {Wang}}, \bibinfo {author} {\bibfnamefont {F.}~\bibnamefont {Wang}}, \bibinfo {author} {\bibfnamefont {Z.}~\bibnamefont {Zhang}}, \bibinfo {author} {\bibfnamefont {Q.}~\bibnamefont {Yang}}, \bibinfo {author} {\bibfnamefont {K.}~\bibnamefont {Watanabe}}, \bibinfo {author} {\bibfnamefont {T.}~\bibnamefont {Taniguchi}}, \bibinfo {author} {\bibfnamefont {X.~C.}\ \bibnamefont {Xie}}, \bibinfo {author} {\bibfnamefont {J.}~\bibnamefont {Wang}}, \bibinfo {author} {\bibfnamefont {K.}~\bibnamefont {Liu}}, \bibinfo {author} {\bibfnamefont {Z.}~\bibnamefont {Song}},\ and\ \bibinfo {author} {\bibfnamefont {X.}~\bibnamefont {Lu}},\ }\bibfield  {title} {\bibinfo {title} {Fractional high-chern insulator in twisted rhombohedral graphene},\ }\href {https://doi.org/10.1038/s41586-026-10762-7} {\bibfield  {journal} {\bibinfo  {journal} {Nature}\ }\textbf {\bibinfo {volume} {655}},\ \bibinfo {pages}
  {892} (\bibinfo {year} {2026})}\BibitemShut {NoStop}%
\bibitem [{\citenamefont {Zhang}\ \emph {et~al.}(2024)\citenamefont {Zhang}, \citenamefont {Li}, \citenamefont {Park}, \citenamefont {Jia}, \citenamefont {Chen}, \citenamefont {Li}, \citenamefont {Liu}, \citenamefont {Bao}, \citenamefont {Leconte}, \citenamefont {Zhou}, \citenamefont {Wang}, \citenamefont {Watanabe}, \citenamefont {Taniguchi}, \citenamefont {Avila}, \citenamefont {Dudin}, \citenamefont {Yu}, \citenamefont {Weng}, \citenamefont {Duan}, \citenamefont {Wu}, \citenamefont {Jung},\ and\ \citenamefont {Zhou}}]{zhang2024observation}%
  \BibitemOpen
  \bibfield  {author} {\bibinfo {author} {\bibfnamefont {H.}~\bibnamefont {Zhang}}, \bibinfo {author} {\bibfnamefont {Q.}~\bibnamefont {Li}}, \bibinfo {author} {\bibfnamefont {Y.}~\bibnamefont {Park}}, \bibinfo {author} {\bibfnamefont {Y.}~\bibnamefont {Jia}}, \bibinfo {author} {\bibfnamefont {W.}~\bibnamefont {Chen}}, \bibinfo {author} {\bibfnamefont {J.}~\bibnamefont {Li}}, \bibinfo {author} {\bibfnamefont {Q.}~\bibnamefont {Liu}}, \bibinfo {author} {\bibfnamefont {C.}~\bibnamefont {Bao}}, \bibinfo {author} {\bibfnamefont {N.}~\bibnamefont {Leconte}}, \bibinfo {author} {\bibfnamefont {S.}~\bibnamefont {Zhou}}, \bibinfo {author} {\bibfnamefont {Y.}~\bibnamefont {Wang}}, \bibinfo {author} {\bibfnamefont {K.}~\bibnamefont {Watanabe}}, \bibinfo {author} {\bibfnamefont {T.}~\bibnamefont {Taniguchi}}, \bibinfo {author} {\bibfnamefont {J.}~\bibnamefont {Avila}}, \bibinfo {author} {\bibfnamefont {P.}~\bibnamefont {Dudin}}, \bibinfo {author} {\bibfnamefont {P.}~\bibnamefont {Yu}}, \bibinfo {author} {\bibfnamefont
  {H.}~\bibnamefont {Weng}}, \bibinfo {author} {\bibfnamefont {W.}~\bibnamefont {Duan}}, \bibinfo {author} {\bibfnamefont {Q.}~\bibnamefont {Wu}}, \bibinfo {author} {\bibfnamefont {J.}~\bibnamefont {Jung}},\ and\ \bibinfo {author} {\bibfnamefont {S.}~\bibnamefont {Zhou}},\ }\bibfield  {title} {\bibinfo {title} {Observation of dichotomic field-tunable electronic structure in twisted monolayer-bilayer graphene},\ }\bibfield  {journal} {\bibinfo  {journal} {Nature Communications}\ }\textbf {\bibinfo {volume} {15}},\ \href {https://doi.org/10.1038/s41467-024-48166-8} {10.1038/s41467-024-48166-8} (\bibinfo {year} {2024})\BibitemShut {NoStop}%
\bibitem [{\citenamefont {Zhang}\ \emph {et~al.}(2022{\natexlab{a}})\citenamefont {Zhang}, \citenamefont {Xu}, \citenamefont {Hou}, \citenamefont {Song}, \citenamefont {Ma}, \citenamefont {Gao}, \citenamefont {Zhu}, \citenamefont {Ma}, \citenamefont {Liu}, \citenamefont {Feng},\ and\ \citenamefont {Li}}]{zhang2022domino}%
  \BibitemOpen
  \bibfield  {author} {\bibinfo {author} {\bibfnamefont {S.}~\bibnamefont {Zhang}}, \bibinfo {author} {\bibfnamefont {Q.}~\bibnamefont {Xu}}, \bibinfo {author} {\bibfnamefont {Y.}~\bibnamefont {Hou}}, \bibinfo {author} {\bibfnamefont {A.}~\bibnamefont {Song}}, \bibinfo {author} {\bibfnamefont {Y.}~\bibnamefont {Ma}}, \bibinfo {author} {\bibfnamefont {L.}~\bibnamefont {Gao}}, \bibinfo {author} {\bibfnamefont {M.}~\bibnamefont {Zhu}}, \bibinfo {author} {\bibfnamefont {T.}~\bibnamefont {Ma}}, \bibinfo {author} {\bibfnamefont {L.}~\bibnamefont {Liu}}, \bibinfo {author} {\bibfnamefont {X.-Q.}\ \bibnamefont {Feng}},\ and\ \bibinfo {author} {\bibfnamefont {Q.}~\bibnamefont {Li}},\ }\bibfield  {title} {\bibinfo {title} {Domino-like stacking order switching in twisted monolayer–multilayer graphene},\ }\href {https://doi.org/10.1038/s41563-022-01232-2} {\bibfield  {journal} {\bibinfo  {journal} {Nat. Mater.}\ }\textbf {\bibinfo {volume} {21}},\ \bibinfo {pages} {621} (\bibinfo {year}
  {2022}{\natexlab{a}})}\BibitemShut {NoStop}%
\bibitem [{\citenamefont {Ding}\ \emph {et~al.}(2024)\citenamefont {Ding}, \citenamefont {Yan}, \citenamefont {Wang}, \citenamefont {Han}, \citenamefont {Yang}, \citenamefont {Chen}, \citenamefont {Zhang}, \citenamefont {Huang}, \citenamefont {Zhao}, \citenamefont {Yang}, \citenamefont {Xue}, \citenamefont {Liu}, \citenamefont {Dai}, \citenamefont {Hou}, \citenamefont {Zhang}, \citenamefont {Xu}, \citenamefont {Wang},\ and\ \citenamefont {Huang}}]{ding2024manipulation}%
  \BibitemOpen
  \bibfield  {author} {\bibinfo {author} {\bibfnamefont {P.}~\bibnamefont {Ding}}, \bibinfo {author} {\bibfnamefont {J.}~\bibnamefont {Yan}}, \bibinfo {author} {\bibfnamefont {J.}~\bibnamefont {Wang}}, \bibinfo {author} {\bibfnamefont {X.}~\bibnamefont {Han}}, \bibinfo {author} {\bibfnamefont {W.}~\bibnamefont {Yang}}, \bibinfo {author} {\bibfnamefont {H.}~\bibnamefont {Chen}}, \bibinfo {author} {\bibfnamefont {D.}~\bibnamefont {Zhang}}, \bibinfo {author} {\bibfnamefont {M.}~\bibnamefont {Huang}}, \bibinfo {author} {\bibfnamefont {J.}~\bibnamefont {Zhao}}, \bibinfo {author} {\bibfnamefont {S.}~\bibnamefont {Yang}}, \bibinfo {author} {\bibfnamefont {T.-t.}\ \bibnamefont {Xue}}, \bibinfo {author} {\bibfnamefont {L.}~\bibnamefont {Liu}}, \bibinfo {author} {\bibfnamefont {Y.}~\bibnamefont {Dai}}, \bibinfo {author} {\bibfnamefont {Y.}~\bibnamefont {Hou}}, \bibinfo {author} {\bibfnamefont {S.}~\bibnamefont {Zhang}}, \bibinfo {author} {\bibfnamefont {X.}~\bibnamefont {Xu}}, \bibinfo {author} {\bibfnamefont
  {Y.}~\bibnamefont {Wang}},\ and\ \bibinfo {author} {\bibfnamefont {Y.}~\bibnamefont {Huang}},\ }\bibfield  {title} {\bibinfo {title} {Manipulation of moiré superlattice in twisted monolayer-multilayer graphene by moving nanobubbles},\ }\href {https://doi.org/10.1021/acs.nanolett.4c02548} {\bibfield  {journal} {\bibinfo  {journal} {Nano Lett.}\ }\textbf {\bibinfo {volume} {24}},\ \bibinfo {pages} {8208} (\bibinfo {year} {2024})}\BibitemShut {NoStop}%
\bibitem [{\citenamefont {Min}\ and\ \citenamefont {MacDonald}(2008)}]{Min2008Chiral}%
  \BibitemOpen
  \bibfield  {author} {\bibinfo {author} {\bibfnamefont {H.}~\bibnamefont {Min}}\ and\ \bibinfo {author} {\bibfnamefont {A.~H.}\ \bibnamefont {MacDonald}},\ }\bibfield  {title} {\bibinfo {title} {Chiral decomposition in the electronic structure of graphene multilayers},\ }\href {https://doi.org/10.1103/PhysRevB.77.155416} {\bibfield  {journal} {\bibinfo  {journal} {Phys. Rev. B}\ }\textbf {\bibinfo {volume} {77}},\ \bibinfo {pages} {155416} (\bibinfo {year} {2008})}\BibitemShut {NoStop}%
\bibitem [{\citenamefont {Koshino}\ and\ \citenamefont {McCann}(2009)}]{Koshino2009Trigonal}%
  \BibitemOpen
  \bibfield  {author} {\bibinfo {author} {\bibfnamefont {M.}~\bibnamefont {Koshino}}\ and\ \bibinfo {author} {\bibfnamefont {E.}~\bibnamefont {McCann}},\ }\bibfield  {title} {\bibinfo {title} {Trigonal warping and berry's phase $n\ensuremath{\pi}$ in abc-stacked multilayer graphene},\ }\href {https://doi.org/10.1103/PhysRevB.80.165409} {\bibfield  {journal} {\bibinfo  {journal} {Phys. Rev. B}\ }\textbf {\bibinfo {volume} {80}},\ \bibinfo {pages} {165409} (\bibinfo {year} {2009})}\BibitemShut {NoStop}%
\bibitem [{\citenamefont {Makov}\ \emph {et~al.}(2024)\citenamefont {Makov}, \citenamefont {Guinea},\ and\ \citenamefont {Stern}}]{Makov2024Flat}%
  \BibitemOpen
  \bibfield  {author} {\bibinfo {author} {\bibfnamefont {R.}~\bibnamefont {Makov}}, \bibinfo {author} {\bibfnamefont {F.}~\bibnamefont {Guinea}},\ and\ \bibinfo {author} {\bibfnamefont {A.}~\bibnamefont {Stern}},\ }\bibfield  {title} {\bibinfo {title} {Flat bands in chiral multilayer graphene},\ }\href {https://doi.org/10.1103/PhysRevB.110.195112} {\bibfield  {journal} {\bibinfo  {journal} {Phys. Rev. B}\ }\textbf {\bibinfo {volume} {110}},\ \bibinfo {pages} {195112} (\bibinfo {year} {2024})}\BibitemShut {NoStop}%
\bibitem [{\citenamefont {Zhang}\ \emph {et~al.}(2025{\natexlab{a}})\citenamefont {Zhang}, \citenamefont {Zhou}, \citenamefont {Zhang}, \citenamefont {Cai}, \citenamefont {Tong}, \citenamefont {Liao}, \citenamefont {Zou}, \citenamefont {Xue}, \citenamefont {Tian}, \citenamefont {Chen}, \citenamefont {Tian}, \citenamefont {Zhang}, \citenamefont {Wang}, \citenamefont {Zou}, \citenamefont {Liu}, \citenamefont {Hu}, \citenamefont {Ren}, \citenamefont {Zhang}, \citenamefont {Zhang}, \citenamefont {Wang}, \citenamefont {He}, \citenamefont {Liao}, \citenamefont {Qin},\ and\ \citenamefont {Yin}}]{Zhang2024Layer}%
  \BibitemOpen
  \bibfield  {author} {\bibinfo {author} {\bibfnamefont {Y.}~\bibnamefont {Zhang}}, \bibinfo {author} {\bibfnamefont {Y.-Y.}\ \bibnamefont {Zhou}}, \bibinfo {author} {\bibfnamefont {S.}~\bibnamefont {Zhang}}, \bibinfo {author} {\bibfnamefont {H.}~\bibnamefont {Cai}}, \bibinfo {author} {\bibfnamefont {L.-H.}\ \bibnamefont {Tong}}, \bibinfo {author} {\bibfnamefont {W.-Y.}\ \bibnamefont {Liao}}, \bibinfo {author} {\bibfnamefont {R.-J.}\ \bibnamefont {Zou}}, \bibinfo {author} {\bibfnamefont {S.-M.}\ \bibnamefont {Xue}}, \bibinfo {author} {\bibfnamefont {Y.}~\bibnamefont {Tian}}, \bibinfo {author} {\bibfnamefont {T.}~\bibnamefont {Chen}}, \bibinfo {author} {\bibfnamefont {Q.}~\bibnamefont {Tian}}, \bibinfo {author} {\bibfnamefont {C.}~\bibnamefont {Zhang}}, \bibinfo {author} {\bibfnamefont {Y.}~\bibnamefont {Wang}}, \bibinfo {author} {\bibfnamefont {X.}~\bibnamefont {Zou}}, \bibinfo {author} {\bibfnamefont {X.}~\bibnamefont {Liu}}, \bibinfo {author} {\bibfnamefont {Y.}~\bibnamefont {Hu}}, \bibinfo {author}
  {\bibfnamefont {Y.-N.}\ \bibnamefont {Ren}}, \bibinfo {author} {\bibfnamefont {L.}~\bibnamefont {Zhang}}, \bibinfo {author} {\bibfnamefont {L.}~\bibnamefont {Zhang}}, \bibinfo {author} {\bibfnamefont {W.-X.}\ \bibnamefont {Wang}}, \bibinfo {author} {\bibfnamefont {L.}~\bibnamefont {He}}, \bibinfo {author} {\bibfnamefont {L.}~\bibnamefont {Liao}}, \bibinfo {author} {\bibfnamefont {Z.}~\bibnamefont {Qin}},\ and\ \bibinfo {author} {\bibfnamefont {L.-J.}\ \bibnamefont {Yin}},\ }\bibfield  {title} {\bibinfo {title} {Layer-dependent evolution of electronic structures and correlations in rhombohedral multilayer graphene},\ }\href {https://doi.org/10.1038/s41565-024-01822-y} {\bibfield  {journal} {\bibinfo  {journal} {Nature Nanotechnology}\ }\textbf {\bibinfo {volume} {20}},\ \bibinfo {pages} {222–228} (\bibinfo {year} {2025}{\natexlab{a}})}\BibitemShut {NoStop}%
\bibitem [{\citenamefont {Dutreix}\ and\ \citenamefont {Katsnelson}(2016)}]{Dutreix2016Friedel}%
  \BibitemOpen
  \bibfield  {author} {\bibinfo {author} {\bibfnamefont {C.}~\bibnamefont {Dutreix}}\ and\ \bibinfo {author} {\bibfnamefont {M.~I.}\ \bibnamefont {Katsnelson}},\ }\bibfield  {title} {\bibinfo {title} {Friedel oscillations at the surfaces of rhombohedral {$N$}-layer graphene},\ }\href {https://doi.org/10.1103/PhysRevB.93.035413} {\bibfield  {journal} {\bibinfo  {journal} {Physical Review B}\ }\textbf {\bibinfo {volume} {93}},\ \bibinfo {pages} {035413} (\bibinfo {year} {2016})}\BibitemShut {NoStop}%
\bibitem [{\citenamefont {Guinea}(2007)}]{Guinea2007Charge}%
  \BibitemOpen
  \bibfield  {author} {\bibinfo {author} {\bibfnamefont {F.}~\bibnamefont {Guinea}},\ }\bibfield  {title} {\bibinfo {title} {Charge distribution and screening in layered graphene systems},\ }\href {https://doi.org/10.1103/PhysRevB.75.235433} {\bibfield  {journal} {\bibinfo  {journal} {Phys. Rev. B}\ }\textbf {\bibinfo {volume} {75}},\ \bibinfo {pages} {235433} (\bibinfo {year} {2007})}\BibitemShut {NoStop}%
\bibitem [{\citenamefont {Koshino}(2010)}]{Koshino2010Interlayer}%
  \BibitemOpen
  \bibfield  {author} {\bibinfo {author} {\bibfnamefont {M.}~\bibnamefont {Koshino}},\ }\bibfield  {title} {\bibinfo {title} {Interlayer screening effect in graphene multilayers with $aba$ and $abc$ stacking},\ }\href {https://doi.org/10.1103/PhysRevB.81.125304} {\bibfield  {journal} {\bibinfo  {journal} {Phys. Rev. B}\ }\textbf {\bibinfo {volume} {81}},\ \bibinfo {pages} {125304} (\bibinfo {year} {2010})}\BibitemShut {NoStop}%
\bibitem [{\citenamefont {Long}\ \emph {et~al.}(2024)\citenamefont {Long}, \citenamefont {Jimeno-Pozo}, \citenamefont {Sainz-Cruz}, \citenamefont {Pantaleón},\ and\ \citenamefont {Guinea}}]{Long2024Evolution}%
  \BibitemOpen
  \bibfield  {author} {\bibinfo {author} {\bibfnamefont {M.}~\bibnamefont {Long}}, \bibinfo {author} {\bibfnamefont {A.}~\bibnamefont {Jimeno-Pozo}}, \bibinfo {author} {\bibfnamefont {H.}~\bibnamefont {Sainz-Cruz}}, \bibinfo {author} {\bibfnamefont {P.~A.}\ \bibnamefont {Pantaleón}},\ and\ \bibinfo {author} {\bibfnamefont {F.}~\bibnamefont {Guinea}},\ }\bibfield  {title} {\bibinfo {title} {Evolution of superconductivity in twisted graphene multilayers},\ }\bibfield  {journal} {\bibinfo  {journal} {Proceedings of the National Academy of Sciences}\ }\textbf {\bibinfo {volume} {121}},\ \href {https://doi.org/10.1073/pnas.2405259121} {10.1073/pnas.2405259121} (\bibinfo {year} {2024})\BibitemShut {NoStop}%
\bibitem [{\citenamefont {Pantaleon}\ \emph {et~al.}(2026)\citenamefont {Pantaleon}, \citenamefont {Jimeno-Pozo},\ and\ \citenamefont {Guinea}}]{Pantaleon2026Interface}%
  \BibitemOpen
  \bibfield  {author} {\bibinfo {author} {\bibfnamefont {P.~A.}\ \bibnamefont {Pantaleon}}, \bibinfo {author} {\bibfnamefont {A.}~\bibnamefont {Jimeno-Pozo}},\ and\ \bibinfo {author} {\bibfnamefont {F.}~\bibnamefont {Guinea}},\ }\href {https://doi.org/10.48550/arXiv.2611.15192} {\bibinfo {title} {Interface-driven remote-surface states in rhombohedral graphene}} (\bibinfo {year} {2026}),\ \bibinfo {note} {(accessed November 24, 2026)}\BibitemShut {NoStop}%
\bibitem [{\citenamefont {Ledwith}\ \emph {et~al.}(2022)\citenamefont {Ledwith}, \citenamefont {Vishwanath},\ and\ \citenamefont {Khalaf}}]{Ledwith2022Family}%
  \BibitemOpen
  \bibfield  {author} {\bibinfo {author} {\bibfnamefont {P.~J.}\ \bibnamefont {Ledwith}}, \bibinfo {author} {\bibfnamefont {A.}~\bibnamefont {Vishwanath}},\ and\ \bibinfo {author} {\bibfnamefont {E.}~\bibnamefont {Khalaf}},\ }\bibfield  {title} {\bibinfo {title} {Family of ideal chern flatbands with arbitrary chern number in chiral twisted graphene multilayers},\ }\bibfield  {journal} {\bibinfo  {journal} {Physical Review Letters}\ }\textbf {\bibinfo {volume} {128}},\ \href {https://doi.org/10.1103/physrevlett.128.176404} {10.1103/physrevlett.128.176404} (\bibinfo {year} {2022})\BibitemShut {NoStop}%
\bibitem [{\citenamefont {Wang}\ and\ \citenamefont {Liu}(2022)}]{Wang2022Hierarchy}%
  \BibitemOpen
  \bibfield  {author} {\bibinfo {author} {\bibfnamefont {J.}~\bibnamefont {Wang}}\ and\ \bibinfo {author} {\bibfnamefont {Z.}~\bibnamefont {Liu}},\ }\bibfield  {title} {\bibinfo {title} {Hierarchy of ideal flatbands in chiral twisted multilayer graphene models},\ }\href {https://doi.org/10.1103/PhysRevLett.128.176403} {\bibfield  {journal} {\bibinfo  {journal} {Phys. Rev. Lett.}\ }\textbf {\bibinfo {volume} {128}},\ \bibinfo {pages} {176403} (\bibinfo {year} {2022})}\BibitemShut {NoStop}%
\bibitem [{\citenamefont {San-Jose}\ \emph {et~al.}(2012)\citenamefont {San-Jose}, \citenamefont {Gonz\'alez},\ and\ \citenamefont {Guinea}}]{SanJose2012NonAbelian}%
  \BibitemOpen
  \bibfield  {author} {\bibinfo {author} {\bibfnamefont {P.}~\bibnamefont {San-Jose}}, \bibinfo {author} {\bibfnamefont {J.}~\bibnamefont {Gonz\'alez}},\ and\ \bibinfo {author} {\bibfnamefont {F.}~\bibnamefont {Guinea}},\ }\bibfield  {title} {\bibinfo {title} {Non-abelian gauge potentials in graphene bilayers},\ }\href {https://doi.org/10.1103/PhysRevLett.108.216802} {\bibfield  {journal} {\bibinfo  {journal} {Phys. Rev. Lett.}\ }\textbf {\bibinfo {volume} {108}},\ \bibinfo {pages} {216802} (\bibinfo {year} {2012})}\BibitemShut {NoStop}%
\bibitem [{\citenamefont {Lopes~dos Santos}\ \emph {et~al.}(2007)\citenamefont {Lopes~dos Santos}, \citenamefont {Peres},\ and\ \citenamefont {Castro~Neto}}]{Lopes2007Graphene}%
  \BibitemOpen
  \bibfield  {author} {\bibinfo {author} {\bibfnamefont {J.~M.~B.}\ \bibnamefont {Lopes~dos Santos}}, \bibinfo {author} {\bibfnamefont {N.~M.~R.}\ \bibnamefont {Peres}},\ and\ \bibinfo {author} {\bibfnamefont {A.~H.}\ \bibnamefont {Castro~Neto}},\ }\bibfield  {title} {\bibinfo {title} {Graphene bilayer with a twist: Electronic structure},\ }\href {https://doi.org/10.1103/PhysRevLett.99.256802} {\bibfield  {journal} {\bibinfo  {journal} {Phys. Rev. Lett.}\ }\textbf {\bibinfo {volume} {99}},\ \bibinfo {pages} {256802} (\bibinfo {year} {2007})}\BibitemShut {NoStop}%
\bibitem [{\citenamefont {Su\'arez~Morell}\ \emph {et~al.}(2011)\citenamefont {Su\'arez~Morell}, \citenamefont {Vargas}, \citenamefont {Chico},\ and\ \citenamefont {Brey}}]{Suarez2011Charge}%
  \BibitemOpen
  \bibfield  {author} {\bibinfo {author} {\bibfnamefont {E.}~\bibnamefont {Su\'arez~Morell}}, \bibinfo {author} {\bibfnamefont {P.}~\bibnamefont {Vargas}}, \bibinfo {author} {\bibfnamefont {L.}~\bibnamefont {Chico}},\ and\ \bibinfo {author} {\bibfnamefont {L.}~\bibnamefont {Brey}},\ }\bibfield  {title} {\bibinfo {title} {Charge redistribution and interlayer coupling in twisted bilayer graphene under electric fields},\ }\href {https://doi.org/10.1103/PhysRevB.84.195421} {\bibfield  {journal} {\bibinfo  {journal} {Phys. Rev. B}\ }\textbf {\bibinfo {volume} {84}},\ \bibinfo {pages} {195421} (\bibinfo {year} {2011})}\BibitemShut {NoStop}%
\bibitem [{\citenamefont {Tong}\ \emph {et~al.}(2022)\citenamefont {Tong}, \citenamefont {Tong}, \citenamefont {Yang}, \citenamefont {Zhou}, \citenamefont {Wu}, \citenamefont {Tian}, \citenamefont {Zhang}, \citenamefont {Zhang}, \citenamefont {Qin},\ and\ \citenamefont {Yin}}]{Tong2022Spectroscopic}%
  \BibitemOpen
  \bibfield  {author} {\bibinfo {author} {\bibfnamefont {L.-H.}\ \bibnamefont {Tong}}, \bibinfo {author} {\bibfnamefont {Q.}~\bibnamefont {Tong}}, \bibinfo {author} {\bibfnamefont {L.-Z.}\ \bibnamefont {Yang}}, \bibinfo {author} {\bibfnamefont {Y.-Y.}\ \bibnamefont {Zhou}}, \bibinfo {author} {\bibfnamefont {Q.}~\bibnamefont {Wu}}, \bibinfo {author} {\bibfnamefont {Y.}~\bibnamefont {Tian}}, \bibinfo {author} {\bibfnamefont {L.}~\bibnamefont {Zhang}}, \bibinfo {author} {\bibfnamefont {L.}~\bibnamefont {Zhang}}, \bibinfo {author} {\bibfnamefont {Z.}~\bibnamefont {Qin}},\ and\ \bibinfo {author} {\bibfnamefont {L.-J.}\ \bibnamefont {Yin}},\ }\bibfield  {title} {\bibinfo {title} {Spectroscopic visualization of flat bands in magic-angle twisted monolayer-bilayer graphene: Coexistence of localization and delocalization},\ }\href {https://doi.org/10.1103/PhysRevLett.128.126401} {\bibfield  {journal} {\bibinfo  {journal} {Physical Review Letters}\ }\textbf {\bibinfo {volume} {128}},\ \bibinfo {pages} {126401} (\bibinfo
  {year} {2022})}\BibitemShut {NoStop}%
\bibitem [{\citenamefont {Li}\ \emph {et~al.}(2023)\citenamefont {Li}, \citenamefont {Wang}, \citenamefont {Xue}, \citenamefont {Cao}, \citenamefont {Watanabe}, \citenamefont {Taniguchi}, \citenamefont {Gao},\ and\ \citenamefont {Mao}}]{Li2023Gate}%
  \BibitemOpen
  \bibfield  {author} {\bibinfo {author} {\bibfnamefont {S.}~\bibnamefont {Li}}, \bibinfo {author} {\bibfnamefont {Z.}~\bibnamefont {Wang}}, \bibinfo {author} {\bibfnamefont {Y.}~\bibnamefont {Xue}}, \bibinfo {author} {\bibfnamefont {L.}~\bibnamefont {Cao}}, \bibinfo {author} {\bibfnamefont {K.}~\bibnamefont {Watanabe}}, \bibinfo {author} {\bibfnamefont {T.}~\bibnamefont {Taniguchi}}, \bibinfo {author} {\bibfnamefont {H.}~\bibnamefont {Gao}},\ and\ \bibinfo {author} {\bibfnamefont {J.}~\bibnamefont {Mao}},\ }\bibfield  {title} {\bibinfo {title} {Gate-controlled localization to delocalization transition of flat band wavefunction in twisted monolayer--bilayer graphene},\ }\href {https://doi.org/10.1088/1674-1056/acc805} {\bibfield  {journal} {\bibinfo  {journal} {Chinese Physics B}\ }\textbf {\bibinfo {volume} {32}},\ \bibinfo {pages} {067304} (\bibinfo {year} {2023})}\BibitemShut {NoStop}%
\bibitem [{\citenamefont {Seewald}\ \emph {et~al.}(2025)\citenamefont {Seewald}, \citenamefont {Ghosh}, \citenamefont {Verma}, \citenamefont {Cenker}, \citenamefont {Dong}, \citenamefont {Yang}, \citenamefont {Basu}, \citenamefont {Taniguchi}, \citenamefont {Watanabe}, \citenamefont {Deshmukh}, \citenamefont {Basov}, \citenamefont {Queiroz}, \citenamefont {Dean},\ and\ \citenamefont {Pasupathy}}]{Seewald2025Mapping}%
  \BibitemOpen
  \bibfield  {author} {\bibinfo {author} {\bibfnamefont {E.}~\bibnamefont {Seewald}}, \bibinfo {author} {\bibfnamefont {S.}~\bibnamefont {Ghosh}}, \bibinfo {author} {\bibfnamefont {N.}~\bibnamefont {Verma}}, \bibinfo {author} {\bibfnamefont {J.}~\bibnamefont {Cenker}}, \bibinfo {author} {\bibfnamefont {Y.}~\bibnamefont {Dong}}, \bibinfo {author} {\bibfnamefont {B.}~\bibnamefont {Yang}}, \bibinfo {author} {\bibfnamefont {A.}~\bibnamefont {Basu}}, \bibinfo {author} {\bibfnamefont {T.}~\bibnamefont {Taniguchi}}, \bibinfo {author} {\bibfnamefont {K.}~\bibnamefont {Watanabe}}, \bibinfo {author} {\bibfnamefont {M.~M.}\ \bibnamefont {Deshmukh}}, \bibinfo {author} {\bibfnamefont {D.~N.}\ \bibnamefont {Basov}}, \bibinfo {author} {\bibfnamefont {R.}~\bibnamefont {Queiroz}}, \bibinfo {author} {\bibfnamefont {C.}~\bibnamefont {Dean}},\ and\ \bibinfo {author} {\bibfnamefont {A.~N.}\ \bibnamefont {Pasupathy}},\ }\href {https://doi.org/10.48550/arXiv.2510.09548} {\bibinfo {title} {Mapping the moir{\'e} potential in
  multi-layer rhombohedral graphene}},\ \bibinfo {howpublished} {arXiv} (\bibinfo {year} {2025}),\ \bibinfo {note} {(accessed November 24, 2026)},\ \Eprint {https://arxiv.org/abs/2510.09548} {arXiv:2510.09548 [cond-mat.mes-hall]} \BibitemShut {NoStop}%
\bibitem [{\citenamefont {Rademaker}\ and\ \citenamefont {Mellado}(2018)}]{Rademaker2018Charge}%
  \BibitemOpen
  \bibfield  {author} {\bibinfo {author} {\bibfnamefont {L.}~\bibnamefont {Rademaker}}\ and\ \bibinfo {author} {\bibfnamefont {P.}~\bibnamefont {Mellado}},\ }\bibfield  {title} {\bibinfo {title} {Charge-transfer insulation in twisted bilayer graphene},\ }\href {https://doi.org/10.1103/PhysRevB.98.235158} {\bibfield  {journal} {\bibinfo  {journal} {Phys. Rev. B}\ }\textbf {\bibinfo {volume} {98}},\ \bibinfo {pages} {235158} (\bibinfo {year} {2018})}\BibitemShut {NoStop}%
\bibitem [{\citenamefont {Guinea}\ and\ \citenamefont {Walet}(2018)}]{Guinea2018Electrostatic}%
  \BibitemOpen
  \bibfield  {author} {\bibinfo {author} {\bibfnamefont {F.}~\bibnamefont {Guinea}}\ and\ \bibinfo {author} {\bibfnamefont {N.~R.}\ \bibnamefont {Walet}},\ }\bibfield  {title} {\bibinfo {title} {Electrostatic effects, band distortions, and superconductivity in twisted graphene bilayers},\ }\href {https://doi.org/10.1073/pnas.1810947115} {\bibfield  {journal} {\bibinfo  {journal} {Proceedings of the National Academy of Sciences}\ }\textbf {\bibinfo {volume} {115}},\ \bibinfo {pages} {13174–13179} (\bibinfo {year} {2018})}\BibitemShut {NoStop}%
\bibitem [{\citenamefont {Pantaleón}\ \emph {et~al.}(2021)\citenamefont {Pantaleón}, \citenamefont {Cea}, \citenamefont {Brown}, \citenamefont {Walet},\ and\ \citenamefont {Guinea}}]{Pantaleon2021Narrow}%
  \BibitemOpen
  \bibfield  {author} {\bibinfo {author} {\bibfnamefont {P.~A.}\ \bibnamefont {Pantaleón}}, \bibinfo {author} {\bibfnamefont {T.}~\bibnamefont {Cea}}, \bibinfo {author} {\bibfnamefont {R.}~\bibnamefont {Brown}}, \bibinfo {author} {\bibfnamefont {N.~R.}\ \bibnamefont {Walet}},\ and\ \bibinfo {author} {\bibfnamefont {F.}~\bibnamefont {Guinea}},\ }\bibfield  {title} {\bibinfo {title} {Narrow bands, electrostatic interactions and band topology in graphene stacks},\ }\href {https://doi.org/10.1088/2053-1583/ac1b6d} {\bibfield  {journal} {\bibinfo  {journal} {2D Materials}\ }\textbf {\bibinfo {volume} {8}},\ \bibinfo {pages} {044006} (\bibinfo {year} {2021})}\BibitemShut {NoStop}%
\bibitem [{\citenamefont {Zhang}\ \emph {et~al.}(2025{\natexlab{b}})\citenamefont {Zhang}, \citenamefont {Chen}, \citenamefont {Watanabe}, \citenamefont {Taniguchi}, \citenamefont {Song},\ and\ \citenamefont {Lu}}]{Zhang2025_R8GhBN}%
  \BibitemOpen
  \bibfield  {author} {\bibinfo {author} {\bibfnamefont {Z.}~\bibnamefont {Zhang}}, \bibinfo {author} {\bibfnamefont {X.}~\bibnamefont {Chen}}, \bibinfo {author} {\bibfnamefont {K.}~\bibnamefont {Watanabe}}, \bibinfo {author} {\bibfnamefont {T.}~\bibnamefont {Taniguchi}}, \bibinfo {author} {\bibfnamefont {Z.}~\bibnamefont {Song}},\ and\ \bibinfo {author} {\bibfnamefont {X.}~\bibnamefont {Lu}},\ }\href {https://doi.org/10.48550/arxiv.2512.21609} {\bibinfo {title} {Entangled moire chern insulator in rhombohedral graphene}},\ \bibinfo {howpublished} {arXiv} (\bibinfo {year} {2025}{\natexlab{b}}),\ \bibinfo {note} {(accessed November 24, 2026)},\ \Eprint {https://arxiv.org/abs/2512.21609} {arXiv:2512.21609} \BibitemShut {NoStop}%
\bibitem [{\citenamefont {Wu}\ \emph {et~al.}(2014)\citenamefont {Wu}, \citenamefont {Zhang}, \citenamefont {Ij\"{a}s}, \citenamefont {Han}, \citenamefont {Qiao}, \citenamefont {Li}, \citenamefont {Jiang}, \citenamefont {Ferrari},\ and\ \citenamefont {Tan}}]{Wu2014}%
  \BibitemOpen
  \bibfield  {author} {\bibinfo {author} {\bibfnamefont {J.-B.}\ \bibnamefont {Wu}}, \bibinfo {author} {\bibfnamefont {X.}~\bibnamefont {Zhang}}, \bibinfo {author} {\bibfnamefont {M.}~\bibnamefont {Ij\"{a}s}}, \bibinfo {author} {\bibfnamefont {W.-P.}\ \bibnamefont {Han}}, \bibinfo {author} {\bibfnamefont {X.-F.}\ \bibnamefont {Qiao}}, \bibinfo {author} {\bibfnamefont {X.-L.}\ \bibnamefont {Li}}, \bibinfo {author} {\bibfnamefont {D.-S.}\ \bibnamefont {Jiang}}, \bibinfo {author} {\bibfnamefont {A.~C.}\ \bibnamefont {Ferrari}},\ and\ \bibinfo {author} {\bibfnamefont {P.-H.}\ \bibnamefont {Tan}},\ }\bibfield  {title} {\bibinfo {title} {Resonant raman spectroscopy of twisted multilayer graphene},\ }\bibfield  {journal} {\bibinfo  {journal} {Nature Communications}\ }\textbf {\bibinfo {volume} {5}},\ \href {https://doi.org/10.1038/ncomms6309} {10.1038/ncomms6309} (\bibinfo {year} {2014})\BibitemShut {NoStop}%
\bibitem [{\citenamefont {Waters}\ \emph {et~al.}(2023)\citenamefont {Waters}, \citenamefont {Thompson}, \citenamefont {Arreguin-Martinez}, \citenamefont {Fujimoto}, \citenamefont {Ren}, \citenamefont {Watanabe}, \citenamefont {Taniguchi}, \citenamefont {Cao}, \citenamefont {Xiao},\ and\ \citenamefont {Yankowitz}}]{Waters2023}%
  \BibitemOpen
  \bibfield  {author} {\bibinfo {author} {\bibfnamefont {D.}~\bibnamefont {Waters}}, \bibinfo {author} {\bibfnamefont {E.}~\bibnamefont {Thompson}}, \bibinfo {author} {\bibfnamefont {E.}~\bibnamefont {Arreguin-Martinez}}, \bibinfo {author} {\bibfnamefont {M.}~\bibnamefont {Fujimoto}}, \bibinfo {author} {\bibfnamefont {Y.}~\bibnamefont {Ren}}, \bibinfo {author} {\bibfnamefont {K.}~\bibnamefont {Watanabe}}, \bibinfo {author} {\bibfnamefont {T.}~\bibnamefont {Taniguchi}}, \bibinfo {author} {\bibfnamefont {T.}~\bibnamefont {Cao}}, \bibinfo {author} {\bibfnamefont {D.}~\bibnamefont {Xiao}},\ and\ \bibinfo {author} {\bibfnamefont {M.}~\bibnamefont {Yankowitz}},\ }\bibfield  {title} {\bibinfo {title} {Mixed-dimensional moiré systems of twisted graphitic thin films},\ }\href {https://doi.org/10.1038/s41586-023-06290-3} {\bibfield  {journal} {\bibinfo  {journal} {Nature}\ }\textbf {\bibinfo {volume} {620}},\ \bibinfo {pages} {750–755} (\bibinfo {year} {2023})}\BibitemShut {NoStop}%
\bibitem [{\citenamefont {Palumbo}(2021)}]{Palumbo2021NonAbelian}%
  \BibitemOpen
  \bibfield  {author} {\bibinfo {author} {\bibfnamefont {G.}~\bibnamefont {Palumbo}},\ }\bibfield  {title} {\bibinfo {title} {Non-abelian tensor berry connections in multiband topological systems},\ }\href {https://doi.org/10.1103/PhysRevLett.126.246801} {\bibfield  {journal} {\bibinfo  {journal} {Phys. Rev. Lett.}\ }\textbf {\bibinfo {volume} {126}},\ \bibinfo {pages} {246801} (\bibinfo {year} {2021})}\BibitemShut {NoStop}%
\bibitem [{\citenamefont {Li}\ \emph {et~al.}(2022)\citenamefont {Li}, \citenamefont {Wang}, \citenamefont {Xue}, \citenamefont {Wang}, \citenamefont {Zhang}, \citenamefont {Liu}, \citenamefont {Zhu}, \citenamefont {Watanabe}, \citenamefont {Taniguchi}, \citenamefont {Gao}, \citenamefont {Jiang},\ and\ \citenamefont {Mao}}]{Li2022Imaging}%
  \BibitemOpen
  \bibfield  {author} {\bibinfo {author} {\bibfnamefont {S.-y.}\ \bibnamefont {Li}}, \bibinfo {author} {\bibfnamefont {Z.}~\bibnamefont {Wang}}, \bibinfo {author} {\bibfnamefont {Y.}~\bibnamefont {Xue}}, \bibinfo {author} {\bibfnamefont {Y.}~\bibnamefont {Wang}}, \bibinfo {author} {\bibfnamefont {S.}~\bibnamefont {Zhang}}, \bibinfo {author} {\bibfnamefont {J.}~\bibnamefont {Liu}}, \bibinfo {author} {\bibfnamefont {Z.}~\bibnamefont {Zhu}}, \bibinfo {author} {\bibfnamefont {K.}~\bibnamefont {Watanabe}}, \bibinfo {author} {\bibfnamefont {T.}~\bibnamefont {Taniguchi}}, \bibinfo {author} {\bibfnamefont {H.-j.}\ \bibnamefont {Gao}}, \bibinfo {author} {\bibfnamefont {Y.}~\bibnamefont {Jiang}},\ and\ \bibinfo {author} {\bibfnamefont {J.}~\bibnamefont {Mao}},\ }\bibfield  {title} {\bibinfo {title} {Imaging topological and correlated insulating states in twisted monolayer-bilayer graphene},\ }\href {https://doi.org/10.1038/s41467-022-31851-x} {\bibfield  {journal} {\bibinfo  {journal} {Nature Communications}\ }\textbf
  {\bibinfo {volume} {13}},\ \bibinfo {pages} {4225} (\bibinfo {year} {2022})}\BibitemShut {NoStop}%
\bibitem [{\citenamefont {Zhang}\ \emph {et~al.}(2010)\citenamefont {Zhang}, \citenamefont {Sahu}, \citenamefont {Min},\ and\ \citenamefont {MacDonald}}]{Zhang2010Band}%
  \BibitemOpen
  \bibfield  {author} {\bibinfo {author} {\bibfnamefont {F.}~\bibnamefont {Zhang}}, \bibinfo {author} {\bibfnamefont {B.}~\bibnamefont {Sahu}}, \bibinfo {author} {\bibfnamefont {H.}~\bibnamefont {Min}},\ and\ \bibinfo {author} {\bibfnamefont {A.~H.}\ \bibnamefont {MacDonald}},\ }\bibfield  {title} {\bibinfo {title} {Band structure of $abc$-stacked graphene trilayers},\ }\href {https://doi.org/10.1103/PhysRevB.82.035409} {\bibfield  {journal} {\bibinfo  {journal} {Phys. Rev. B}\ }\textbf {\bibinfo {volume} {82}},\ \bibinfo {pages} {035409} (\bibinfo {year} {2010})}\BibitemShut {NoStop}%
\bibitem [{\citenamefont {Bistritzer}\ and\ \citenamefont {MacDonald}(2011)}]{Bistritzer2011Moire}%
  \BibitemOpen
  \bibfield  {author} {\bibinfo {author} {\bibfnamefont {R.}~\bibnamefont {Bistritzer}}\ and\ \bibinfo {author} {\bibfnamefont {A.~H.}\ \bibnamefont {MacDonald}},\ }\bibfield  {title} {\bibinfo {title} {Moir{\'e} bands in twisted double-layer graphene},\ }\href {https://doi.org/10.1073/pnas.1108174108} {\bibfield  {journal} {\bibinfo  {journal} {Proceedings of the National Academy of Sciences}\ }\textbf {\bibinfo {volume} {108}},\ \bibinfo {pages} {12233} (\bibinfo {year} {2011})}\BibitemShut {NoStop}%
\bibitem [{\citenamefont {Koshino}(2015)}]{Koshino2015Interlayer}%
  \BibitemOpen
  \bibfield  {author} {\bibinfo {author} {\bibfnamefont {M.}~\bibnamefont {Koshino}},\ }\bibfield  {title} {\bibinfo {title} {Interlayer interaction in general incommensurate atomic layers},\ }\href {https://doi.org/10.1088/1367-2630/17/1/015014} {\bibfield  {journal} {\bibinfo  {journal} {New Journal of Physics}\ }\textbf {\bibinfo {volume} {17}},\ \bibinfo {pages} {015014} (\bibinfo {year} {2015})}\BibitemShut {NoStop}%
\bibitem [{\citenamefont {Koshino}\ \emph {et~al.}(2018)\citenamefont {Koshino}, \citenamefont {Yuan}, \citenamefont {Koretsune}, \citenamefont {Ochi}, \citenamefont {Kuroki},\ and\ \citenamefont {Fu}}]{Koshino2018Maximally}%
  \BibitemOpen
  \bibfield  {author} {\bibinfo {author} {\bibfnamefont {M.}~\bibnamefont {Koshino}}, \bibinfo {author} {\bibfnamefont {N.~F.~Q.}\ \bibnamefont {Yuan}}, \bibinfo {author} {\bibfnamefont {T.}~\bibnamefont {Koretsune}}, \bibinfo {author} {\bibfnamefont {M.}~\bibnamefont {Ochi}}, \bibinfo {author} {\bibfnamefont {K.}~\bibnamefont {Kuroki}},\ and\ \bibinfo {author} {\bibfnamefont {L.}~\bibnamefont {Fu}},\ }\bibfield  {title} {\bibinfo {title} {Maximally localized wannier orbitals and the extended hubbard model for twisted bilayer graphene},\ }\href {https://doi.org/10.1103/PhysRevX.8.031087} {\bibfield  {journal} {\bibinfo  {journal} {Phys. Rev. X}\ }\textbf {\bibinfo {volume} {8}},\ \bibinfo {pages} {031087} (\bibinfo {year} {2018})}\BibitemShut {NoStop}%
\bibitem [{\citenamefont {Parra‐Martínez}\ \emph {et~al.}(2025)\citenamefont {Parra‐Martínez}, \citenamefont {Jimeno‐Pozo}, \citenamefont {Silva‐Guillén},\ and\ \citenamefont {Guinea}}]{ParraMartinez2025}%
  \BibitemOpen
  \bibfield  {author} {\bibinfo {author} {\bibfnamefont {G.}~\bibnamefont {Parra‐Martínez}}, \bibinfo {author} {\bibfnamefont {A.}~\bibnamefont {Jimeno‐Pozo}}, \bibinfo {author} {\bibfnamefont {J.~A.}\ \bibnamefont {Silva‐Guillén}},\ and\ \bibinfo {author} {\bibfnamefont {F.}~\bibnamefont {Guinea}},\ }\bibfield  {title} {\bibinfo {title} {Nematic and partially polarized phases in rhombohedral graphene with varying number of layers: An extensive hartree–fock study},\ }\bibfield  {journal} {\bibinfo  {journal} {Advanced Materials}\ }\textbf {\bibinfo {volume} {38}},\ \href {https://doi.org/10.1002/adma.202514008} {10.1002/adma.202514008} (\bibinfo {year} {2025})\BibitemShut {NoStop}%
\bibitem [{\citenamefont {Zhou}\ \emph {et~al.}(2021)\citenamefont {Zhou}, \citenamefont {Xie}, \citenamefont {Ghazaryan}, \citenamefont {Holder}, \citenamefont {Ehrets}, \citenamefont {Spanton}, \citenamefont {Taniguchi}, \citenamefont {Watanabe}, \citenamefont {Berg}, \citenamefont {Serbyn},\ and\ \citenamefont {Young}}]{Zhou2021}%
  \BibitemOpen
  \bibfield  {author} {\bibinfo {author} {\bibfnamefont {H.}~\bibnamefont {Zhou}}, \bibinfo {author} {\bibfnamefont {T.}~\bibnamefont {Xie}}, \bibinfo {author} {\bibfnamefont {A.}~\bibnamefont {Ghazaryan}}, \bibinfo {author} {\bibfnamefont {T.}~\bibnamefont {Holder}}, \bibinfo {author} {\bibfnamefont {J.~R.}\ \bibnamefont {Ehrets}}, \bibinfo {author} {\bibfnamefont {E.~M.}\ \bibnamefont {Spanton}}, \bibinfo {author} {\bibfnamefont {T.}~\bibnamefont {Taniguchi}}, \bibinfo {author} {\bibfnamefont {K.}~\bibnamefont {Watanabe}}, \bibinfo {author} {\bibfnamefont {E.}~\bibnamefont {Berg}}, \bibinfo {author} {\bibfnamefont {M.}~\bibnamefont {Serbyn}},\ and\ \bibinfo {author} {\bibfnamefont {A.~F.}\ \bibnamefont {Young}},\ }\bibfield  {title} {\bibinfo {title} {Half- and quarter-metals in rhombohedral trilayer graphene},\ }\href {https://doi.org/10.1038/s41586-021-03938-w} {\bibfield  {journal} {\bibinfo  {journal} {Nature}\ }\textbf {\bibinfo {volume} {598}},\ \bibinfo {pages} {429–433} (\bibinfo {year}
  {2021})}\BibitemShut {NoStop}%
\bibitem [{\citenamefont {Fukui}\ \emph {et~al.}(2005)\citenamefont {Fukui}, \citenamefont {Hatsugai},\ and\ \citenamefont {Suzuki}}]{Fukui2005}%
  \BibitemOpen
  \bibfield  {author} {\bibinfo {author} {\bibfnamefont {T.}~\bibnamefont {Fukui}}, \bibinfo {author} {\bibfnamefont {Y.}~\bibnamefont {Hatsugai}},\ and\ \bibinfo {author} {\bibfnamefont {H.}~\bibnamefont {Suzuki}},\ }\bibfield  {title} {\bibinfo {title} {Chern numbers in discretized brillouin zone: Efficient method of computing (spin) hall conductances},\ }\href {https://doi.org/10.1143/jpsj.74.1674} {\bibfield  {journal} {\bibinfo  {journal} {J. Phys. Soc. Jpn.}\ }\textbf {\bibinfo {volume} {74}},\ \bibinfo {pages} {1674} (\bibinfo {year} {2005})}\BibitemShut {NoStop}%
\bibitem [{\citenamefont {Phong}\ and\ \citenamefont {Lewandowski}(2026)}]{Phong2026Coulomb}%
  \BibitemOpen
  \bibfield  {author} {\bibinfo {author} {\bibfnamefont {V.~T.}\ \bibnamefont {Phong}}\ and\ \bibinfo {author} {\bibfnamefont {C.}~\bibnamefont {Lewandowski}},\ }\bibfield  {title} {\bibinfo {title} {Coulomb interaction-stabilized isolated narrow bands with chern numbers $\mathcal{C} > 1$ in twisted rhombohedral trilayer-bilayer graphene},\ }\bibfield  {journal} {\bibinfo  {journal} {Nature Communications}\ }\textbf {\bibinfo {volume} {17}},\ \href {https://doi.org/10.1038/s41467-026-74428-8} {10.1038/s41467-026-74428-8} (\bibinfo {year} {2026})\BibitemShut {NoStop}%
\bibitem [{\citenamefont {Mikitik}\ and\ \citenamefont {Sharlai}(2008)}]{Mikitik2008_BerryPhase}%
  \BibitemOpen
  \bibfield  {author} {\bibinfo {author} {\bibfnamefont {G.~P.}\ \bibnamefont {Mikitik}}\ and\ \bibinfo {author} {\bibfnamefont {Y.~V.}\ \bibnamefont {Sharlai}},\ }\bibfield  {title} {\bibinfo {title} {The berry phase in graphene and graphite multilayers},\ }\href {https://doi.org/10.1063/1.2981389} {\bibfield  {journal} {\bibinfo  {journal} {Low Temperature Physics}\ }\textbf {\bibinfo {volume} {34}},\ \bibinfo {pages} {794–800} (\bibinfo {year} {2008})}\BibitemShut {NoStop}%
\bibitem [{\citenamefont {Zhang}\ \emph {et~al.}(2022{\natexlab{b}})\citenamefont {Zhang}, \citenamefont {Dai},\ and\ \citenamefont {Liu}}]{Zhang2022_topoTrans}%
  \BibitemOpen
  \bibfield  {author} {\bibinfo {author} {\bibfnamefont {S.}~\bibnamefont {Zhang}}, \bibinfo {author} {\bibfnamefont {X.}~\bibnamefont {Dai}},\ and\ \bibinfo {author} {\bibfnamefont {J.}~\bibnamefont {Liu}},\ }\bibfield  {title} {\bibinfo {title} {Spin-polarized nematic order, quantum valley hall states, and field-tunable topological transitions in twisted multilayer graphene systems},\ }\bibfield  {journal} {\bibinfo  {journal} {Physical Review Letters}\ }\textbf {\bibinfo {volume} {128}},\ \href {https://doi.org/10.1103/physrevlett.128.026403} {10.1103/physrevlett.128.026403} (\bibinfo {year} {2022}{\natexlab{b}})\BibitemShut {NoStop}%
\end{thebibliography}

\clearpage
\onecolumngrid
\begin{center}
{\large\bfseries Supplemental Material for ``Topological Narrow Bands in Twisted Rhombohedral Multilayer Graphene''}\\[0.5em]
Alejandro Jimeno-Pozo, Danna Liu, Francisco Guinea, and Pierre A. Pantale\'on
\end{center}
\setcounter{equation}{0}
\setcounter{figure}{0}
\setcounter{table}{0}
\renewcommand{\theequation}{S\arabic{equation}}
\renewcommand{\thefigure}{S\arabic{figure}}
\renewcommand{\thetable}{S\arabic{table}}
\setcounter{section}{0}
\renewcommand{\thesection}{S\arabic{section}}

\let\addcontentsline\origaddcontentsline
\tableofcontents

\section{Full continuum model and conventions}\label{sec:S1}

We consider a twisted graphene multilayer composed of a single graphene monolayer stacked on top of an $N$-layer rhombohedral, or ABC-stacked, graphene system, resulting in a total of $N+1$ layers. We also include $N=2$, corresponding to an AB-stacked bilayer, as the simplest reference system. In the non-rotated configuration, the rhombohedral slab consists of $N$ graphene layers labeled $l=1,\dots,N$, with atomic sites $A_l$ and $B_l$ arranged in the standard ABC stacking sequence such that the sites $B_l$ and $A_{l+1}$ are vertically aligned between neighbouring layers. The rotated monolayer is labeled by $l=0$.

The in-plane lattice of graphene is defined by the primitive vectors
\begin{equation}
\boldsymbol{a}_1=a(1,0),\qquad
\boldsymbol{a}_2=a\left(\frac12,\frac{\sqrt3}{2}\right),
\end{equation}
with $a\simeq0.246\,\mathrm{nm}$. The corresponding reciprocal lattice vectors are
\begin{equation}
\boldsymbol{b}_1=\frac{2\pi}{a}\left(1,-\frac{1}{\sqrt3}\right),\qquad
\boldsymbol{b}_2=\frac{2\pi}{a}\left(0,\frac{2}{\sqrt3}\right).
\end{equation}
The top monolayer is rotated by $+\theta/2$, while the entire rhombohedral stack is rotated by $-\theta/2$. The reciprocal lattice vectors of each layer are $\boldsymbol{b}_i^{(l)}=R(\theta_l)\boldsymbol{b}_i$, with $\theta_0=+\theta/2$ and $\theta_l=-\theta/2$ for $l\ge1$. The moir\'e reciprocal vectors are
\begin{equation}
\boldsymbol{G}_i=\boldsymbol{b}_i^{(0)}-\boldsymbol{b}_i^{(1)},\qquad i=1,2,
\end{equation}
and the Dirac point of layer $l$ is
\begin{equation}
\boldsymbol{K}^{(l)}=\frac{2\boldsymbol{b}_1^{(l)}+\boldsymbol{b}_2^{(l)}}{3}.
\end{equation}
Within a single-valley continuum description, the full Hamiltonian is written in layer-block form as
\begin{equation}
H(\boldsymbol{q})=
\begin{pmatrix}
H_{00}(\boldsymbol{q}) & T_{01}(\boldsymbol{r}) & 0 & 0 & \cdots & 0 \\
T_{01}^{\dagger}(\boldsymbol{r}) & H_{11}(\boldsymbol{q}) & U_{12}(\boldsymbol{q}) & W_{13} & \cdots & 0 \\
0 & U_{21}(\boldsymbol{q}) & H_{22}(\boldsymbol{q}) & U_{23}(\boldsymbol{q}) & \ddots & \vdots \\
0 & W_{13}^{\dagger} & U_{32}(\boldsymbol{q}) & H_{33}(\boldsymbol{q}) & \ddots & W_{(N-2)N} \\
\vdots & \vdots & \ddots & \ddots & \ddots & U_{(N-1)N}(\boldsymbol{q}) \\
0 & 0 & \cdots & W_{(N-2)N}^{\dagger} & U_{N(N-1)}(\boldsymbol{q}) & H_{NN}(\boldsymbol{q})
\end{pmatrix}.
\label{eq:supp_full_H}
\end{equation}
The only term that directly couples the rotated monolayer to the rhombohedral slab is the interfacial moir\'e tunnelling $T_{01}(\boldsymbol{r})$. The layer-resolved Dirac operators are
\begin{equation}
\pi^{(l)}=\eta q_x^{(l)}+iq_y^{(l)},\qquad
(\pi^{(l)})^\dagger=\eta q_x^{(l)}-iq_y^{(l)},
\end{equation}
where $\eta=\pm1$ is the valley index and $\boldsymbol{q}^{(l)}=R(\mp\theta/2)(\boldsymbol{k}-\boldsymbol{K}^{(l)})$, with the upper sign for the rotated monolayer and the lower sign for the rhombohedral slab. In units where $\hbar=1$, the intralayer blocks are
\begin{equation}
H_{ll}(\boldsymbol{q})=
\begin{pmatrix}
\Delta^{(l)} & v_f(\pi^{(l)})^\dagger\\
v_f\pi^{(l)} & \Delta^{(l)}
\end{pmatrix},
\label{eq:supp_H_intra}
\end{equation}
with $v_f=(\sqrt3/2)a\gamma_0$. A perpendicular displacement field is included through an onsite potential acting on the two outer layers,
\begin{equation}
\Delta^{(l)}=
\begin{cases}
-V/2, & l=0,\\
+V/2, & l=N,\\
0, & \mathrm{otherwise}.
\end{cases}
\label{eq:supp_bias}
\end{equation}
This profile us used because it provides a reasonable low-energy description for moderate potential differences~\cite{Koshino2010Interlayer}. It retains the electrostatic detuning responsible for the field-driven redistribution of wave-function weight. A potential varying across the internal layers would refine the dispersion and hybridization, but resolving this profile is not essential for identifying the surface-coupling mechanism considered in this work. The approximation is intended for surface-dominated central states at moderate bias; internal-layer potentials become more relevant at strong fields or when these states mix substantially with remote bands; see, for example, Ref.~\cite{Phong2026Coulomb}.  Sec.~\ref{sec:S9} compares this profile with a uniform field that drops linearly across all layers. For $l=1,\dots,N-1$, the coupling between adjacent rhombohedral layers is
\begin{equation}
U_{l,l+1}(\boldsymbol{q})=
\begin{pmatrix}
v_4(\pi^{(l)})^\dagger & v_3\pi^{(l)}\\
\gamma_1 & v_4(\pi^{(l)})^\dagger
\end{pmatrix},
\label{eq:supp_U}
\end{equation}
where $v_j=(\sqrt3/2)a\gamma_j$ for $j=3,4$, and $U_{l+1,l}=U_{l,l+1}^{\dagger}$. When $N\ge3$, the next-nearest-layer hopping is
\begin{equation}
W_{l,l+2}=\begin{pmatrix}0 & \gamma_2/2\\0&0\end{pmatrix},\qquad l=1,\dots,N-2,
\label{eq:supp_W}
\end{equation}
with $W_{l+2,l}=W_{l,l+2}^{\dagger}$. The moir\'e tunnelling between the rotated monolayer and the first rhombohedral layer is
\begin{equation}
T_{01}(\boldsymbol{r})=T_1+T_2e^{i\eta\boldsymbol{G}_1\cdot\boldsymbol{r}}+T_3e^{i\eta(\boldsymbol{G}_1+\boldsymbol{G}_2)\cdot\boldsymbol{r}},
\label{eq:supp_T01}
\end{equation}
where
\begin{equation}
T_j=\begin{pmatrix}u & u'e^{-i\phi_j}\\ u'e^{i\phi_j} & u\end{pmatrix},\qquad
\phi_j=\eta(j-1)\frac{2\pi}{3}.
\label{eq:supp_Tj}
\end{equation}
The parameters used in the calculations are the interlayer hoppings $\gamma_0=3.16$ eV, $\gamma_1=0.502$ eV, $\gamma_2=-0.0171$ eV, $\gamma_3=-0.377$ eV, and $\gamma_4=-0.099$ eV~\cite{Zhang2010Band}, together with the moir\'e tunnelling amplitudes $u=0.0797$ eV and $u'=0.0975$ eV~\cite{Koshino2018Maximally,Bistritzer2011Moire,Koshino2015Interlayer,Pantaleon2021Narrow}.  The sensitivity of the results to this choice is examined in Sec.~\ref{sec:S9} using an alternative set of rhombohedral hopping parameters.

For numerical diagonalization, each layer and sublattice component is expanded in moir\'e plane waves,
\begin{equation}
\psi_{l\alpha}^{(n)}(\boldsymbol r;\boldsymbol k)=\sum_{\boldsymbol G\in\mathcal G}C_{l\alpha,\boldsymbol G}^{(n)}(\boldsymbol k)e^{i(\boldsymbol k+\boldsymbol G)\cdot\boldsymbol r},
\label{eq:supp_plane_wave_expansion}
\end{equation}
where $\mathcal G$ is a finite set of reciprocal vectors generated by $\boldsymbol G_1$ and $\boldsymbol G_2$. The same converged reciprocal-space truncation is used for the spectra, layer weights, and real-space densities compared at a given set of parameters. Sec.~\ref{sec:S2} derives the low-energy surface theory from this full Hamiltonian. Secs.~\ref{sec:S3}--\ref{sec:S6} then use the full-model eigenstates to analyse layer hybridization and real-space density, while Secs.~\ref{sec:S7} and~\ref{sec:S8} give the topological calculation and the upper-band phase diagrams. Sec.~\ref{sec:S9} tests the robustness of the results against the hopping parametrization and the displacement-field profile.

\section{Path-projected surface model and chiral limit}\label{sec:S2}

In this section we derive a surface theory in order to retain the leading propagation processes of rhombohedral graphene while eliminating the high-energy dimerized sites. We retain $\gamma_1$, $\gamma_2$, and $\gamma_3$, set $\gamma_4=0$, and keep both moir\'e tunnelling amplitudes $u$ and $u'$. Omitting $\gamma_4$ permits the compact off-diagonal surface form below, because $\gamma_4$ couples equal sublattices and generates additional diagonal self-energy terms. The construction extends the path expansion of rhombohedral graphene described in Ref.~\cite{Koshino2009Trigonal} to a twisted monolayer coupled to the interface and remote surface sites of an $N$-layer slab.

For $N\ge2$, the strong vertical hopping $\gamma_1$ pairs $B_l$ with $A_{l+1}$, leaving $A_1$ and $B_N$ as the two surface sites. Including the rotated monolayer, the retained sector is
\begin{equation}
\mathcal L=\{A_0,B_0,A_1,B_N\},
\label{eq:supp_low_sector}
\end{equation}
and the eliminated sector is
\begin{equation}
\mathcal H=\{B_1,A_2,B_2,A_3,\ldots,B_{N-1},A_N\}.
\label{eq:supp_high_sector}
\end{equation}
We consider $N\ge2$. For completeness, when $N=1$ the eliminated sector $\mathcal H$ is empty and no projection is required. In the ordered basis $(\mathcal L,\mathcal H)$,
\begin{equation}
H=\begin{pmatrix}H_{\mathcal L\mathcal L}&H_{\mathcal L\mathcal H}\\H_{\mathcal H\mathcal L}&H_{\mathcal H\mathcal H}\end{pmatrix}.
\label{eq:supp_block_form}
\end{equation}
The exact downfolded Hamiltonian is energy dependent. For $|E|\ll\gamma_1$, we retain the leading L\"owdin term
\begin{equation}
H_{\mathrm{eff}}=H_{\mathcal L\mathcal L}-H_{\mathcal L\mathcal H}H_{\mathcal H\mathcal H}^{-1}H_{\mathcal H\mathcal L}.
\label{eq:supp_lowdin}
\end{equation}
All layers of the rhombohedral slab share the same rotated Dirac operator,
\begin{equation}
\pi=\eta q_x+iq_y,\qquad \pi^\dagger=\eta q_x-iq_y.
\label{eq:supp_pi_definition}
\end{equation}
It is convenient to define
\begin{equation}
\xi_1=v_f\pi,\qquad \xi_2=v_3\pi^\dagger,\qquad \xi_3=\frac{\gamma_2}{2},
\label{eq:supp_path_steps}
\end{equation}
where $\xi_1$, $\xi_2$, and $\xi_3$ advance an effective path from $A_1$ toward $B_N$ by one, two, and three layer indices, respectively. The ordinary intralayer hop supplies $\xi_1$, the skew $\gamma_3$ hop supplies $\xi_2$, and the next-nearest-layer hop supplies $\xi_3$. The surface-to-surface matrix element $X_N(\boldsymbol q)$ is defined by
\begin{equation}
X_N=-\frac{1}{\gamma_1}\left(\xi_1X_{N-1}+\xi_2X_{N-2}+\xi_3X_{N-3}\right),
\qquad N\ge1,
\label{eq:supp_X_recursion}
\end{equation}
with
\begin{equation}
X_0=-\gamma_1,\qquad X_{N<0}=0.
\label{eq:supp_X_initial}
\end{equation}
Equivalently,
\begin{equation}
X_N(\boldsymbol q)=
\sum_{\substack{n_1,n_2,n_3\ge0\\n_1+2n_2+3n_3=N}}
\frac{(n_1+n_2+n_3)!}{n_1!n_2!n_3!}
\frac{\xi_1^{n_1}\xi_2^{n_2}\xi_3^{n_3}}
{(-\gamma_1)^{n_1+n_2+n_3-1}}.
\label{eq:supp_X_closed}
\end{equation}
as obtained in Ref.~\cite{Koshino2009Trigonal}. The constraint $n_1+2n_2+3n_3=N$ counts the total advance between the two surface sites, while the multinomial coefficient counts the distinct orderings of the elementary steps. The first few amplitudes are
\begin{align}
X_1&=\xi_1,\\
X_2&=-\frac{\xi_1^2}{\gamma_1}+\xi_2,\\
X_3&=\frac{\xi_1^3}{\gamma_1^2}-\frac{2\xi_1\xi_2}{\gamma_1}+\xi_3,
\end{align}
which reproduce the usual chiral term, trigonal warping, and the direct $\gamma_2/2$ contribution for rhombohedral trilayer graphene~\cite{Min2008Chiral,Koshino2009Trigonal,Mikitik2008_BerryPhase}.

At zero displacement field, the projected rhombohedral Hamiltonian in the surface basis $(A_1,B_N)$ is
\begin{equation}
H_{\mathrm{ABC}}^{(N),\mathrm{path}}(\boldsymbol q)=
\begin{pmatrix}
0&X_N^\dagger(\boldsymbol q)\\
X_N(\boldsymbol q)&0
\end{pmatrix},
\label{eq:supp_HABC_path}
\end{equation}
with eigenvalues $E_\pm=\pm|X_N|$. The strict chiral power is therefore one contribution to a more general coherent path sum. The same projection dresses the moir\'e tunnelling that initially terminates on $B_1$. Define
\begin{equation}
D_m(\boldsymbol q)=-\frac{X_m(\boldsymbol q)}{\gamma_1},\qquad D_0=1.
\label{eq:supp_D_definition}
\end{equation}
Writing the original interface matrix as
\begin{equation}
T_{01}(\boldsymbol r)=
\begin{pmatrix}
T_{A_0A_1}(\boldsymbol r)&T_{A_0B_1}(\boldsymbol r)\\
T_{B_0A_1}(\boldsymbol r)&T_{B_0B_1}(\boldsymbol r)
\end{pmatrix},
\end{equation}
the projected tunnelling between $(A_0,B_0)$ and $(A_1,B_N)$ is
\begin{equation}
\widetilde T_{01}^{\mathrm{path}}(\boldsymbol r,\boldsymbol q)=
\begin{pmatrix}
T_{A_0A_1}(\boldsymbol r)&T_{A_0B_1}(\boldsymbol r)D_{N-1}^\dagger(\boldsymbol q)\\[6pt]
T_{B_0A_1}(\boldsymbol r)&T_{B_0B_1}(\boldsymbol r)D_{N-1}^\dagger(\boldsymbol q)
\end{pmatrix}.
\label{eq:supp_Teff_path}
\end{equation}
The first column is direct because $A_1$ is retained. The second column represents tunnelling into $B_1$ followed by virtual propagation to $B_N$. At $V=0$, the complete projected Hamiltonian is
\begin{equation}
H_{\mathrm{eff}}^{\mathrm{path}}=
\begin{pmatrix}
H_{00}^{0}&\widetilde T_{01}^{\mathrm{path}}\\[3pt]
\left(\widetilde T_{01}^{\mathrm{path}}\right)^\dagger&H_{\mathrm{ABC}}^{(N),\mathrm{path}}
\end{pmatrix},
\label{eq:supp_Hpath_V0}
\end{equation}
where
\begin{equation}
H_{00}^{0}(\boldsymbol q)=
\begin{pmatrix}
0&v_f(\pi^{(0)})^\dagger\\
v_f\pi^{(0)}&0
\end{pmatrix}.
\end{equation}
We next include the outer-layer displacement field, with $\Delta=V/2$. The monolayer $L_0$ carries $-\Delta$, and both sublattices of the remote layer $L_N$ carry $+\Delta$. For $N\ge2$, $B_N$ is retained but $A_N$ is eliminated. The latter therefore generates an additional contribution that would be missed by adding onsite energies only to the retained basis. Within the leading energy-independent projection, the eliminated block can be inverted exactly in $\Delta$, giving
\begin{equation}
H_{\mathrm{eff},\Delta}^{\mathrm{path}}=
\begin{pmatrix}
H_{00}^{\Delta}&\widetilde T_{01}^{\mathrm{path}}\\[3pt]
\left(\widetilde T_{01}^{\mathrm{path}}\right)^\dagger&
H_{\mathrm{ABC},\Delta}^{(N),0}
\end{pmatrix}
+\Delta\Lambda\Lambda^\dagger,
\qquad N\ge2,
\label{eq:supp_Hpath_finite}
\end{equation}
with
\begin{equation}
H_{00}^{\Delta}=
\begin{pmatrix}-\Delta&v_f(\pi^{(0)})^\dagger\\v_f\pi^{(0)}&-\Delta\end{pmatrix},
\qquad
H_{\mathrm{ABC},\Delta}^{(N),0}=
\begin{pmatrix}0&X_N^\dagger\\X_N&\Delta\end{pmatrix},
\end{equation}
and
\begin{equation}
\Lambda=
\begin{pmatrix}
\gamma_1^{-1}T_{A_0B_1}D_{N-2}^\dagger\\[4pt]
\gamma_1^{-1}T_{B_0B_1}D_{N-2}^\dagger\\[4pt]
-D_{N-1}^\dagger\\[4pt]
0
\end{pmatrix}.
\label{eq:supp_Lambda}
\end{equation}
The correction $\Delta\Lambda\Lambda^\dagger$ has the sign of $\Delta$. For the isolated projected slab, it gives
\begin{equation}
H_{\mathrm{ABC},\Delta}^{(N),\mathrm{isolated}}=
\begin{pmatrix}
\Delta|D_{N-1}|^2&X_N^\dagger\\
X_N&\Delta
\end{pmatrix}.
\label{eq:supp_HABC_finite}
\end{equation}
For $N=1$, no site is eliminated: the basis is $(A_0,B_0,A_1,B_1)$, both sites of $L_1$ receive $+\Delta$, and the correction is absent. The strict chiral limit is
\begin{equation}
\gamma_2=\gamma_3=\gamma_4=0,\qquad u=0, 
\label{eq:supp_strict_chiral}
\end{equation}
with these conditions, only the path containing $N$ ordinary Dirac steps survives, and
\begin{equation}
X_N^{\mathrm{ch}}=(-1)^{N-1}\frac{(v_f\pi)^N}{\gamma_1^{N-1}},
\qquad
D_{N-1}^{\dagger,\mathrm{ch}}=(-1)^{N-1}\frac{(v_f\pi^\dagger)^{N-1}}{\gamma_1^{N-1}}.
\label{eq:supp_X_chiral}
\end{equation}
The reduced model used in the numerical comparisons sets $\gamma_2=\gamma_3=\gamma_4=0$ while retaining $\gamma_1$, $u$, and $u'$. Its rhombohedral surface amplitude therefore has the same pure-power form as Eq.~\eqref{eq:supp_X_chiral}, but its moir\'e tunnelling retains both amplitudes. The strict chiral model further sets $u=0$ while retaining $\gamma_1$ and $u'$. In this limit, the same-sublattice moir\'e elements $T_{A_0A_1}$ and $T_{B_0B_1}$ vanish, and Eq.~\eqref{eq:supp_Teff_path} becomes
\begin{equation}
\widetilde T_{01}^{\mathrm{ch}}=
\begin{pmatrix}
0&T_{A_0B_1}D_{N-1}^{\dagger,\mathrm{ch}}\\[5pt]
T_{B_0A_1}&0
\end{pmatrix}.
\label{eq:supp_Teff_chiral}
\end{equation}
At $u=V=0$ and $\gamma_4=0$, both the path-projected and strict chiral Hamiltonians anticommute with
\begin{equation}
\Gamma=\mathrm{diag}(1,-1,1,-1),\qquad \{\Gamma,H_{\mathrm{eff}}\}=0.
\label{eq:supp_chiral_symmetry}
\end{equation}
Thus retaining $\gamma_2$ and $\gamma_3$ preserves sublattice symmetry and $E\leftrightarrow-E$. The path constraint also produces a local small-momentum $N\bmod3$ structure when $\gamma_2\ne0$. For $N=3m$, the lowest-order path contains $m$ factors of $\xi_3$ and is momentum independent. For $N=3m+1$, the lowest-order contribution contains one factor of $\xi_1$ and is proportional to $\pi$. For $N=3m+2$, the lowest-order contribution contains one factor of $\xi_2$ and is proportional to $\pi^\dagger$. This local asymptotic classification is distinct from the moir\'e-scale recurrence of the central-band charge textures discussed in the main text.

\begin{figure}[!htb]
\centering
\includegraphics[width=1\textwidth]{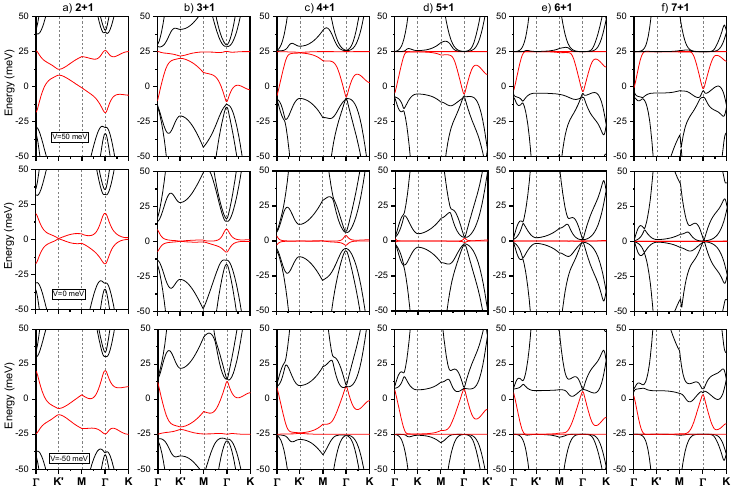}
\caption{\textbf{Electronic structure of the reduced model.} Band structure of the reduced $(N+1)$-layer rhombohedral graphene model for $N=2$ to $N=7$ at $\theta=1.05^\circ$. The top row corresponds to $V=50$ meV, the middle row to $V=0$, and the bottom row to $V=-50$ meV. Red curves highlight the two central bands. The energy scale changes with the number of layers.}
\label{fig:reduced_bands}
\end{figure}

The path-projected Hamiltonian is a low-energy theory. It omits the energy dependence of $(E-H_{\mathcal H\mathcal H})^{-1}$, the high-energy dimer bands, and the same-sublattice self-energies generated by $\gamma_4$. The approximation is controlled when the energies and surface-path amplitudes are small compared with the interlayer $\gamma_1$. The full continuum model remains necessary near remote-band crossings, large displacement fields and for quantitative phase boundaries. The reduced- and strict-chiral-model spectra are collected here as numerical checks of the model hierarchy derived above. The general definitions and physical role of these limits are given in the main text; the purpose of Supplementary Figs.~\ref{fig:reduced_bands} and~\ref{fig:supp_chiral_bands} is to show which spectral features survive after successive hopping channels are removed.

Supplementary Fig.~\ref{fig:reduced_bands} shows the reduced model, in which $\gamma_2=\gamma_3=\gamma_4=0$ while $\gamma_1$, $u$, and $u'$ remain finite. The progressive narrowing of the two central bands with increasing $N$ remains visible in all three field configurations. This confirms numerically that the dominant layer-number dependence does not require the further-neighbour hopping parameters: it is already generated by the pure-power rhombohedral surface propagation retained in the reduced model. The separation between the monolayer-dominated states near $K'$ and the rhombohedral-surface states near $K$ also persists. The displacement field acts differently on these two sectors. Near $K'$, where the low-energy states have stronger monolayer character, the field mainly changes their relative energy. Near $K$, where the states derive from the two rhombohedral surfaces, it changes the surface detuning and reconstructs the local gap. Removing $\gamma_2$, $\gamma_3$, and $\gamma_4$ suppresses much of the sign asymmetry and many of the additional avoided crossings visible in the full model, but the spectra are not exactly symmetric under $E\rightarrow-E$ because the same-sublattice moir\'e amplitude $u$ remains finite.

Supplementary Fig.~\ref{fig:supp_chiral_bands} shows the strict chiral limit, obtained by additionally setting $u=0$. At $V=0$, the numerical spectra display the exact symmetry implied by Eq.~\eqref{eq:supp_chiral_symmetry}: every state at energy $E$ has a partner at $-E$. The central-band narrowing with increasing $N$ is retained, demonstrating that it follows from coupling the monolayer Dirac cone to the strict $N$th-order rhombohedral surface path rather than from particle--hole-asymmetric hopping processes. A finite displacement field adds diagonal outer-layer potentials and breaks this spectral symmetry, producing the asymmetric top and bottom rows. This interpretation connects the present construction to earlier theories of chiral rhombohedral graphene, ideal chiral twisted bilayers, and general twisted multilayers~\cite{Tarnopolsky2019_origin,Liu2019,Khalaf2019_magic,Xie2022_alternating,Ding2023Mirror,Zhang2022_topoTrans}.

\begin{figure}[!htb]
\centering
\includegraphics[width=0.95\textwidth]{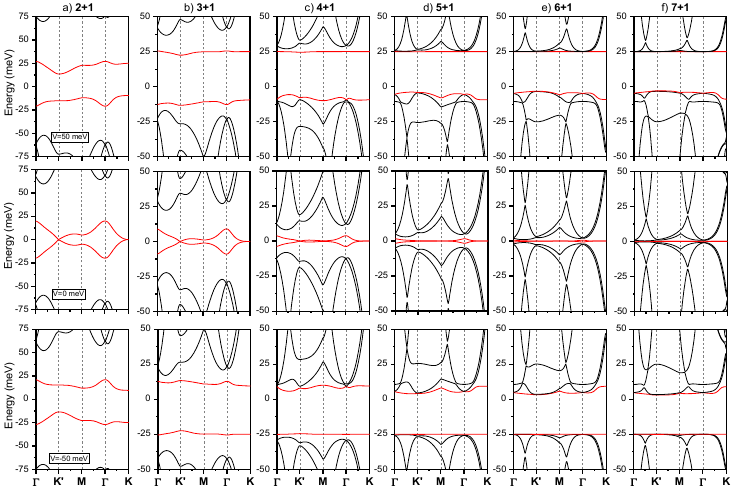}
\caption{\textbf{Electronic structure of the strict chiral model.} Band structure of the strict chiral $(N+1)$-layer rhombohedral graphene model for $N=2$ to $N=7$ at $\theta=1.05^\circ$. The top row corresponds to $V=50$ meV, the middle row to $V=0$, and the bottom row to $V=-50$ meV. Red curves highlight the two central bands. The spectra show the idealized limit in which the central-band narrowing is controlled by the coupling between a monolayer Dirac cone and the strict $N$th-order rhombohedral surface path.}
\label{fig:supp_chiral_bands}
\end{figure}

Taken together, the two figures separate robust from model-dependent features. The central-band narrowing and the distinction between the $K'$ monolayer sector and the $K$ rhombohedral sector survive in both limits. Same-sublattice moir\'e tunnelling $u$ accounts for the residual spectral asymmetry of the reduced model, while $\gamma_2$, $\gamma_3$, and $\gamma_4$ in the full model generate additional warping, particle--hole asymmetry, and remote-band avoided crossings. The strict chiral model isolates the exact symmetry and surface winding; the reduced and full models show how those structures are deformed without changing their low-energy origin.

\section{Layer-hybridization evolution for different layer numbers}\label{sec:S3}

We compare the coupling evolution of the AB-bilayer ($N=2$) with that of the rhombohedral slabs ($N=3,\ldots,6$), demonstrating the persistence of the hybridization mechanism across the family. For an eigenstate of band $n$ at momentum $\boldsymbol{k}$, the weight on layer $L_l$ is
\begin{equation}
W_l^{(n)}(\boldsymbol{k})=\sum_{\alpha=A,B}\sum_{\boldsymbol{G}\in\mathcal{G}}\left|C^{(n)}_{l\alpha,\boldsymbol G}(\boldsymbol{k})\right|^2,
\end{equation}
where $C^{(n)}_{l\alpha,\boldsymbol G}(\boldsymbol{k})$ is the plane-wave coefficient introduced in Eq.~\eqref{eq:supp_plane_wave_expansion} for the normalized eigenvector on layer $l$, sublattice $\alpha$, and moir\'e harmonic $\boldsymbol{G}$. The reciprocal-space set $\mathcal G$ is chosen sufficiently large to converge the displayed bands and layer weights.

\begin{figure}[!htb]
\centering
\includegraphics[width=0.99\textwidth]{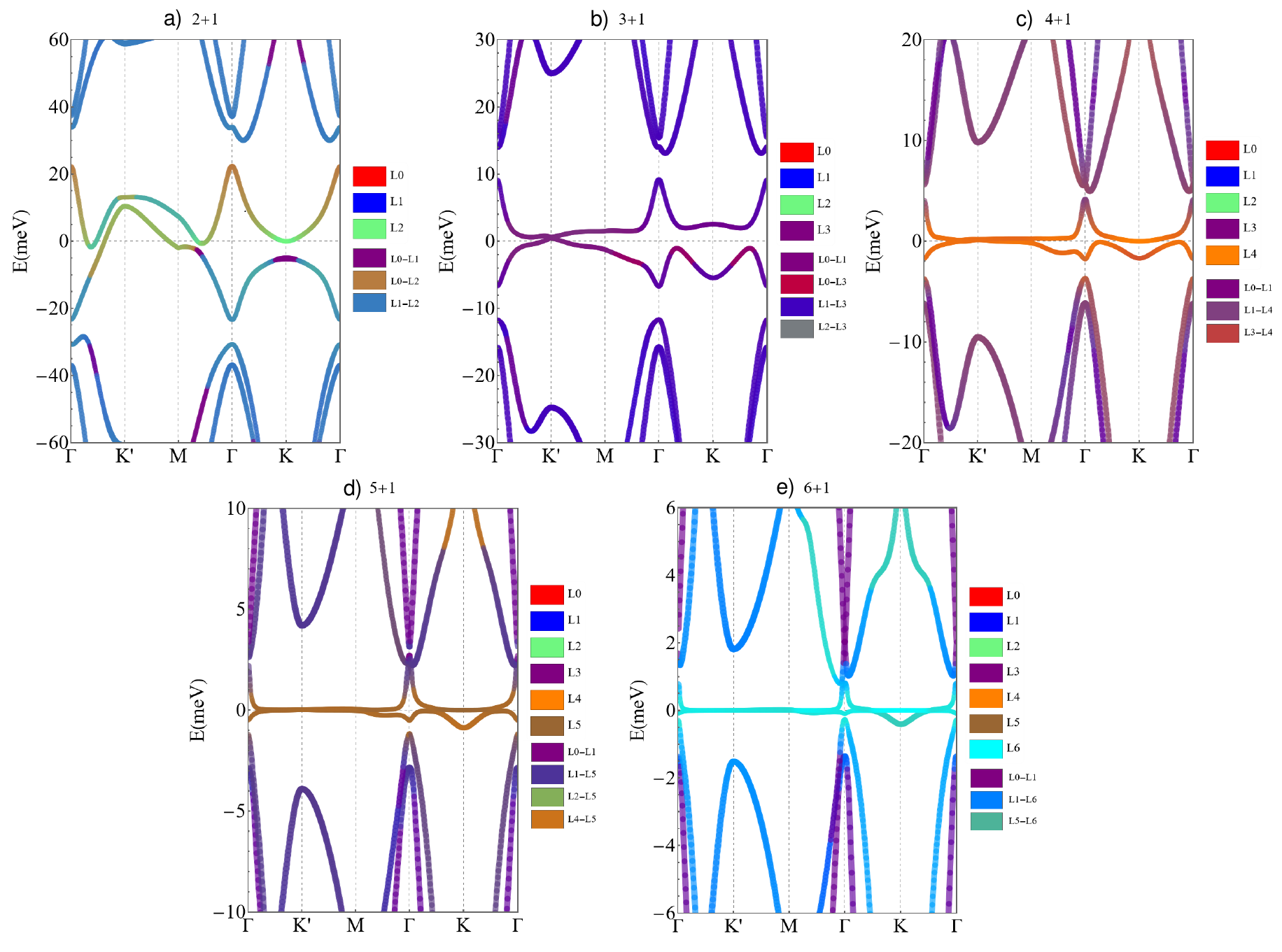}
\caption{\textbf{Layer hybridization across the monolayer--multilayer family.} Coupling evolution for the AB-bilayer  (\textbf{a}, $N=2$) and rhombohedral slabs (\textbf{b--e}, $N=3,\ldots,6$). Colour, thickness, and opacity follow the conventions defined in the Methods section of the main text. The layer-colour convention is indicated in each panel. For the plots we use the projected model described in Sec.~\ref{sec:S2}.}
\label{fig:supp_hybridization_N2}
\end{figure}

Supplementary Fig.~\ref{fig:supp_hybridization_N2} summarizes the evolution for $N=2,\ldots,6$. Colour, line thickness, and opacity encode the layer composition following the convention defined in the Methods section of the main text; a thin grey skeleton indicates the underlying dispersion. Mixed colours identify comparable leading contributions; they do not imply that all remaining layer weights vanish.

The basis-colour order is fixed by layer index and indicated in each panel. For smaller $N$, the sequence is truncated at the last available layer. Because each $N$ is normalized independently, the comparison indicates a clear evidence for the persistence of the same hybridization mechanism. In every case, increasing the moir\'e coupling transfers central-band weight from the rotated monolayer into the rhombohedral sector and ultimately toward the remote surface $L_N$, while retaining momentum-dependent interface--surface mixing.

\section{Layer-resolved band structure under displacement field}\label{sec:S4}

\begin{figure}[!htb]
\centering
\includegraphics[width=0.95\textwidth]{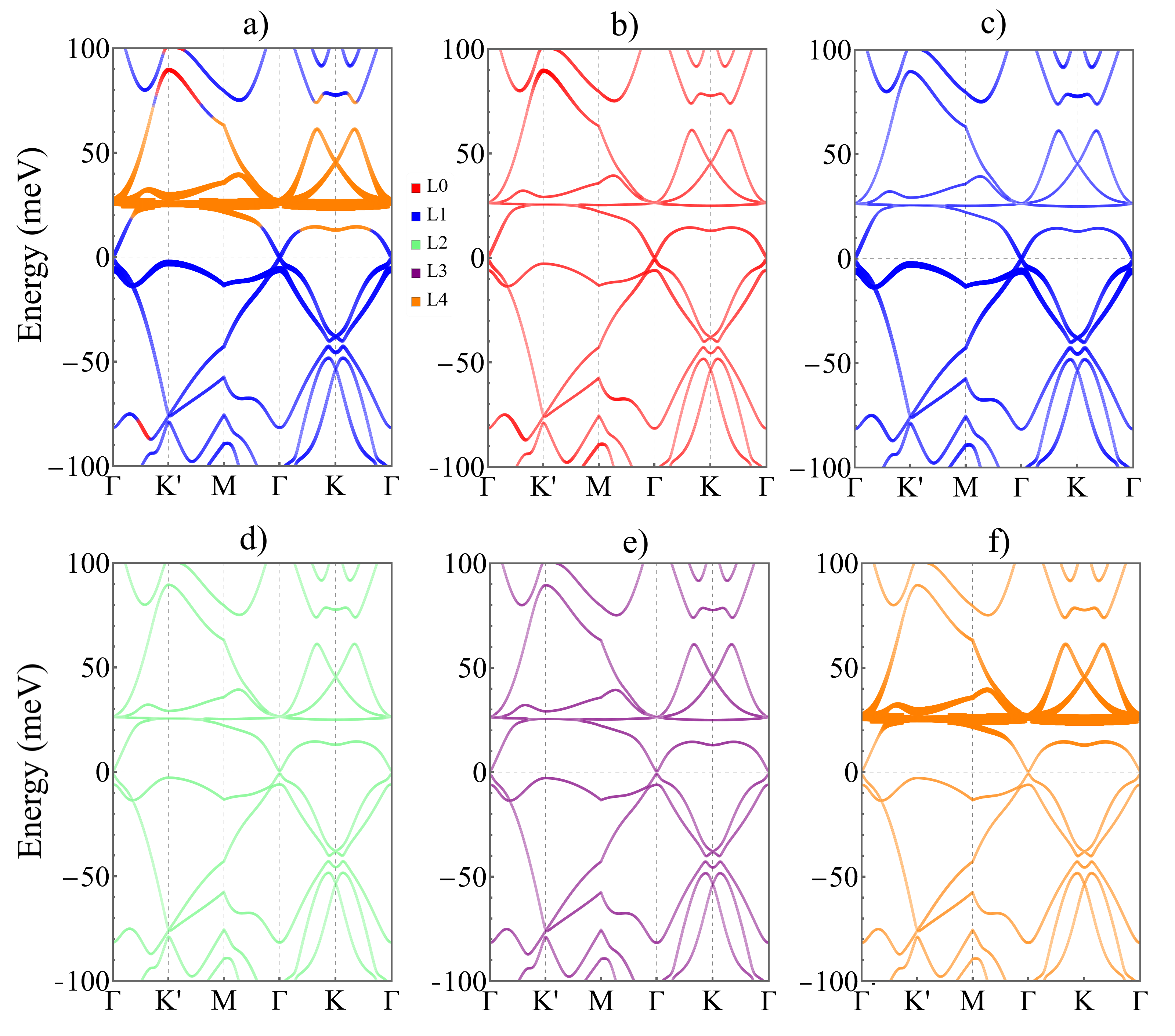}
\caption{\textbf{Layer-resolved band structure for $N=4$ and $V=+50$~meV.} \textbf{a}, Layer-composition-resolved band structure. \textbf{b--f}, Individual layer weights on $L_0$, $L_1$, $L_2$, $L_3$, and $L_4$, respectively. The upper central band and nearby upper remote bands carry strong remote-surface $L_4$ character, while the lower central band has predominantly twisted-interface character with momentum-dependent remote-surface admixture.}
\label{fig:supp_layerproj_vp}
\end{figure}

\begin{figure}[!htb]
\centering
\includegraphics[width=0.95\textwidth]{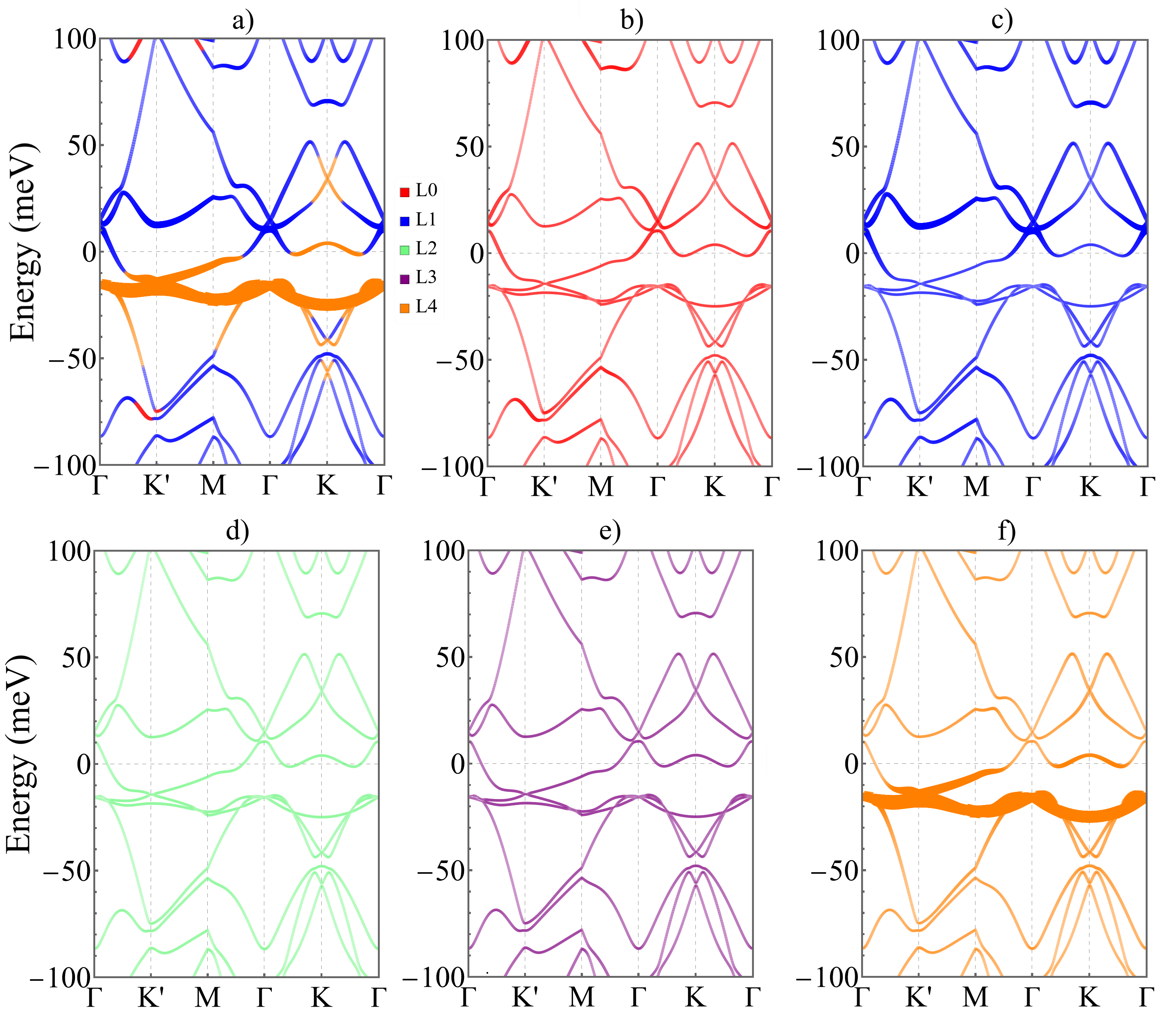}
\caption{\textbf{Layer-resolved band structure for $N=4$ and $V=-50$~meV.} Same representation as in Supplementary Fig.~\ref{fig:supp_layerproj_vp}. Reversing the displacement field reverses the layer polarization: the lower central band and nearby lower bands become dominated by the remote surface $L_4$, while the upper central band and nearby upper bands are concentrated mainly in the twisted-interface sector.}
\label{fig:supp_layerproj_vm}
\end{figure}

Supplementary Figs.~\ref{fig:supp_layerproj_vp} and~\ref{fig:supp_layerproj_vm} provide a detailed layer decomposition of the displacement-field-induced reconstruction for the representative $N=4$ system. Panel \textbf{a} of each figure shows the layer-composition-resolved band structure using the same colour, thickness, and opacity convention as Sec.~\ref{sec:S3}. Panels \textbf{b--f} show the individual layer weights on $L_0,L_1,L_2,L_3$, and $L_4$.

For $V=+50$~meV, the upper central band and nearby upper remote bands carry strong remote-surface $L_4$ character, whereas the lower central band is concentrated mainly in the twisted-interface sector formed by $L_0$ and $L_1$. The layer composition remains momentum dependent, and the nominally interface-dominated band retains remote-surface admixture near selected avoided crossings. For $V=-50$~meV, the layer polarization is reversed: the lower central band and nearby lower bands acquire predominantly remote-surface character, while the upper central and nearby upper bands are concentrated mainly in the twisted-interface sector. These figures provide the momentum-space counterpart of the real-space layer switching discussed in Sec.~\ref{sec:S6}.

\section{Layer-resolved real-space densities}\label{sec:S5}

\begin{figure}[!htb]
\centering
\includegraphics[width=0.90\textwidth]{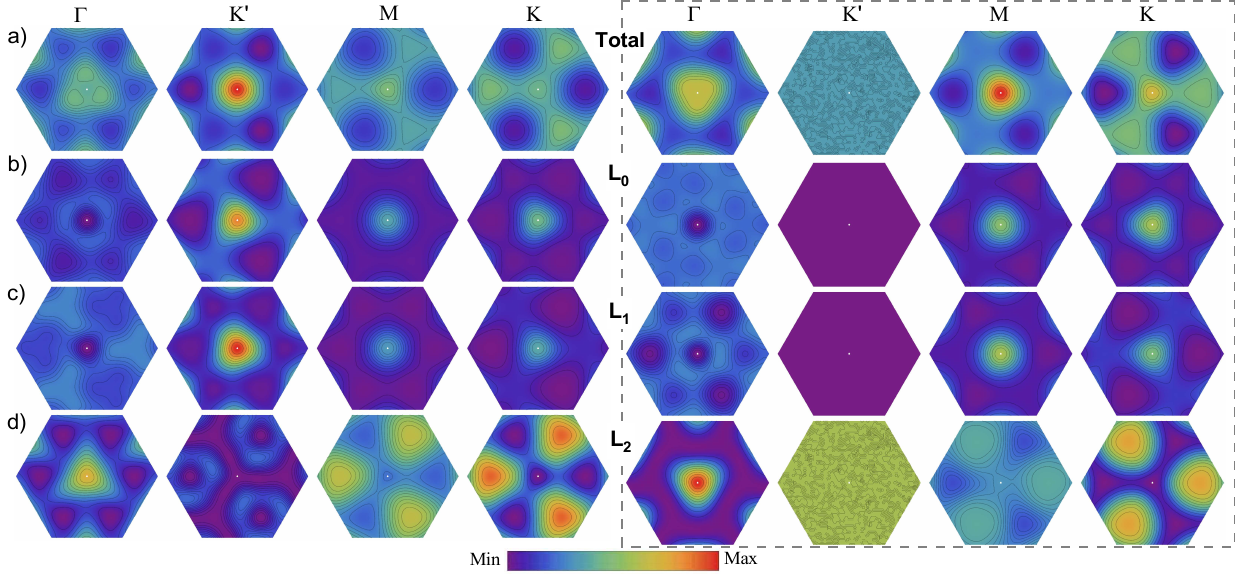}
\caption{\textbf{Layer-resolved real-space density for $N=2$.} Total and layer-resolved densities of the lower and upper central bands evaluated at $\Gamma$, $K'$, $M$, and $K$. The left group corresponds to the lower central band and the right group to the upper central band. $L_0$ denotes the rotated monolayer, while $L_1$ and $L_2$ are the two layers of the AB bilayer. A common colour scale is fixed by the global minimum and maximum of the total density over all displayed momenta.}
\label{fig:supp_layer_N2}
\end{figure}

\begin{figure}[!htb]
\centering
\includegraphics[width=0.90\textwidth]{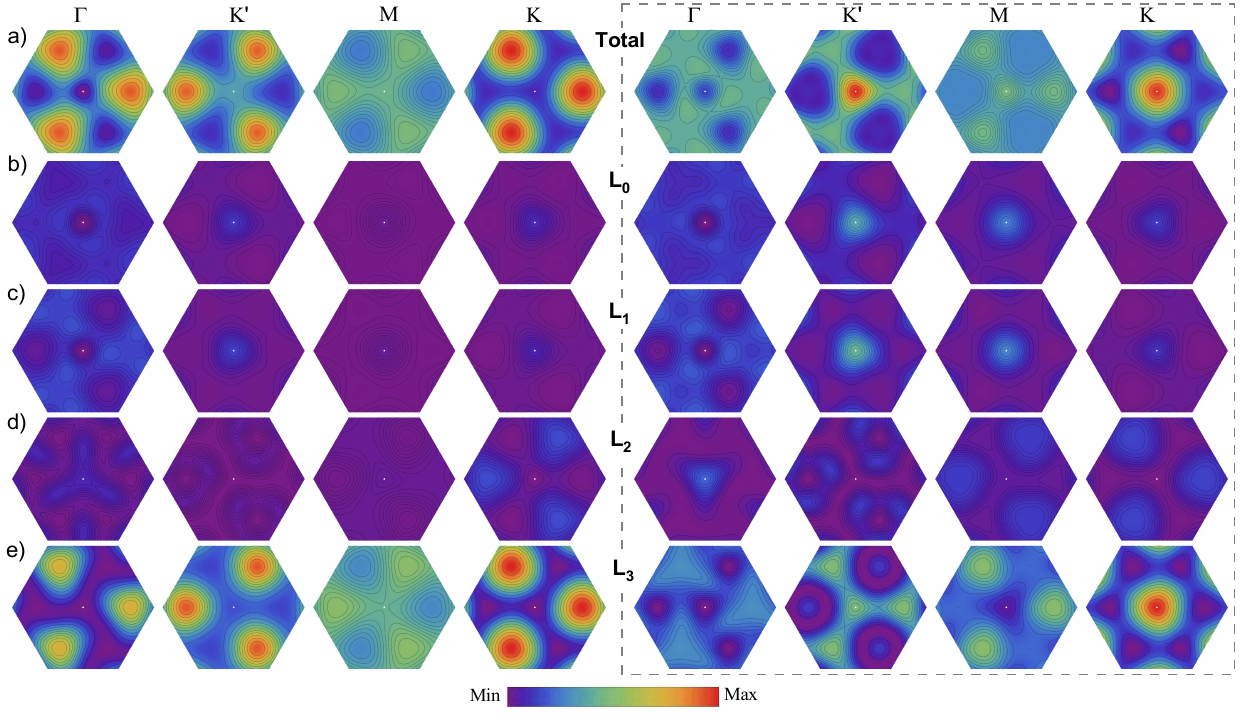}
\caption{\textbf{Layer-resolved real-space density for $N=3$.} Same representation and momentum order as in Supplementary Fig.~\ref{fig:supp_layer_N2}, for the $N=3$ system. The interface layers carry the strongest spatial texture, while the central states also develop substantial remote-surface weight on $L_3$.}
\label{fig:supp_layer_N3}
\end{figure}

\begin{figure}[!htb]
\centering
\includegraphics[width=0.90\textwidth]{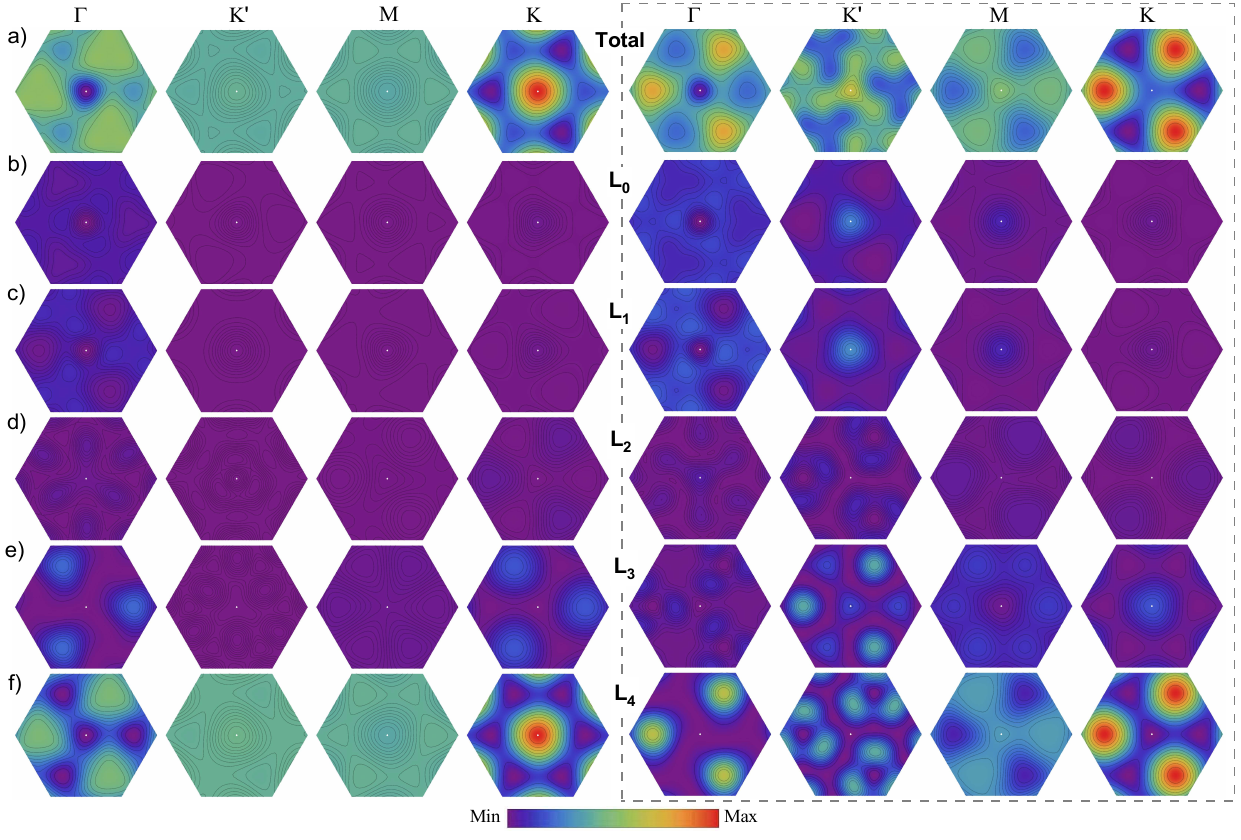}
\caption{\textbf{Layer-resolved real-space density for $N=4$.} Same representation and momentum order as in Supplementary Fig.~\ref{fig:supp_layer_N2}, for the $N=4$ system. The separation between the strongly modulated twisted-interface density and the more weakly modulated remote-surface contribution becomes increasingly clear.}
\label{fig:supp_layer_N4}
\end{figure}

The main text shows the total real-space density across layer number. In Figs.~\ref{fig:supp_layer_N2},~\ref{fig:supp_layer_N3}, and~\ref{fig:supp_layer_N4} we resolve the  zero-field central states into individual layers for $N=2,3,4$. The momenta are ordered as $\Gamma$, $K'$, $M$, and $K$, consistently with the main-text figures, and a common colour normalization is fixed by the global minimum and maximum of the total density over all displayed high-symmetry points. Each map represents the probability density of a central-band state at the indicated momentum. The layer decomposition identifies both its spatial modulation within the moir\'e unit cell and its weight on the interface or remote surface.

The layer decomposition distinguishes spatial modulation from integrated layer weight. The rotated monolayer $L_0$ and the interface layer $L_1$ generally carry the strongest moir\'e texture because they are directly connected by $T_{01}$. The remaining layers respond through the internal ABC propagation. Consequently, a state may have substantial integrated weight on the remote surface $L_N$ while displaying only weak real-space modulation there~\cite{Pantaleon2026Interface}. In the path-projected description, this remote component is carried mainly by $B_N$ and is reached through the dressed amplitude $D_{N-1}$ rather than by direct tunnelling from the rotated monolayer.

Supplementary Fig.~\ref{fig:supp_layer_N2} shows the AB-bilayer for which the interface and remote surface are distinct. The $L_0$--$L_1$ pair carries the strongest triangular texture, while $L_2$ completes the two-surface low-energy sector of the AB bilayer. The system is therefore neither a conventional twisted bilayer nor an isolated AB bilayer, but a hybrid of a directly perturbed interface and an AB-bilayer boundary sector. Supplementary Figs.~\ref{fig:supp_layer_N3} and~\ref{fig:supp_layer_N4} show how this distinction becomes sharper with thickness: the interface modulation remains concentrated on $L_0$ and $L_1$, whereas an appreciable but smoother contribution develops on $L_3$ and $L_4$, respectively. These panels provide the layer-resolved information that is absent from the total-density maps in the main text.

The triangular interface textures are related to those generated by other moir\'e perturbations of multilayer graphene. Rhombohedral graphene on hBN and related superlattice platforms likewise show that a periodic perturbation can reconstruct a multilayer spectrum and generate triangular charge textures~\cite{Pantaleon2021Narrow,Zhang2025_R8GhBN}. Here, however, the periodic perturbation is produced by interfacial monolayer--rhombohedral hybridization rather than by an external hBN potential.

\section{Displacement-field-dependent real-space densities}\label{sec:S6}

\begin{figure}[!htb]
\centering
\includegraphics[width=0.90\textwidth]{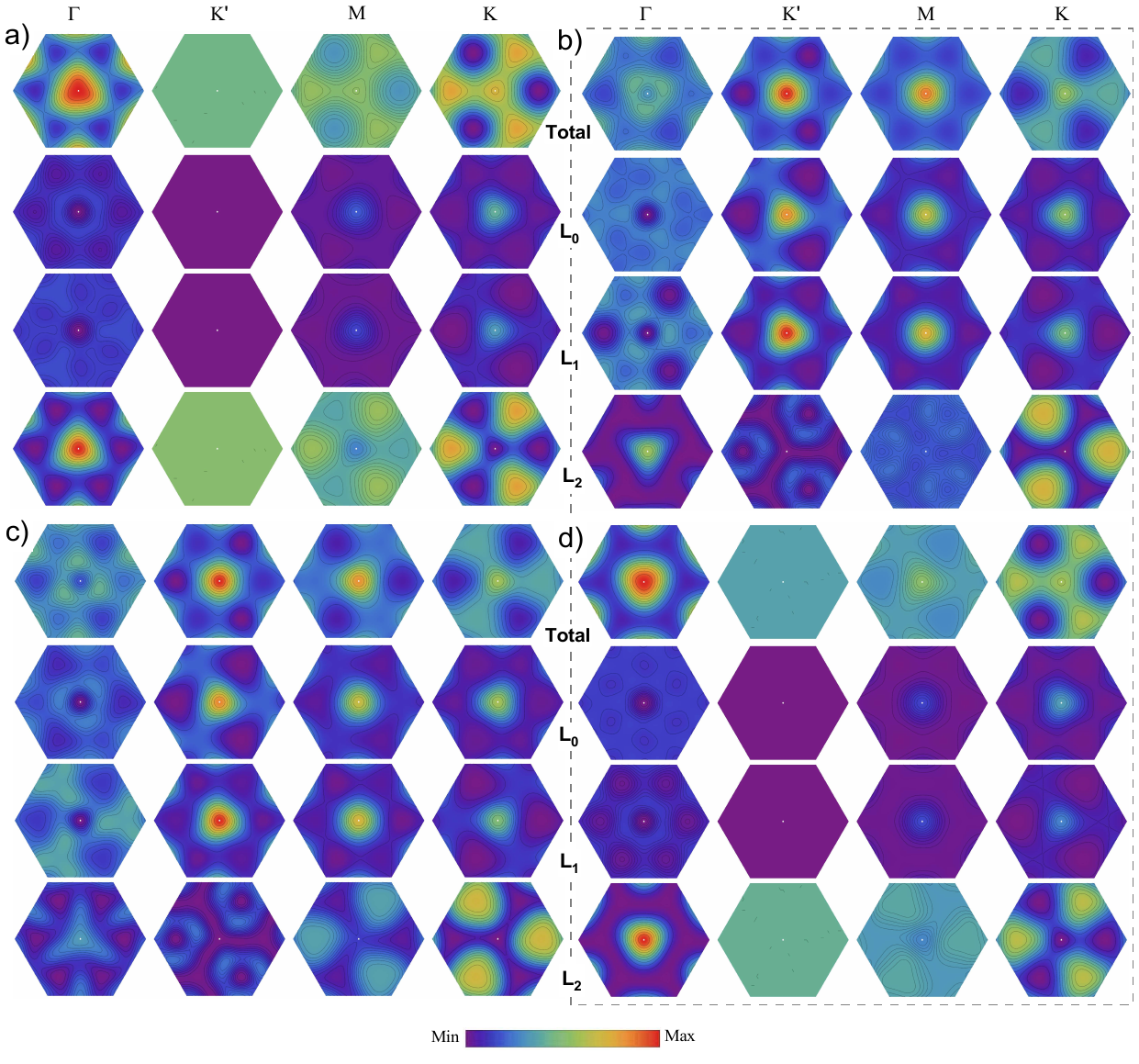}
\caption{\textbf{Displacement-field dependence of the layer-resolved density for $N=2$.} Total and layer-resolved real-space densities evaluated at $\Gamma$, $K'$, $M$, and $K$. The left and right groups correspond to the lower and upper central bands, respectively. The panels compare $V=-50$~meV and $V=+50$~meV. A common colour scale is fixed by the global minimum and maximum of the total density over all displayed momenta. Reversing $V$ exchanges the interface and remote-surface character of the two central bands.}
\label{fig:supp_bias_N2}
\end{figure}

\begin{figure}[!htb]
\centering
\includegraphics[width=0.95\textwidth]{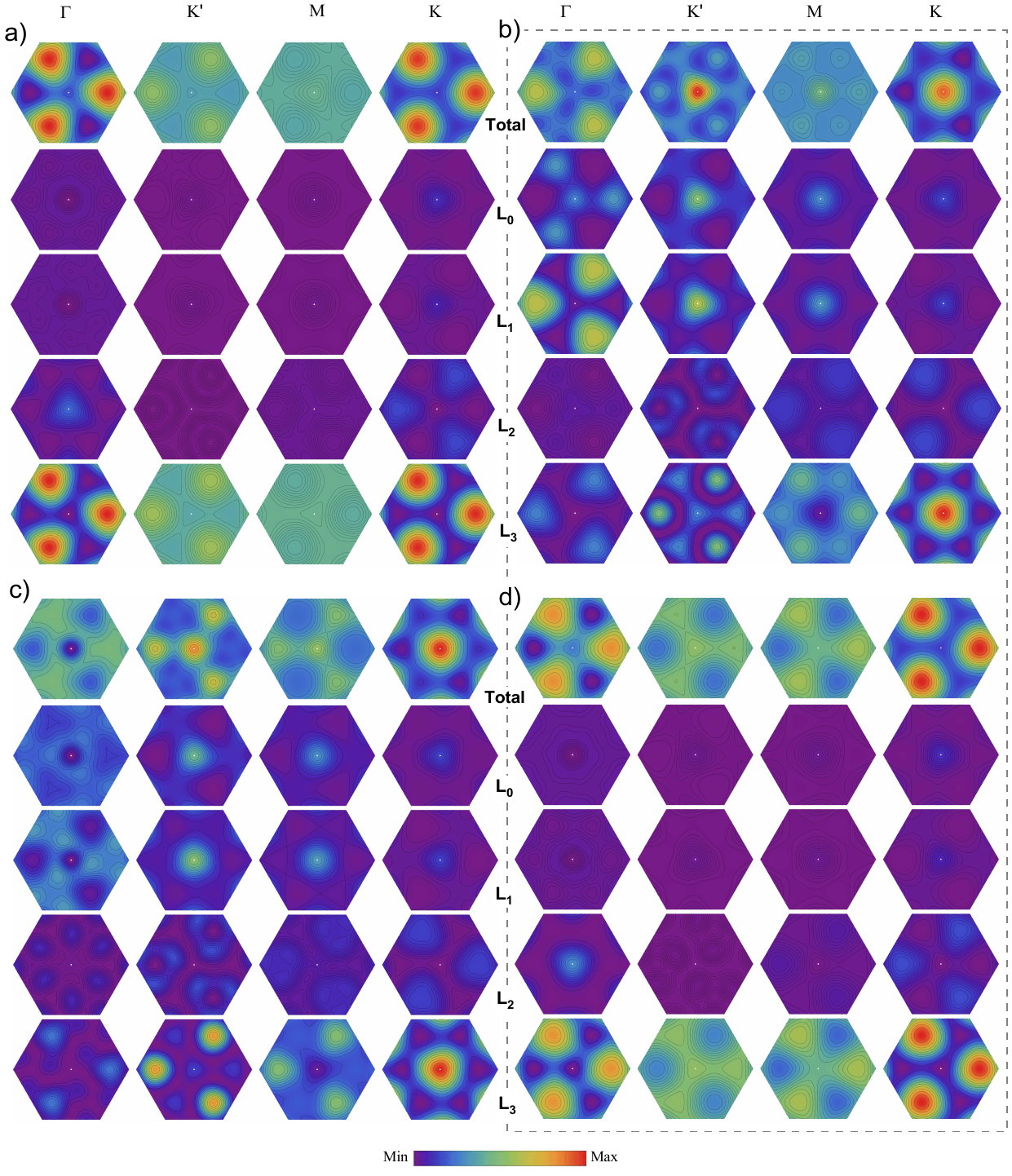}
\caption{\textbf{Displacement-field dependence of the layer-resolved density for $N=3$.} Total and layer-resolved densities at $\Gamma$, $K'$, $M$, and $K$ for the lower and upper central bands at $V=-50$~meV and $V=+50$~meV. At negative field, the lower band carries the dominant remote-surface $L_3$ contribution; at positive field, this contribution is transferred to the upper band. A common colour scale, fixed by the global minimum and maximum of the total density over all displayed high-symmetry points, is used for all maps.}
\label{fig:bias_density_N3}
\end{figure}

\begin{figure}[!htb]
\centering
\includegraphics[width=0.90\textwidth]{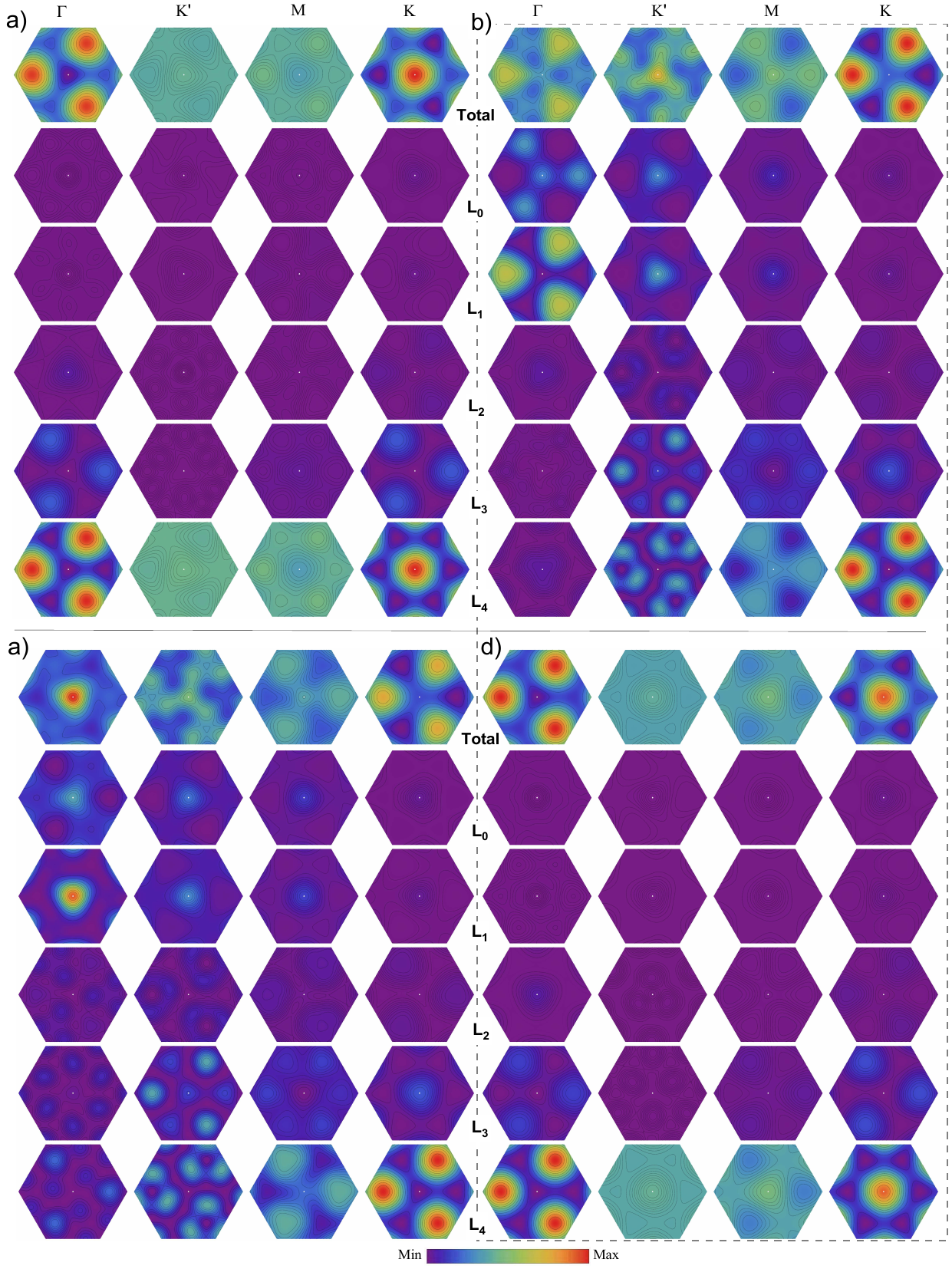}
\caption{\textbf{Displacement-field dependence of the layer-resolved density for $N=4$.} Same representation and momentum order as in Supplementary Fig.~\ref{fig:supp_bias_N2}. Field reversal transfers the dominant remote-surface $L_4$ contribution between the lower and upper central bands, while momentum-dependent interface--surface mixing remains visible.}
\label{fig:supp_bias_N4}
\end{figure}

Supplementary Figs.~\ref{fig:supp_bias_N2},~\ref{fig:bias_density_N3}, and~\ref{fig:supp_bias_N4} show the displacement-field-dependent layer-resolved densities for $N=2,3,4$, respectively. In every case, $V=-50$~meV places the dominant remote-surface character in the lower central band, while $V=+50$~meV transfers it to the upper central band. The complementary central band remains more strongly concentrated in the twisted-interface sector formed by $L_0$ and $L_1$. Residual weight on the opposite side of the stack persists at selected momenta because the central states remain coherently hybridized rather than becoming pure interface and remote-surface.

For $N=2$ [Supplementary Fig.~\ref{fig:supp_bias_N2}], field reversal exchanges the contribution of the remote bilayer surface $L_2$ between the two central bands, while the directly coupled $L_0$--$L_1$ interface retains the strongest spatial modulation. For $N=3$ [Supplementary Fig.~\ref{fig:bias_density_N3}], the lower band at $V=-50$~meV is concentrated predominantly on $L_3$, with a smaller contribution from $L_2$. The upper band remains distributed over several layers and displays stronger structure, particularly at the momenta where several moir\'e harmonics interfere. At $V=+50$~meV, the dominant $L_3$ contribution moves to the upper central band, while the lower band becomes more strongly associated with the twisted-interface region. Supplementary Fig.~\ref{fig:supp_bias_N4} shows the same exchange for $N=4$, with the remote contribution carried by $L_4$ and with momentum-dependent admixture retained near avoided crossings.

These real-space maps complement the layer-composition-resolved band structures of Sec.~\ref{sec:S4}. The band structures measure integrated layer weights along the high-symmetry path, whereas the density maps additionally reveal interference among the moir\'e harmonics within each layer. The two diagnostics therefore answer different questions: the first identifies where the probability is located across the stack, and the second shows how coherently mixed that probability is inside the moir\'e unit cell. Together they establish that reversing $V$ exchanges the dominant remote-surface character of the two central bands without completely eliminating interface--surface hybridization.

\section{Chern-number calculation}\label{sec:S7}

Here we specify the numerical implementation of the topological classification used in the main-text Chern maps. The defining expressions for the Abelian Berry curvature of an isolated band, the Wilson loops of the isolated two-band subspace, and the resulting Chern numbers are given in the Methods section of the main text. Numerically, the moir\'e Brillouin zone is discretized into a mesh consistent with the triangular moir\'e lattice. At each momentum point, the eigenvectors of the selected central band are computed; when the two central bands touch while remaining isolated from the remote spectrum, both orthonormal eigenvectors are retained as a common subspace. The overlap matrices between neighbouring momenta are unitarized by singular-value decomposition, and the Berry flux of each triangular plaquette is obtained from the argument of the determinant of the ordered product of its three link matrices. The Chern number is the sum of the plaquette fluxes divided by $2\pi$, following the gauge-invariant lattice formulation of Ref.~\cite{Fukui2005} and its non-Abelian extension.

The coloured regions in the Chern maps correspond to parameter points where the selected central band is isolated from both the other central band and remote bands. Hatched regions correspond to cases where the two central bands touch each other but remain isolated from remote bands; these regions are assigned the Chern number of the isolated two-band subspace calculated using the non-Abelian Berry connection. Black regions denote parameter points where the central-band sector is not isolated from remote bands.

\section{Topological phases in the upper central band}\label{sec:S8}

\begin{figure}[!htb]
\centering
\includegraphics[width=0.95\textwidth]{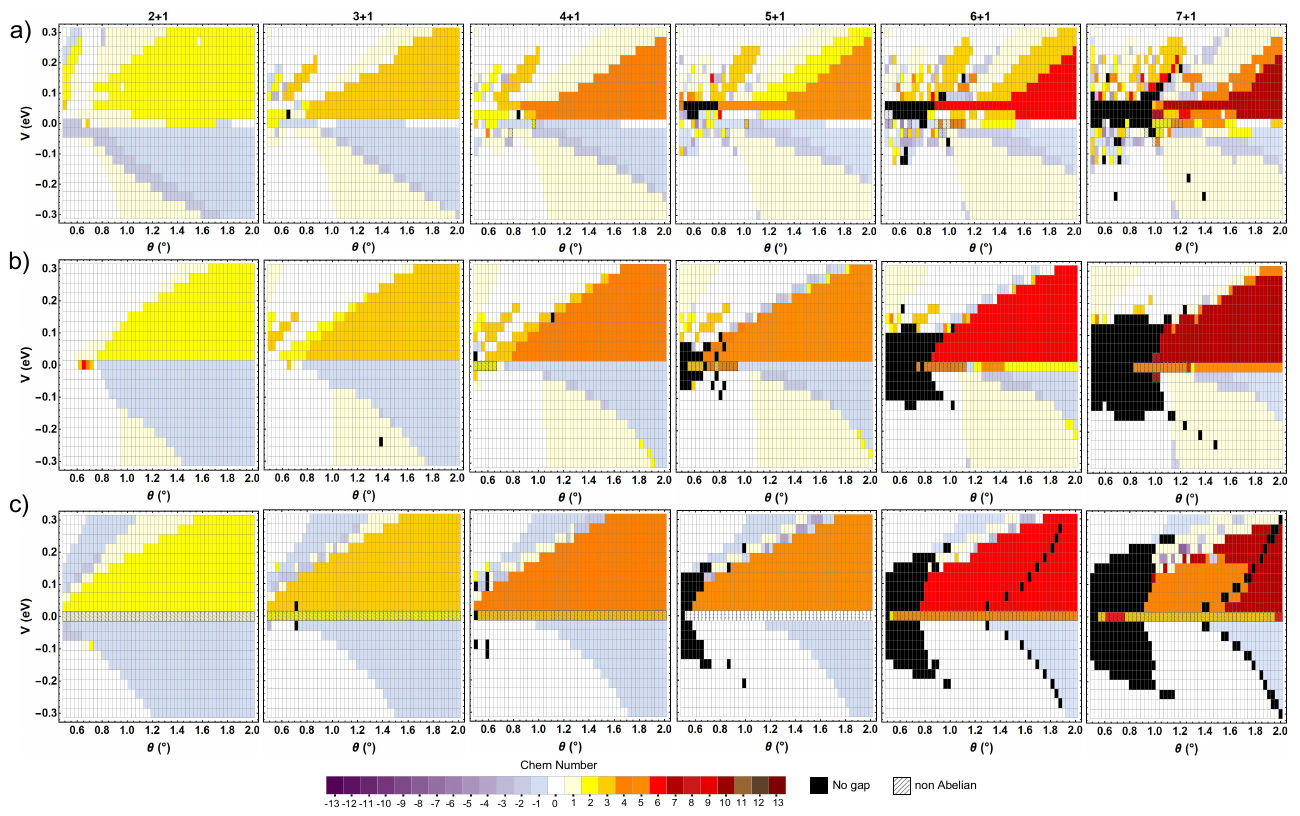}
\caption{\textbf{Chern-number phase diagram for the upper central band.} Comparison between the full, reduced, and strict chiral models for $(N+1)$-layer rhombohedral graphene from $N=2$ to $N=7$. The Chern number is encoded in the colour map. Coloured regions indicate parameter points where the upper central band is isolated and assigned an Abelian Chern number. Hatched regions indicate that the two central bands touch but remain isolated from the remote spectrum, so the Chern number is assigned to the isolated two-band subspace using the non-Abelian Berry connection, and black regions correspond to parameter regimes where the central bands are not isolated from remote bands.}
\label{fig:supp_chern_upper}
\end{figure}

Supplementary Fig.~\ref{fig:supp_chern_upper} shows the Chern-number phase diagrams for the upper central band. These results complement the lower-band maps discussed in the main text. In the large-angle regime, where the monolayer and rhombohedral sectors are only weakly hybridized, the upper central band carries the topological character opposite to that of the lower central band: the displacement field exchanges the monolayer-like and rhombohedral-like sectors between the two bands. Thus both central bands are organized by the exchange between the unit-winding monolayer sector and the rhombohedral sector that is adiabatically connected to the strict $N$th-order chiral surface state. At smaller twist angles, the upper-band phase diagrams show the same qualitative trends found for the lower band: stronger moir\'e hybridization, reduced isolation from remote bands, and additional band inversions produce fragmented Chern regions, two-band regions requiring a non-Abelian Berry-connection treatment, and black regions where the central-band sector is no longer isolated.

The comparison between the full, reduced, and strict chiral models again shows that the strict chiral limit provides the organizing reference, while the path amplitude $X_N$ explains how $\gamma_2$ and $\gamma_3$ deform that surface sector and the full hopping parametrization introduces additional particle-hole asymmetry and phase-boundary complexity. Our upper- and lower-band maps connect the present theory to recent experiments in which different twisted rhombohedral configurations show high-Chern or interaction-reconstructed topological phases controlled by layer number and displacement field~\cite{Wang2025_TDTG,Liu2025_TwistMonoPenta,Wang2026High,Su2025,Dong2025_TwistBiTri,chen2026layer}.

\section{Robustness to the hopping parametrization and the displacement-field profile}\label{sec:S9}

The results in the main text use a single set of rhombohedral hopping parameters and the outer-layer potential profile of Eq.~\eqref{eq:supp_bias}. To separate the robust features from those that depend on this microscopic input, we repeat the full-model calculations with two independent modifications, keeping the moir\'e tunnelling amplitudes $u$ and $u'$, the plane-wave truncation, and the numerical procedure of Sec.~\ref{sec:S7} unchanged. We replace the parameters of Ref.~\cite{Zhang2010Band} by those used for rhombohedral multilayers in Ref.~\cite{ParraMartinez2025}, which were adopted in Ref.~\cite{Zhou2021} to describe rhombohedral trilayer graphene:
\begin{equation}
\gamma_0=3.10~\mathrm{eV},\quad \gamma_1=0.380~\mathrm{eV},\quad \gamma_2=-0.015~\mathrm{eV},\quad \gamma_3=-0.290~\mathrm{eV},\quad \gamma_4=-0.141~\mathrm{eV}.
\label{eq:supp_params_alt}
\end{equation}
The signs of $\gamma_3$ and $\gamma_4$ correspond to the matrix convention of Eq.~\eqref{eq:supp_U}, which coincides with that of Ref.~\cite{Zhou2021}. We also replace Eq.~\eqref{eq:supp_bias} by a uniform field, for which the potential drops linearly across all $N+1$ layers,
\begin{equation}
\Delta^{(l)}=-\frac{V}{2}+\frac{l}{N}\,V,\qquad l=0,\dots,N.
\label{eq:supp_bias_linear}
\end{equation}
Equation~\eqref{eq:supp_bias} corresponds to complete screening of the field inside the rhombohedral slab, whereas Eq.~\eqref{eq:supp_bias_linear} corresponds to no screening. A self-consistent profile lies between these two limits~\cite{Koshino2010Interlayer}. Both profiles assign the same potential difference $V$ between the rotated monolayer $L_0$ and the remote surface $L_N$, so the corresponding maps can be compared directly at equal $V$.

Using the different parametrization, in supplementary Fig.~\ref{fig:supp_robust_chern} we compare the Chern maps of the lower (\textbf{a}) and upper (\textbf{b}) central bands for $N=4$ obtained with the main-text model (left column), with the alternative hopping parameters (central column), and with the linear profile (right column). In all three cases, the large-angle region has the same organization. For one sign of $V$, the central band belongs to the rhombohedral sector and carries $\mathcal C=N=4$; for the opposite sign, it belongs to the monolayer-like sector with $|\mathcal C|=1$. Reversing $V$ exchanges these sectors between the lower and upper central bands. This is the behaviour expected from the mechanism of the main text: as long as the selected band remains isolated, its Chern number cannot change under a continuous deformation of the parameters, so the large-angle sectors persist.

These modifications, however, may shift phase boundaries quantitatively and rearrange the small-angle region. This is consistent with the regime of validity discussed in Sec.~\ref{sec:S1}: internal-layer potentials and further-neighbour hoppings are important mainly where the central states approach and mix with the remote spectrum.

\begin{figure}[!htb]
\centering
\includegraphics[width=\textwidth]{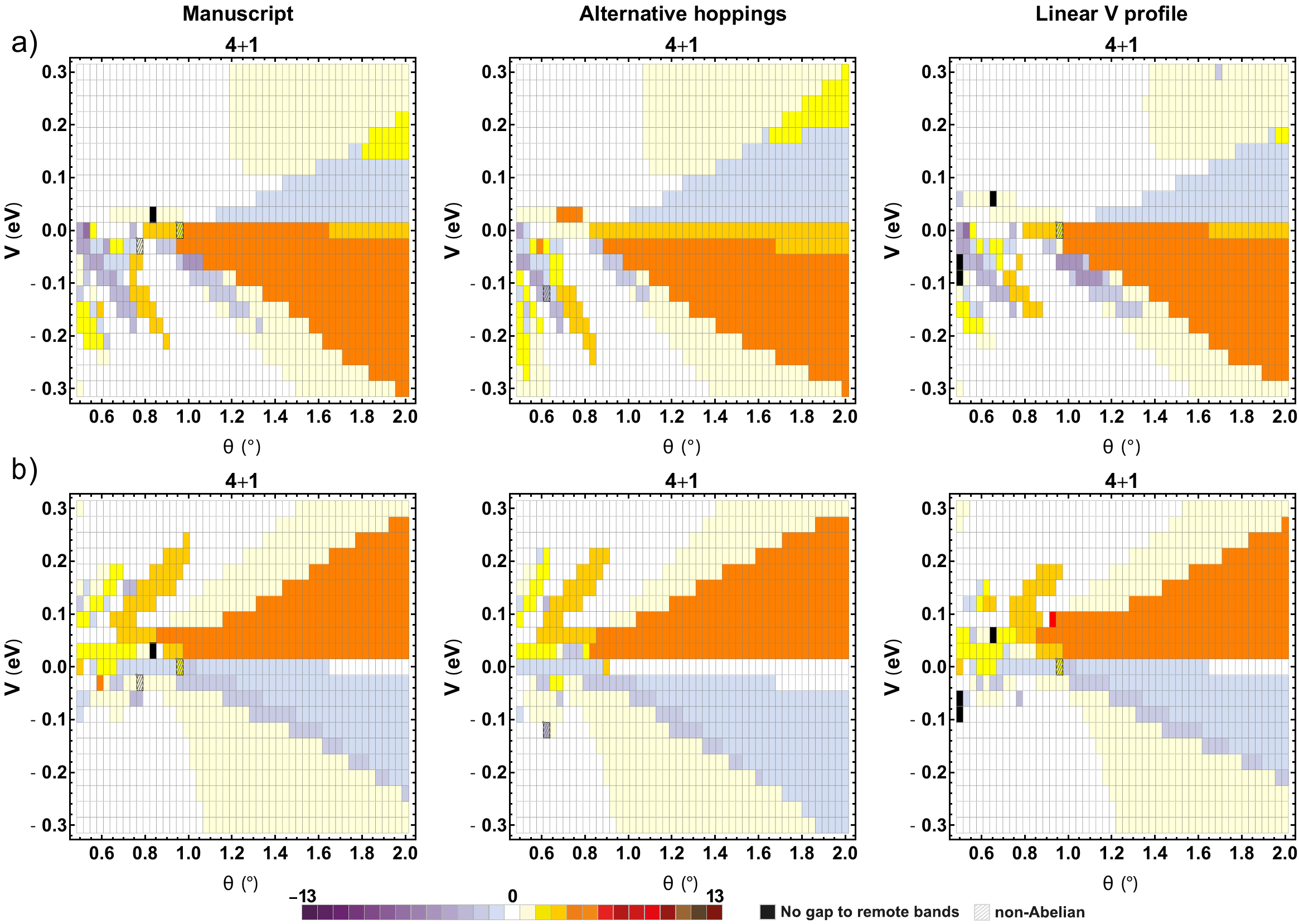}
\caption{\textbf{Robustness of the Chern-number maps for $N=4$.} Full-model Chern maps for the lower (\textbf{a}) and upper (\textbf{b}) central bands. Left column (Manuscript): main-text hopping parameters and outer-layer potential profile, Eq.~\eqref{eq:supp_bias}. Central column (Alternative hoppings): alternative hopping parameters of Eq.~\eqref{eq:supp_params_alt}~\cite{ParraMartinez2025,Zhou2021} with the outer-layer profile. Right column (Linear $V$ profile): main-text hopping parameters with the linear profile of Eq.~\eqref{eq:supp_bias_linear}. The moir\'e tunnelling amplitudes are the same in all panels. Colours, hatching, and black cells follow the conventions of Sec.~\ref{sec:S7}.}
\label{fig:supp_robust_chern}
\end{figure}

\end{document}